\documentclass[superscriptaddress,prc,preprint,onecolumn,showpacs,preprintnumbers,amsmath,amssymb,bibnotes,secnumarabic,altaffilletter,floatfix,aps]{revtex4-1}

\usepackage{graphicx}
\usepackage{longtable}
\usepackage{natbib}
\usepackage{dcolumn}

\newcommand{\hto}{H$_2$O}
\newcommand{\dto}{D$_2$O}

\newcommand{\NS} {\chem{16}{N}}

\newcommand{\chem}[2]{\mbox{$\rm ^{#1}{#2}$}}
\newcommand{\xvec}[1]{\ensuremath{{\vec{r}_{#1}}}}
\newcommand{\tind}[1]{\ensuremath{{t_{#1}}}}
\newcommand{\xfit}{\xvec{\mathrm{fit}}}
\newcommand{\xpmt}{\xvec{\mathrm{PMT}}}
\newcommand{\tfit}{\tind{\mathrm{fit}}}
\newcommand{\tpmt}{\tind{\mathrm{PMT}}}
\newcommand{\tres}{\tind{\mathrm{res}}}

\newcommand{\ddrift}{\ensuremath{\delta_\text{drift}}}

\newcommand{\fecal}{\ensuremath{\mathcal{F}_T}}
\newcommand{\hw}{D\ensuremath{_2}O}
\newcommand{\ith}{\ensuremath{i^{\text{\tiny{th}}}}}
\newcommand{\lw}{H\ensuremath{_2}O}
\newcommand{\naught}{\ensuremath{_\circ}}
\newcommand{\neff}{\ensuremath{N_{\text{eff}}}}
\newcommand{\ngamma}{\ensuremath{N_\gamma}}
\newcommand{\nh}{\ensuremath{\hat{n}}}
\newcommand{\nhit}{\ensuremath{N_{\text{{hit}}}}}
\newcommand{\npmts}{\ensuremath{N_{\text{PMTs}}}}
\newcommand{\npred}{\ensuremath{N_{\text{predicted}}}}

\newcommand{\nst}{\ensuremath{^{16}}N}
\newcommand{\pv}{\ensuremath{\vec{p}}}

\newcommand{\rv}{\ensuremath{\vec{r}}}
\newcommand{\te}{\ensuremath{T_{\text{e}}}}
\newcommand{\teff}{\ensuremath{T_{\text{eff}}}}
\newcommand{\rfit}{\ensuremath{R_{\text{fit}}}}
\newcommand{\rav}{\ensuremath{R_{\text{AV}}}}
\newcommand{\rrfit}{\vec{r}_{\text{fit}}}
\newcommand{\uh}{\ensuremath{\hat{u}}}
\newcommand{\uhfit}{\ensuremath{\hat{u}_{\text{fit}}}}
\newcommand{\epelec}{\ensuremath{E_i^{\text{electronic}}}}

\newcommand{\epncd}{\ensuremath{\epsilon^{\text{NCD}}}}
\newcommand{\epnot}{\ensuremath{\epsilon_{\textrm{PCE}}}}
\newcommand{\epopt}{\ensuremath{E_i^{\text{optical}}}}
\newcommand{\snodmsquared}{7.59^{+0.19}_{-0.21}\times 10^{-5}\mbox{ eV}^2}
\newcommand{\snothetaonetwo}{34.4^{+1.3}_{-1.2}\mbox{ degrees}}
\newcommand{\snoccfluxunc}{1.67^{+0.05}_{-0.04}\mbox{(stat.)}^{+0.07}_{-0.08}\mbox{(syst.)}} 
\newcommand{\snoesfluxunc}{1.77^{+0.24}_{-0.21}\mbox{(stat.)}^{+0.09}_{-0.10}~\mbox{(syst.)}} 
\newcommand{\snoncfluxunc}{5.54^{+0.33}_{-0.31}~\mbox{(stat.)}^{+0.36}_{-0.34}~\mbox{(syst.)}} 
\newcommand{\snoccncratio}{0.301\pm0.033~\mbox{(total)}}

\newcommand{\alta}{Dept.\,of Physics, University of Alberta, Edmonton, Alberta, T6G 2R3, Canada}
\newcommand{\ubc}{Dept.\,of Physics and Astronomy, University of British Columbia, Vancouver, BC V6T 1Z1, Canada}
\newcommand{\bnl}{Chemistry Dept., Brookhaven National Laboratory,  Upton, NY 11973-5000}
\newcommand{\carleton}{Ottawa-Carleton Institute for Physics, Dept.\, of Physics, Carleton University, Ottawa, Ontario K1S 5B6, Canada}
\newcommand{\uog}{Physics Dept., University of Guelph, Guelph, Ontario N1G 2W1, Canada}
\newcommand{\lu}{Dept.\,of Physics and Astronomy, Laurentian University, Sudbury, Ontario P3E 2C6, Canada}
\newcommand{\lbnl}{Institute for Nuclear and Particle Astrophysics and Nuclear Science Division, Lawrence Berkeley National Laboratory, Berkeley, CA 94720}
\newcommand{\lanl}{Los Alamos National Laboratory, Los Alamos, NM 87545}

\newcommand{\oxford}{Dept.\,of Physics, University of Oxford, Denys Wilkinson Building, Keble Road, Oxford OX1 3RH, UK}
\newcommand{\penn}{Dept.\,of Physics and Astronomy, University of Pennsylvania, Philadelphia, PA 19104-6396}
\newcommand{\queens}{Dept.\,of Physics, Queen's University, Kingston, Ontario K7L 3N6, Canada}
\newcommand{\uw}{Center for Experimental Nuclear Physics and Astrophysics, and Dept.\,of Physics, University of Washington, Seattle, WA 98195}
\newcommand{\uta}{Dept.\,of Physics, University of Texas at Austin, Austin, TX 78712-0264}
\newcommand{\triumf}{TRIUMF, 4004 Wesbrook Mall, Vancouver, BC V6T 2A3, Canada}

\newcommand{\snoi}{SNOLAB, Sudbury, ON P3Y 1M3, Canada}
\newcommand{\lbla}{Lawrence Berkeley National Laboratory, Berkeley, CA}

\newcommand{\hu}{Dept.\,of Physics, Hiroshima University, Hiroshima, Japan}

\newcommand{\cwru}{Dept.\,of Physics, Case Western Reserve University, Cleveland, OH}
\newcommand{\pnnl}{Pacific Northwest National Laboratory, Richland, WA}

\newcommand{\mitt}{Laboratory for Nuclear Science, Massachusetts Institute of Technology, Cambridge, MA 02139} 

\newcommand{	\lsu	}{Dept.\,of Physics and Astronomy, Louisiana State University, Baton Rouge, LA 70803}

\newcommand{\suss}{ Dept.\,of Physics and Astronomy, University of Sussex, Brighton, UK} 
\newcommand{	\lifep	}{Laborat\'{o}rio de Instrumenta\c{c}\~{a}o e F\'{\i}sica Experimental de Part\'{\i}culas, Av. Elias Garcia 14, 1$^{\circ}$,  Lisboa, Portugal} 

\newcommand{\princeton}{Dept.\,of Physics, Princeton University, Princeton, NJ} 

\newcommand{\liverpool}{Dept.\,of Physics, University of Liverpool, Liverpool, UK}

\newcommand{\uwisc}{Dept.\,of Physics, University of Wisconsin, Madison, WI}

\newcommand{\unc}{Dept.\,of Physics, University of North Carolina, Chapel Hill, NC}
\newcommand{\dresden}{Institut f\"{u}r Kern- und Teilchenphysik, Technische Universit\"{a}t Dresden,   Dresden, Germany} 

\newcommand{\qmul}{Dept.\,of Physics, Queen Mary University, London, UK}
\newcommand{\ucsb}{Dept.\,of Physics, University of California, Santa Barbara, CA}
\newcommand{\cern}{CERN, Geneva, Switzerland}

\newcommand{\casa}{Center for Astrophysics and Space Astronomy, University of Colorado, Boulder, CO}  
\newcommand{\aasu}{Dept.\,of Chemistry and Physics, Armstrong Atlantic State University, Savannah, GA}
\newcommand{\susel}{Sanford Laboratory at Homestake, Lead, SD}  
\newcommand{\queensa}{Dept.\,of Physics, Queen's University, Kingston, Ontario, Canada} 
\newcommand{\ntu}{Center of Cosmology and Particle Astrophysics, National Taiwan University, Taiwan}
\newcommand{\berlin}{Institute for Space Sciences, Freie Universit\"{a}t Berlin, Leibniz-Institute of Freshwater Ecology and Inland Fisheries, Germany}
\newcommand{\bhsu}{Black Hills State University, Spearfish, SD} 

\begin{document}

\affiliation{\alta}
\affiliation{\ubc}
\affiliation{\bnl}
\affiliation{\carleton}
\affiliation{\uog}
\affiliation{\lu}
\affiliation{\lbnl}
\affiliation{\lifep}
\affiliation{\lanl}
\affiliation{\lsu}
\affiliation{\mitt}
\affiliation{\oxford}
\affiliation{\penn}
\affiliation{\queens}
\affiliation{\snoi}
\affiliation{\uta}
\affiliation{\triumf}
\affiliation{\uw}

\author{B.~Aharmim}\affiliation{\lu}
\author{S.\,N.~Ahmed}\affiliation{\queens}
\author{J.\,F.~Amsbaugh}\affiliation{\uw}
\author{J.\,M.~Anaya}\affiliation{\lanl}
\author{A.\,E.~Anthony}\altaffiliation{Present address: \casa}\affiliation{\uta}
\author{J.~Banar}\affiliation{\lanl}
\author{N.~Barros}\affiliation{\lifep}\affiliation{\dresden}
\author{E.\,W.~Beier}\affiliation{\penn}
\author{A.~Bellerive}\affiliation{\carleton}
\author{B.~Beltran}\affiliation{\alta}
% ...
			
\collaboration{SNO Collaboration}\noaffiliation

\title{Measurement of the $\nu_e$ and Total $^{8}$B Solar Neutrino Fluxes with
the Sudbury Neutrino Observatory Phase-III Data Set}


\begin{abstract}

This paper details the solar neutrino analysis of the 385.17-day Phase-III data set acquired by the Sudbury Neutrino Observatory (SNO).   An array of $^3$He proportional counters was installed in the heavy-water target to measure precisely the rate of neutrino-deuteron neutral-current interactions.  This technique to determine the total active $^8$B solar neutrino flux was largely independent of the methods employed in previous phases.  The total flux of active neutrinos was measured to be $\snoncfluxunc$ $\times~10^{6}$~cm$^{-2}$~s$^{-1}$, consistent with previous measurements and standard solar models.  A global analysis of solar and reactor neutrino mixing parameters yielded the best-fit values of  $\Delta m^2 = \snodmsquared$ and $\theta = \snothetaonetwo$.

\end{abstract}


\maketitle


\section{Introduction\label{sec:intro}}

The Sudbury Neutrino Observatory (SNO) experiment~\cite{NIM} has firmly established that electron-type neutrinos ($\nu_e$) produced in the solar core transform into other active flavors while in transit to the Earth~\cite{snocc,snonc,snodn,longd2o,nsp,ncdprl}.  This direct observation of neutrino flavor transformation, through the simultaneous observation of the disappearance of $\nu_e$ and the appearance of other active neutrino types, confirmed the total solar neutrino flux predicted by solar models~\cite{BP01,TC}, and explained the deficit of solar neutrinos that was seen by other pioneering experiments~\cite{cl,sage,gallex,SK,gno}.  The SNO results, when combined with other solar neutrino experiments and reactor antineutrino results from the KamLAND experiment~\cite{kam}, demonstrated that neutrino oscillations~\cite{pontecorvo,wolfenstein,ms} are the cause of this flavor change.

In the first two phases of the SNO experiment the determination of the total active $^8$B solar neutrino flux and its $\nu_e$ component required a statistical separation of the Cherenkov signals observed by the detector's photomultiplier tube (PMT) array.  In the third phase of the experiment, an array of $^3$He proportional counters~\cite{ncdcountersnim} was deployed in the detector's heavy-water target.  The neutron signal in the inclusive total active neutrino flux measurement was detected predominantly by this ``Neutral-Current Detection'' (NCD) array, and was separate from the Cherenkov light signals observed by the PMT array in the $\nu_e$ flux measurement.  This technique to measure the total active  $^8$B solar neutrino flux was largely independent of the methods employed by SNO in previous phases.  

The results from the third phase of the SNO experiment were reported in a letter~\cite{ncdprl}, and confirmed those from previous phases.  We present in this article the details of this measurement and an analysis of the neutrino oscillation parameters.  In Sec.~\ref{sec:overview} we present an overview of the SNO experiment and the solar neutrino measurement with the NCD array.  The Phase-III data set that was used in this measurement is described in Sec.~\ref{sec:dataset}.  Details of the optical response of the PMT array and the reconstruction of its data are provided in Sec.~\ref{sec:pmtresponse}.  The electronic and energy response of the NCD array will be discussed in Sec.~\ref{sec:ncdresponse}.  The determination of the neutron detection efficiencies for both the PMT and the NCD arrays, which are crucial to the measurement of the total active solar neutrino flux, is presented in Sec.~\ref{sec:neutron}.  The evaluation of backgrounds in the measurement is summarized in Sec.~\ref{sec:backgrounds}.  Alpha decays in the construction materials of the NCD array were a non-negligible background for the detection of signal neutrons.  We developed an extensive pulse-shape simulation and applied it to understand the response of the NCD counters to these alpha decays.   The pulse-shape simulation model is presented in Sec.~\ref{sec:pulsesim}.  In Sec.~\ref{sec:sigex}, we discuss the analysis that determined the total active solar neutrino flux and the electron-type neutrino flux.   These measured fluxes, along with results from previous SNO measurements and other solar and reactor neutrino experiments, were then used in the determination of the neutrino mixing parameters as described in Sec.~\ref{sec:physint}.   A description of the cuts we used to remove instrumental backgrounds in the NCD array data can be found in Appendix~\ref{sec:apdxa}.  A discussion of the parameterization of nuisance parameters in the neutrino flux analysis is provided in Appendix~\ref{sec:apdxb}.

\section{Overview of the SNO experiment\label{sec:overview}}

\subsection{The SNO detector}

The SNO detector was located in Vale's Creighton Mine ($46 ^\circ 28' 30''$ N latitude, $81 ^\circ 12' 04''$ W longitude) near Sudbury, Ontario, Canada.  The center of this real-time heavy-water  ($^2$H$_2$O, \dto\ hereafter) Cherenkov detector was at a depth of 2092~m (5890$\pm$94 meters of water equivalent).  At this depth, the rate of cosmic-ray muons entering the detector was approximately three per hour.  The solar neutrino target was 1000 metric tons (tonnes) of 99.92\% isotopically pure \dto\ contained inside a 12-m-diameter acrylic vessel (AV).  An array of 9456 20-cm Hamamatsu R1408 PMTs, installed on an 18-m diameter stainless steel geodesic structure (PSUP), was used to detect Cherenkov radiation in the target.  A non-imaging light concentrator~\cite{concentrator} was mounted on each PMT to increase the effective photocathode coverage to nearly 55\% of 4$\pi$.  The AV and the PSUP were suspended in an underground cavity filled with approximately 7 kilotonnes of  ultra-pure light water (\hto), which shielded the \dto\ volume against radioactive backgrounds from the cavity rock.  The inner 1.7 kilotonnes of \hto\ between the AV and the PSUP also shielded the target against radioactive backgrounds from the geodesic structure and PMTs.  On the outer surface of the PSUP, 91 outward-facing PMTs were installed to tag cosmic-ray events.  An array of 23 PMTs were mounted in a rectangular frame that was suspended facing inwards in the outer \hto\ region.  These PMTs, along with the 8 PMTs installed in the neck region of the AV, were used to reject instrumental background light.  A full description of the SNO detector can be found in Ref.~\cite{NIM}.  In the third phase of the SNO experiment, an array of $^3$He proportional counters was deployed in the \dto\ volume.  Details of this array are presented in Sec.~\ref{sec:ncddescr} and in Ref.~\cite{ncdcountersnim}.

The SNO detector detected solar neutrinos through the following processes:
\begin{center}
\begin{tabular}{rcl}
CC & : & $ \nu_e + d \rightarrow p + p + e^- - 1.442\mbox{ MeV}$ \\
NC & : & $ \nu_x + d \rightarrow p + n + \nu_x - 2.224\mbox{ MeV}$ \\
ES & : & $ \nu_x + e^- \rightarrow \nu_x + e^- $\\
\end{tabular}
\end{center}
where $\nu_x$ refers to any active neutrino flavor ($x=e,\mu,\tau$).  The charged-current (CC) reaction is sensitive exclusively to $\nu_e$, whereas the neutral-current (NC) reaction is equally sensitive to all active neutrino flavors.  Chen~\cite{hhchen} realized that the NC measurement of the total active solar neutrino flux tests the solar model predictions independently of the neutrino-oscillation hypothesis, and a comparison of this flux to the CC measurement of $\nu_e$ flux tests neutrino flavor transformation independently of solar models.  The neutrino-electron elastic scattering (ES) reaction is used to observe neutrinos of all active flavors in SNO and other real-time water Cherenkov and liquid scintillator detectors.  Its cross section for $\nu_e$ is approximately six times larger than $\nu_\mu$ and $\nu_\tau$ for $^8$B solar neutrinos, but is smaller than the CC or NC cross sections in the energy region of interest.

In the first phase of the SNO experiment, which used an unadulterated \dto\ target, NC interactions were observed by detecting the 6.25-MeV gamma ray following the capture of the neutron by a deuteron.  Under the assumption of an undistorted $^8$B neutrino spectrum, the hypothesis of the observed CC, NC and ES rates due solely to $\nu_e$ interactions was rejected at 5.3$\sigma$.  Details of the solar neutrino analysis in this phase can be found in Ref.~\cite{longd2o}.  Approximately two tonnes of sodium chloride (NaCl) were added to the \dto\ in the second phase of the SNO experiment.   This addition enhanced the neutron detection efficiencies and allowed the statistical separation of CC and NC signals without making any assumption about the energy dependence of neutrino flavor change.  As a result the accuracy of the $\nu_e$ and the total active neutrino flux measurements were significantly improved.  A full description of the solar neutrino analysis in this phase can be found in Ref.~\cite{nsp}.  Recently the results of a solar neutrino analysis that combined the Phase-I and Phase-II data sets were reported in Ref.~\cite{leta}.

\subsection{The Neutral-Current Detection (NCD) array\label{sec:ncddescr}}

\begin{figure}
\begin{center}
\includegraphics[width=0.65\textwidth]{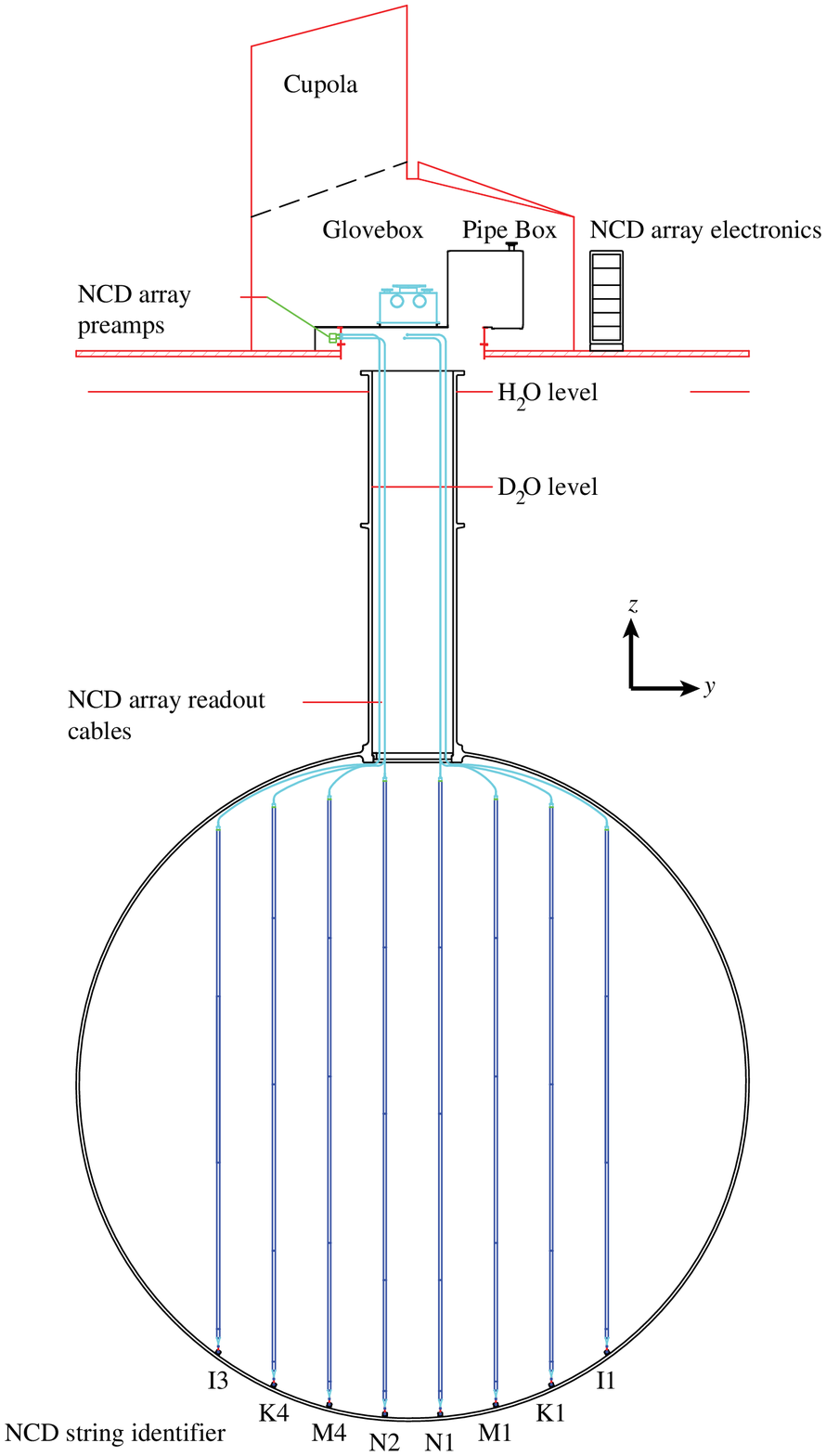}
\caption{Side view of the SNO detector in Phase III.  The center of the acrylic vessel is the origin of the Cartesian coordinate system used in this paper.   The NCD counter strings were arranged on a square grid with 1-m spacing (shown in Fig.~\ref{fig:ncdgrid}).  In this figure only the first row of NCD strings from the $y-z$ plane are displayed.\label{fig:SNO_ZYview} }
\end{center}
\end{figure}

\begin{figure}
\begin{center}
\includegraphics[width=0.80\textwidth]{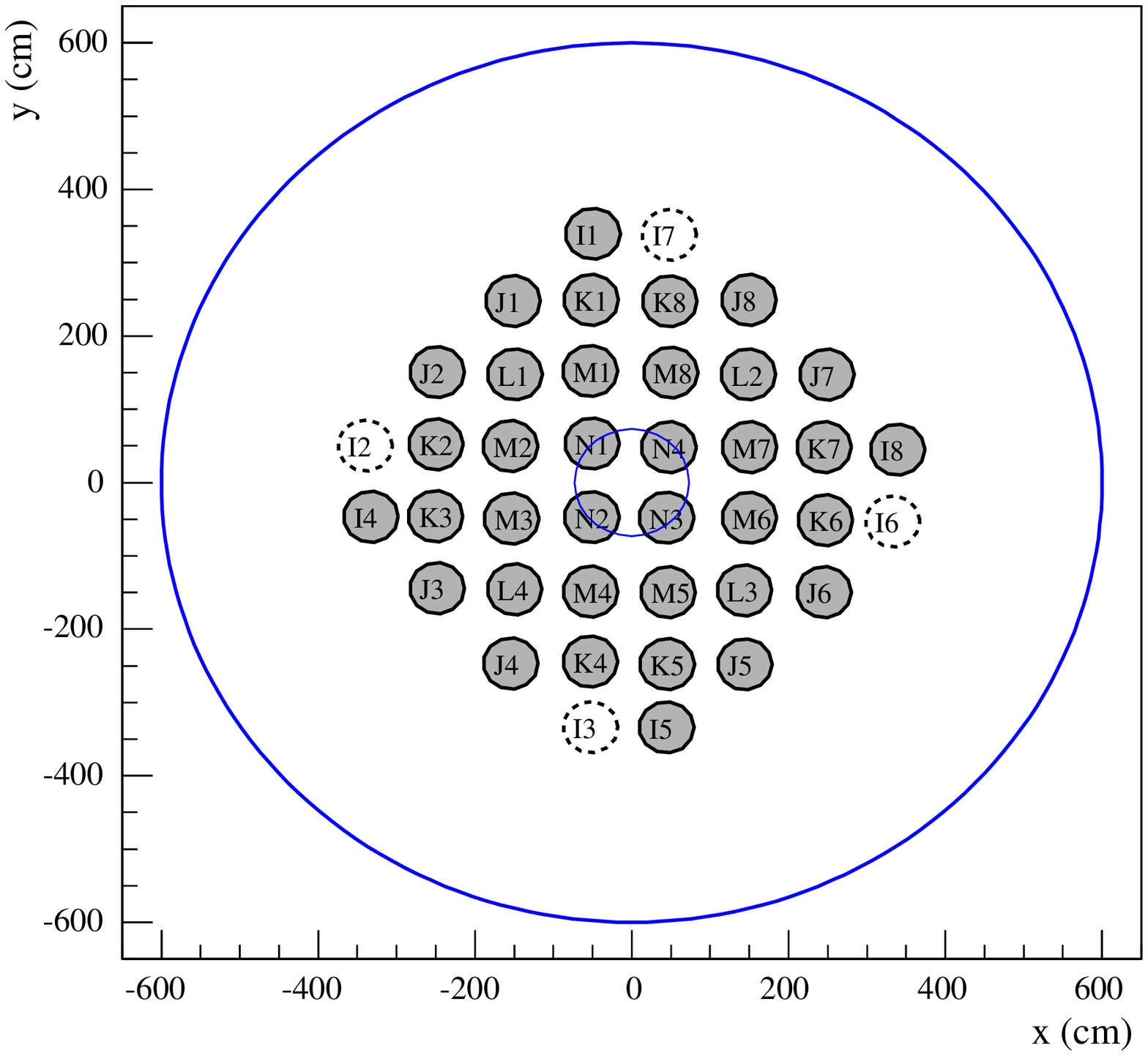}
\caption{The positions of the strings in the NCD array projected onto the plane of the AV equator ($x-y$ plane). The array was anchored on a square lattice with a 1-m grid constant.  The strings labeled with the same letter denote strings of the same length and distance from the center of the AV. Strings I2, I3, I6 and I7 contained $^4$He instead of $^3$He. The outer circle is the AV equator and the inner circle is the neck of the AV through which the NCD strings were deployed. The NCD string markers are not drawn to scale.\label{fig:ncdgrid}}
\end{center}
\end{figure}

The NCD array, consisting of 36 strings of $^3$He and 4 strings of $^4$He proportional counters, was deployed in the \dto\ target, after the removal of NaCl, in the third phase of the experiment.  Figure~\ref{fig:SNO_ZYview} shows a side view of the SNO detector with the NCD array in place.  The NCD counter strings were arranged on a square grid with 1-m spacing as shown in Fig.~\ref{fig:ncdgrid}.   The acrylic anchors to which the NCD strings were attached were bonded to the AV during its construction, and their positions were surveyed precisely by laser theodolite and were taken as reference.  Details of the deployment of the NCD string can be found in Ref.~\cite{ncdcountersnim}.

Each NCD string was 9 to 11 meters in length, and was made up of three or four individual 5-cm-diameter counters that were laser-welded together.  Ultra-low radioactivity nickel, produced by a chemical vapor deposition (CVD) process, was used in the construction of the counter bodies and end-caps.   This process suppressed all but trace amounts of impurities.  The nominal thickness of the counter wall was $\sim$370~$\mu$m.  Each counter was strung with a 50-$\mu$m-diameter low-background copper wire that was pretensioned with a 30-g mass.  The gas in the counters was a mixture of 85\% $^3$He (or $^4$He) and 15\% CF$_4$ (by pressure) at 2.5 atmospheres (1900 Torr). 

The coordinates of the top of the NCD strings were not a priori known because they were pulled slightly out of the anchors' reference positions by the cables. There were altogether three campaigns to measure the positions.   Before and after data-taking in Phase III, laser range-finder (LRF) surveys were made of the counter tops using a custom-built LRF that could be introduced through the AV neck and immersed in the \dto.  During data-taking, optical reconstruction of the average positions was obtained from the shadowing of calibration sources~(see Sec.~\ref{sec:opticalres}).  These three measurements were in good agreement, with the shadowing results being about two times more precise than the LRF results.

Neutrons from the NC reaction were detected in the NCD array via the reaction 
\begin{displaymath}
n + \mbox{$^3$He} \rightarrow p + t + 764\mbox{ keV}.
\end{displaymath}
The $^4$He strings were not sensitive to neutrons and were used to characterize non-neutron backgrounds in the array.   

During normal operation the anode wires were maintained at 1950~V, resulting in a gas gain of $\sim$220.  A primary ionization of the counter gas would trigger an avalanche of secondary ionizations, which led to a current pulse on the anode wire.  This pulse traveled in both directions, up and down a counter string. The delay line at the bottom of each string added approximately 90~ns to the travel time of the portion of the pulse that traveled down the string. The termination at the end of the delay line was open, so the pulse was reflected without inversion. The direct and reflected portions of the pulse were separated by approximately 90-350 ns, depending on the origin of the pulse along the length of the NCD string.   At the top of each counter string was a 93-$\Omega$ impedance coaxial cable that led to a current preamplifier.  This preamplifier linearly transformed the signal to a voltage amplitude with a gain of 27.5 mV/$\mu$A.

The NCD array had two independently triggered readout systems.  The ``Shaper-ADC'' system used a pulse-shaping and peak-detection network to integrate the signal pulse and measure its energy.  This fast system was triggered by the pulse integral crossing a threshold and could handle the kilohertz event rates expected from a galactic supernova.  The ``Multiplexer-Scope'' (MUX-scope) system digitized and recorded the entire 15-$\mu$s pulse. It consisted of four independent sets of electronics, or MUX boxes, each of which could accept signals from up to twelve strings.  Each channel was triggered by the pulse amplitude crossing a threshold.  Pulses were amplified by the logarithmic amplifier (``log-amp'') in a MUX box to increase the range of pulse sizes that could be digitized at a 1-GHz sampling rate  by the 8-bit digitizer in one of the two digital oscilloscopes.   The multiplexer controller triggered the oscilloscope that was not busy (or toggled between them when neither was busy), allowing for a maximum digitization rate of 1.8~Hz.  If the oscilloscopes were busy and the MUX system triggered, a ``partial MUX'' event was recorded without the digitized pulse.  The MUX-scope system adopted in the 1990s is not a solution to be recommended today, but it was sufficient to handle typical solar-neutrino signal and background event rates.  Signals from the PMT array and from these two readout systems of the NCD array were integrated in a global trigger system that combined the data streams with event timing information.  Further details of the design and construction of these counters and their associated electronic systems can be found in Ref.~\cite{ncdcountersnim}.

\section{Data set\label{sec:dataset}}

\subsection{Data selection\label{sec:dataselect}}

The measurements reported here are based on analysis of $385.17 \pm 0.14$~live days of data recorded between November 27, 2004 and November 28, 2006.  The selection of solar neutrino data runs for analysis was based on evaluation as outlined in previous papers~\cite{longd2o,nsp}.  In addition to the offline inspections of run data from the PMT array, data-quality checks of the NCD array data were implemented.  These checks validated the running condition, such as the trigger thresholds, of the NCD array. 

To accurately determine the total active solar neutrino flux using the NCD array it was essential to utilize only data from strings that were operating properly.  Six $^3$He strings were defective and their data were excluded in the analysis.  One of the counters in the string K5 was slowly leaking $^3$He into an inter-counter space.  Two strings, K2 and M8, had mechanical problems with the resistive coupling to the top of the counter string, as confirmed in postmortem examination at the end of the experiment, resulting in unstable responses.  The string K7 showed similar behavior, but a physical examination of this string at the end of the experiment did not indicate a loose coupling.  The strings J3 and N4 were observed to produce anomalous instrumental background events in the neutron signal window.  A loose resistive coupling was found during a physical examination of J3, but not in N4.

\begin{figure}
\centering
\includegraphics[width=0.80\textwidth]{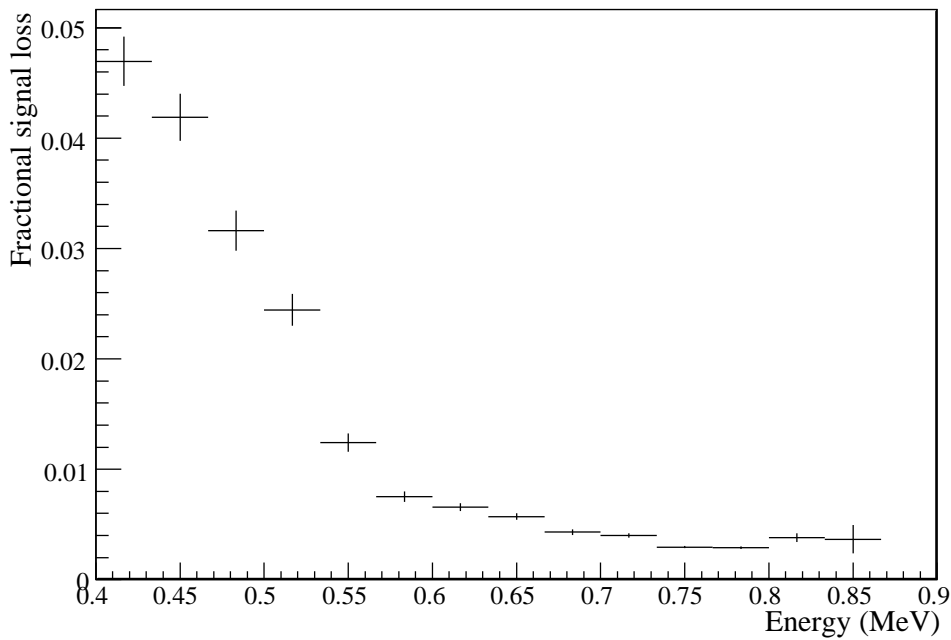}
\caption[Energy dependence of fractional signal loss]
{Energy dependence of fractional signal loss of NCD array data due to instrumental background cuts, as measured with $^{252}$Cf and Am-Be neutron calibration data. The abscissa is the event energy recorded by the shaper-ADC system. }
\label{fig:sac_energy}
\end{figure}

A variety of other kinds of instrumental events were recorded in the shaper and digitized-data paths.  Data reduction cuts were developed to remove these instrumental backgrounds.  In Appendix~\ref{sec:apdxa} examples of these background events and a summary of the cuts to remove them are provided.  Physics events in a counter would trigger both the shaper-ADC and the MUX-scope subsystems; thus, a large fraction of instrumental backgrounds was removed simply by accepting only events with both triggers present.  NCD array events that passed this selection criterion were subsequently analyzed by algorithms that were designed to identify non-ionization pulses such as micro-discharges and oscillatory noise.  Two independent sets of  cuts were developed.  One of these sets examined the logarithmically amplified digitized waveforms in the time domain, while the other set utilized the frequency domain.  The two sets of cuts were shown to overlap substantially, with 99.46\% of cut events removed by both sets of cuts, 0.02\% removed only by the time-domain cuts, and 0.52\% removed only by the frequency-domain cuts.   Both sets of cuts were used in reducing the data set.  The number of raw triggers from the NCD array data stream was 1,417,811, and the data set was reduced to 91,636 ``NCD events'' after application of data reduction cuts.  Figure~\ref{fig:sac_energy} shows the energy dependence of the fractional signal loss determined from $^{252}$Cf and Am-Be neutron calibration sources.  
A suite of instrumental background cuts for the PMT array data was developed in previous phases of the experiment~\cite{longd2o,nsp}.   These cuts were re-evaluated and re-calibrated to ensure their robustness in the analysis of the third-phase data.   The number of raw triggers from the PMT array data stream was 146,431,347, with 2,381 ``PMT events'' passing data reduction and analysis selection requirements similar to those in Ref.~\cite{longd2o}.  These selected PMT events have reconstructed radial distance $\rfit \leq 550$~cm and reconstructed electron effective kinetic energies $\teff\geq 6.0$~MeV. 

\subsection{Live time}

The raw live time of selected runs was calculated from the time differences of the first and last triggered event using the main trigger system's 10-MHz clock, which was synchronized to a global positioning system.  Due to the combination of the data streams from the PMT and NCD arrays, a run boundary cut was applied to ensure that both systems were taking data by defining the start of the run as 1.1 seconds after the first event of either array, whichever came later.  A reverse order cut was applied to define the end of each run.  These calculated times were verified by comparing the results against those measured by a 50-MHz detector-system clock and a 10-MHz clock used by the NCD array's trigger system.  

Several data selection cuts removed small periods of time from the data set during normal data taking conditions.  The largest of these removed time intervals following high-energy cosmic-ray events and intervals containing time-correlated instrumental events.  These cuts removed events from both the PMT and the NCD arrays.  The final live time for the neutrino analysis was calculated by subtracting the total time removed by these cuts from the raw live time.  This resulted in a reduction  of 1.96\% of the raw live time.

The final live time was checked with analyses of data from two detector diagnostic triggers: the pulsed global trigger (PGT) and the NCD system's random pulser (NRP).  The PGT was a detector-wide trigger issued at a frequency of 5~Hz based on timing from the 50-MHz system clock.  The NRP randomly pulsed a spare channel on the NCD system at an average rate of 7.75~mHz.  Systematic uncertainties in live time were evaluated by comparing the PGT and NRP measurements to the 10-MHz clock measurement.  The total live time uncertainty was calculated to be 0.036\%.

The ``day'' data set, which was acquired when the solar zenith angle $\cos\theta_z >0$, has a live time of 176.59~days.  The night data set, for which $\cos\theta_z\leq 0$, has a live time of 208.58 days.

\section{Response of the PMT array\label{sec:pmtresponse}}

In the solar neutrino measurement, the SNO PMT array observed Cherenkov radiation from high energy electrons resulting from direct neutrino interactions, $\beta$ decays of radioactive backgrounds, and Compton scattering of gamma rays from nuclear de-excitations and radiative captures.  A thorough understanding of the propagation and the detection of Cherenkov photons in the SNO detector was vital to reconstruct the observables, such as energy and vertex position, of each triggered event in the PMT array.   Details on the extensive optical and energy calibration of the PMT array in previous phases of SNO can be found in Refs.~\cite{longd2o,nsp}.

In the third phase of SNO,  optical and energy calibration procedures were modified from those in previous phases in order to account for complexities that did not exist before.  The nickel body of the NCD strings scattered and absorbed Cherenkov photons.  The orientation of the NCD array and its signal cables also accentuated the vertical ($z$) asymmetry in the response of the PMT array.  These effects had to be incorporated in the reconstruction of Cherenkov events in order to precisely determine the energy and spatial distributions of neutrino signals and radioactive backgrounds in the heavy-water detector.     In this section, we will first present the response of the PMT array to optical photons.  This will be followed by a study of event vertex reconstruction, which required the arrival time of the detected photons as input, and a study of the energy response of the PMT array, which depended on reconstructed event vertex position.

\subsection{Optical response\label{sec:opticalres}}

The measurement of the SNO optical model parameters (see Sec.~IV.A of
Ref.~\cite{nsp}) was done by a $\chi^2$ fit that minimized the differences
between the measured and predicted light intensities at each PMT, for
a set of calibration runs (``scan'') taken with the ``laserball'' source
\cite{laserball} at a number of positions inside the D$_2$O volume.
These calibration methods had to be significantly modified with
respect to the previous phases, in order to take into account the
optical effects of the NCD array.  The construction of a high-isotropy
laserball source and the optimization of the calibration plans enabled
good sampling of the PMT array despite partial shadowing from the NCD
array.  The changes in the analysis of the calibration data, and its
results, are described below.

\subsubsection{Source and NCD string positions}

The source position in individual laserball runs was determined by
minimizing the differences between the calculated and measured PMT
prompt peak time, similarly to what was done previously, adding cuts
to remove PMTs with very few counts.

\begin{figure}[t]
\centering
\includegraphics[width=0.80\textwidth]{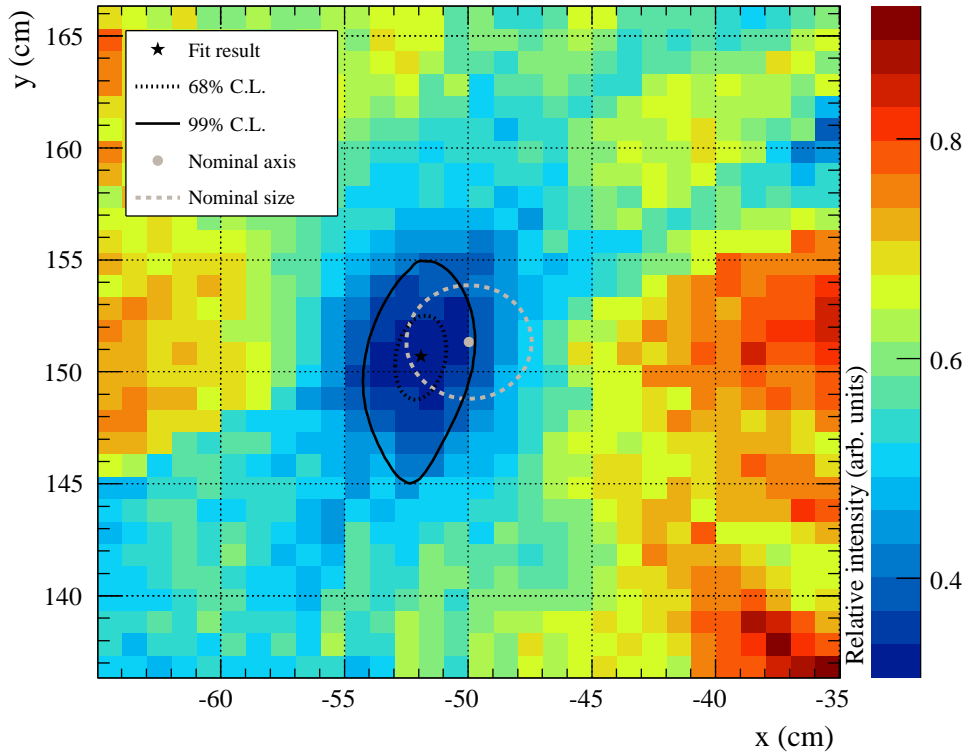}
\caption[Example of NCD position reconstruction from one of the
  optical scans.]  { An example of NCD string position reconstruction
  from the data of one of the laserball scans at 500~nm.  The contours
  represent the 68\% and 99\% CL of the fitted position of NCD string
  M1.  Also shown in this figure is the nominal position of the
  string.  The difference between the fitted position and the nominal
  position is $\sim$2~cm in this case.  }
\label{fig:optics:ncd_38}
\end{figure}

In SNO Phase III, the laserball source positions and the shadowing
patterns observed in the PMT array were used to obtain the NCD string
positions, which were in turn compared to the installation reference, or
nominal, positions.  These reference positions were the locations of the NCD string anchors that were surveyed by laser theodolite during the construction of the acrylic vessel.  The method selected an ensemble of about
30~source positions in order to triangulate and reconstruct the
position of each NCD string~\cite{simard}.  For each NCD string, sets
of PMTs were selected so that the light path between them and the
laserball lay near to the string, in $x$ and $y$ coordinates.  The PMT
occupancy for that run was filled in a two-dimensional map
corresponding to the $x-y$ line including the point of closest
approach of the source-PMT path to the NCD string, and all the other
points along that source-PMT direction (projected in the $x-y$ plane).
Since the count rate depended on the conditions of a run (such as
laser pulse rate, source stability, PMT thresholds, source position,
etc.), the occupancy of the selected PMTs in each run was normalized
by their mean occupancy.  A map of these relative occupancies was built
and fitted to a two-dimensional Gaussian function.
Figure~\ref{fig:optics:ncd_38} shows an example of a reconstructed
string position.
The extraction of all NCD string positions gave an average difference between the fitted and nominal coordinates, the average horizontal displacement, of about 2~cm in the $x$ and $y$ directions, which was consistent with the estimated average uncertainty of the string positions.

The NCD string position fit was repeated by dividing the various
trajectories into three $z$-bins, such that the $(x,y)$-coordinates
were obtained as a function of $z$.  The slope of $(x,y)$~vs~$z$ gave
the deviation of the counters from their nominal vertical position.
The best-fit angular deviation of all the counters was found to be
less than one degree, and was consistent with the measured average displacements.
Therefore the rest of the analysis assumed the counters were perfectly vertical but had an average horizontal displacement.  This choice simplified the numerical simulation model.

The various systematic uncertainties of source position
reconstruction, that summed up to approximately 2~cm, were propagated
to get an estimated resolution of the $(x,y)$-coordinates in the
triangulation method.  The method yielded an average uncertainty of
$2.2 \pm 0.3$~cm on the individual NCD string positions, projected at $z=0$.
The spread of 0.3~cm arose partly because of the geometry of the NCD array with respect to the calibration planes that limited the laserball positioning to $(x,0,z)$ and $(0,y,z)$ coordinates.
Thus strings in the outer rings were sometimes shadowed by other strings in inner rings, resulting in larger uncertainties for those strings.
In addition, the uncertainty contained a small scan-to-scan variation which was taken into account in the spread.
In the optical analysis, the 2.2~cm uncertainty and its spread were input parameters to the shadow-removal code that treated all laserball and string pairs in the same way, independently of their position, to remove the shadowed PMTs with an estimated efficiency of 99\%.  

\subsubsection{Optical effects of the NCD array}

The attenuation lengths of various optical media and the PMT angular response were determined
by analyzing the PMT data that were not affected by unwanted optical
effects, including those caused by the NCD array.  Given the fitted
position of the laserball and NCD strings, the shadowed PMTs were removed from the
analysis on a run-by-run basis.

In addition to the shadowing effects from the NCD strings and their
signal cables, the anchors that held them down to the bottom of the AV
were also taken into account since they were made of UV-absorbing
acrylic.  The implementation of the anchor cut showed improvements in
the determination of the efficiency for PMTs located at the bottom of
the detectors.

In order to handle PMT-to-PMT variations in efficiency in previous
phases, the light intensity at a PMT for a given run was always
normalized to the intensity of that PMT in a run where the laserball
was deployed at the center of the detector.  In the third phase, this
technique would result in the systematic removal of the shadowed PMTs
in that central run.  Therefore, the optical model used the PMT
relative efficiencies measured in the ``preparatory'' phase, a run
period after the removal of salt but prior to the installation of the
NCD array, taking into account changes in PMT gain and threshold.
Even when the NCD string shadow and anchor cuts removed many PMTs in
the analysis, the new fit method and optimally chosen laserball
positions compensated for this loss, resulting in the overall
statistics in a given scan at most 50\% lower than in previous phases.

Light reflections off the surface of NCD strings were predominantly
diffuse, and it was impossible to associate specific trajectories
between the source and the PMTs.  Therefore, the effect of such
reflections on the PMT counts must be estimated and corrected for runs
at various source positions on a run-by-run basis.  An analytic
correction was derived for each source-PMT trajectory by calculating
the fraction of solid angle corresponding to the optical paths between
the source and a given PMT that included a reflection in one NCD with
respect to the direct paths.  The correction for diffuse reflection
off the NCD strings was found to be less than 5\% of a PMT's occupancy on average.
The same correction could be inferred using
various MC scenarios, employing the ratio of PMT calibrations from
reflection-on and reflection-off simulations.  The analytic and
MC-based corrections agreed to within 10\% and the difference between
the two was applied as a systematic uncertainty on the optical
parameters.

\subsubsection{Determination of the optical parameters\label{sec:opticalparam}}

In addition to an optical data set taken during the
preparatory phase as a reference of the detector state, eight optical
calibration scans were performed during the third phase, among which
five were selected for use in the analysis.  For each scan, data were
taken at six different wavelengths (337, 365, 386, 420, 500 and
620~nm) at multiple source positions in the SNO detector.  The time
span of the calibration sets allowed us to monitor the stability of
the optical model parameters, which included the heavy- and
light-water attenuation lengths, the PMT angular response, and the
laserball's light isotropy.

\begin{figure}[!ht]
\begin{center}
\includegraphics[width=0.80\textwidth]{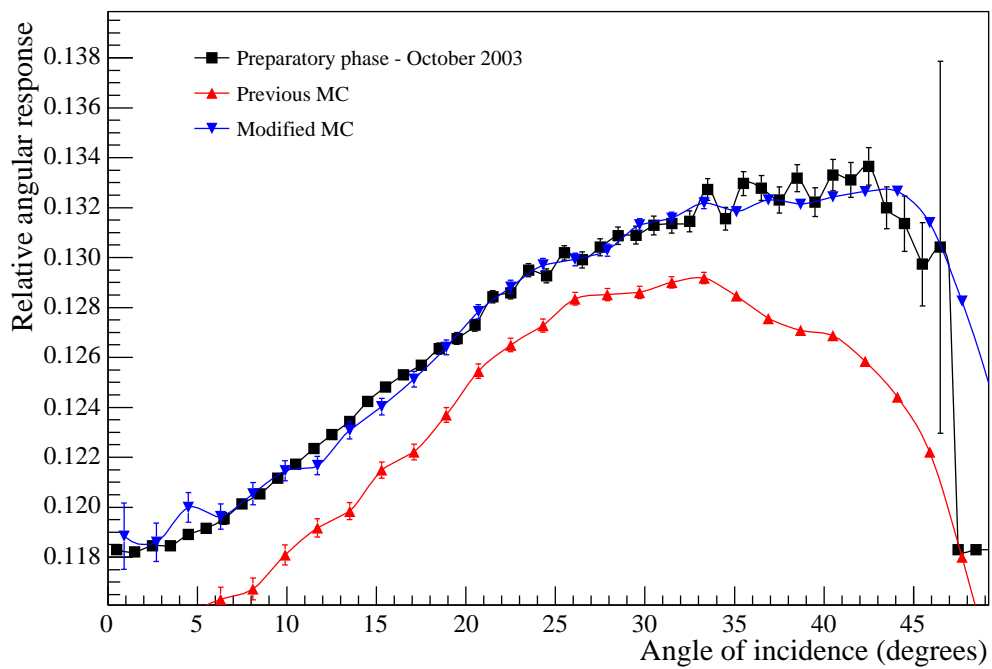}
\caption{Angular response curves at 386~nm for the PMT-reflector
  assembly generated by the MC simulation both before and after the
  optimizations discussed in the text, in comparison to the optical
  scan in the preparatory phase in October 2003.  Note that the $y$
  axis zero is suppressed.}
\label{fig:angresp}
\end{center}
\end{figure}

The modeling of the angular response of the PMTs was improved using
optical scan data from the preparatory phase.  An empirical collection
efficiency function in the simulation modifies the response of the PMT
as a function of the position at which a photon strikes the
photocathode, thus altering the angular response.  This function has
five tunable parameters, which were previously optimized to reproduce
laserball scans at 386$\,$nm, the most probable wavelength for registering
a hit in the detector.  In this phase, a joint $\chi^2$ fit was
performed at all six wavelengths at which laserball calibration data were
taken, fitting for both the shape of the response at each wavelength
and the relative normalizations.  The calibration data at each
wavelength were first normalized to the amplitude at normal incidence,
and then scaled by the quantum efficiency for a typical PMT at that
wavelength.  The $\chi^2$ of the empirical collection efficiency
function fit at each wavelength was weighted by the relative
likelihood of a successful hit being caused by a photon at that
wavelength in a Cherenkov light event, in order to optimize the fit at
the most probable wavelengths.  This resulted in a greater weighting
for the more probable wavelengths and a very small weighting for the
data at 620~nm, for example, where the probability of a photon triggering 
a PMT was very low.  The angular response shape thus produced by
the simulation showed a significant improvement in the agreement with
calibration data at all the most probable wavelengths.  Figure
\ref{fig:angresp} illustrates this shape and scale modeling
improvement at 386~nm.

\begin{figure}
\centering
\includegraphics[width=0.80\textwidth]{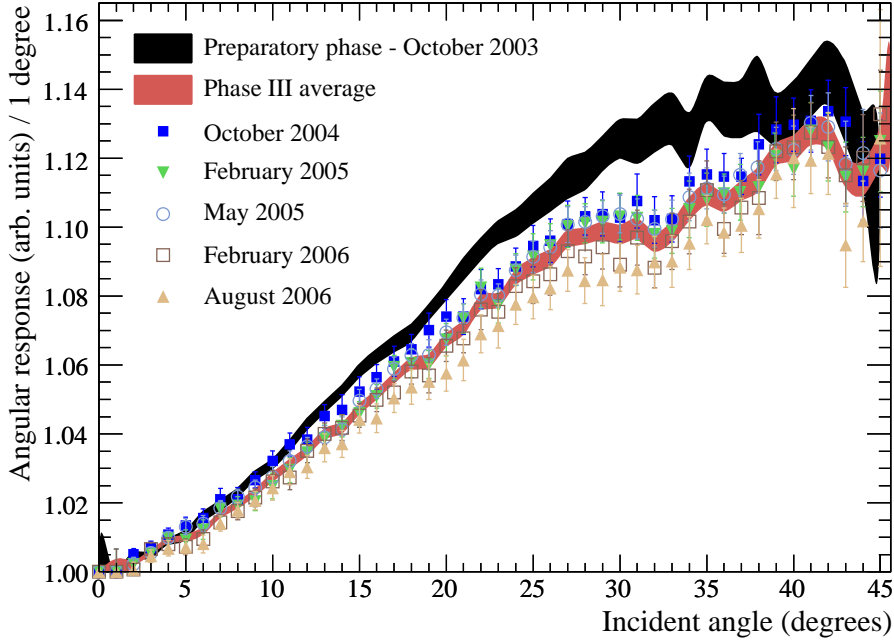}
\caption[PMT relative angular response at 421 nm for the five analyzed
  optical scans.]  { PMT relative angular response at 421 nm.  }
\label{fig:optics:pmtr_420}
\end{figure}

Figure~\ref{fig:optics:pmtr_420} shows the relative PMT angular
response for the five scans in the third phase.  The measurement from
the preparatory phase is also shown for reference.  The average change
in the response from the preparatory phase to the third phase, after
analyzing the data at all six wavelengths, was about 4\% at higher
incidence angles.  The change from the first to the last scan in
Phase~III was around 2\%.  The decrease in the response was consistent
with observations in previous phases, where it was attributed
primarily to aging of the light concentrators.  %

\begin{figure}
\centering
\includegraphics[width=0.80\textwidth]{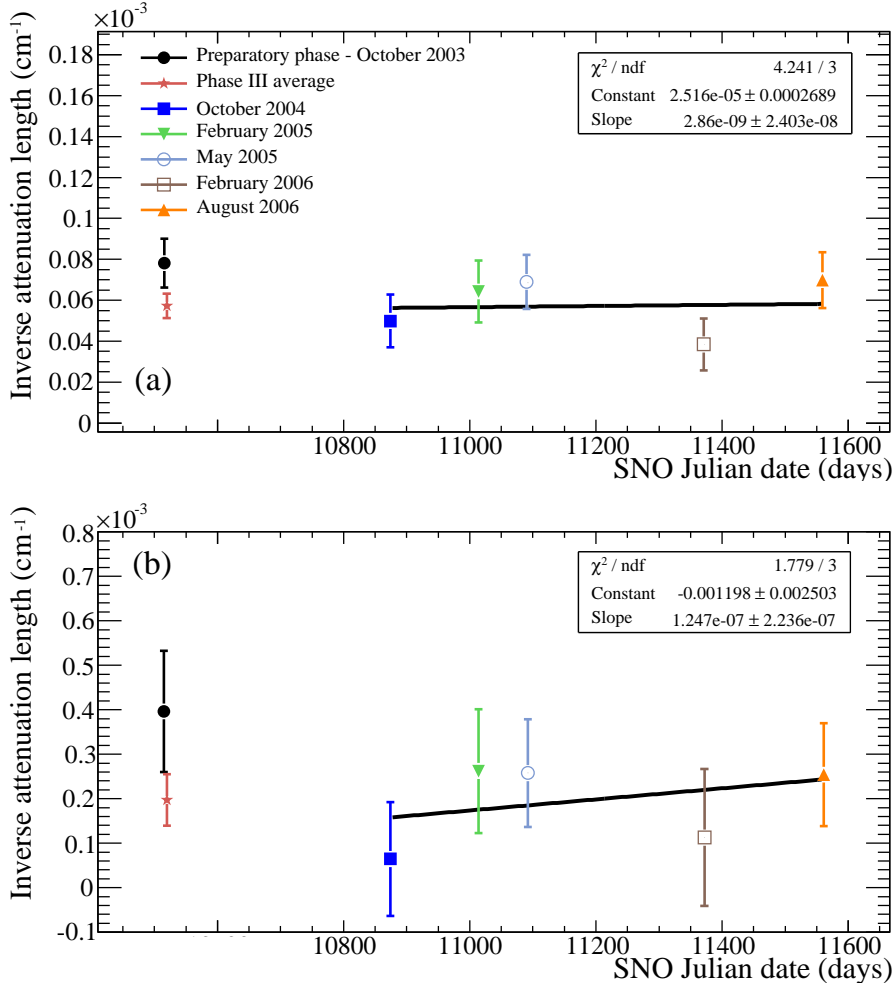}
\caption[D$_2$O and H$_2$O inverse attenuation lengths as a function
  of time at 421 nm.]  { (a) D$_2$O and (b) H$_2$O inverse attenuation
  lengths as a function of time at 421 nm.  The lines show the linear fits covering the commissioning and data-taking periods of Phase III.}
\label{fig:optics:attdrift_420}
\end{figure}

Figure~\ref{fig:optics:attdrift_420} shows the heavy- and light-water
attenuation coefficients obtained at 421 nm.  The time dependence of
the attenuation lengths was very small, showing better stability than
in Phase~II.  Therefore an average value of the optical parameters was
used in all Monte Carlo simulations in the third phase.  %

The new systematic uncertainties evaluated for the optical parameters
included the precision of the PMT efficiency estimations, the NCD
string shadow cut efficiency, the NCD string reflection corrections,
and the up-down asymmetry in the PMT array response.  The new
systematic uncertainties contributed to a 10-25\% increase of the uncertainties in
the optical parameters compared to previous phases, and since the
total uncertainties were dominated by the systematic component, the
loss in statistics induced a negligible increase in the total
uncertainties of the optical parameters.  After averaging the results
from all the analyzed scans, the uncertainties on the D$_{2}$O and H$_{2}$O
attenuation coefficients and PMT response were generally well below
10\%, 15\%, and 1.5\%, respectively.  The effect of these
uncertainties on event vertex position reconstruction accuracy and
energy estimation was estimated to be less than 0.1\% and 0.25\% 
respectively.

\subsection{Vertex reconstruction of Cherenkov events}

Algorithms that maximize the likelihood of event vertex position and direction, given the distribution of PMT trigger times and positions, were used to reconstruct Cherenkov events in SNO.  The following sections describe event reconstruction in Phase~III and the methodology used to determine the associated systematic uncertainties.

\subsubsection{Event vertex}

Vertex reconstruction in the third phase was performed by maximizing the likelihood function
\begin{equation}
\mathcal{L}=\prod_{i=1}^{N_\mathrm{hits}}f(\tres|\mathrm{hit};\xpmt,\xfit),
\end{equation}
where  $\tres$ is the time-of-flight corrected PMT trigger time
\begin{equation}
\tres=\tpmt-\tfit-\frac{|\xfit-\xpmt|}{c_\mathrm{avg}},
\end{equation}
\xfit\ and \tfit\ are the reconstructed event position and time respectively, $N_\mathrm{hits}$ is the number of selected PMT hits, and $c_\mathrm{avg}$ (=21.87~cm/ns) is the group velocity of the mean detected photon wavelength at 380~nm.  The function $f(\tres|\mathrm{hit};\xpmt,\xfit)$ is the probability density function (PDF) that a particular PMT fires at time $\tres$, given its position $\xpmt$ and the reconstructed event position.  In SNO's third phase, the $\tres$ PDF was dependent on $\xpmt$ and $\xfit$ due to partial or complete shadowing by the NCD array.  This shadowing effect was incorporated by generating $\tres$ distributions for non-shadowed and completely shadowed PMTs using Monte Carlo (MC) simulations and by interpolating between the two for partially shadowed PMTs using an algorithm that computed shadowing analytically.  To reduce the effects of reflected photons, the PDF was approximated as a constant for $\tres$ values greater than 15~ns, and a time cut of $\pm50$~ns around the median PMT hit time was imposed.  This determination of the PDF with shadowing effects allowed an overall improvement of $5\%$ in spatial resolution.

\begin{figure}
\begin{center}
\includegraphics[width=0.60\textwidth]{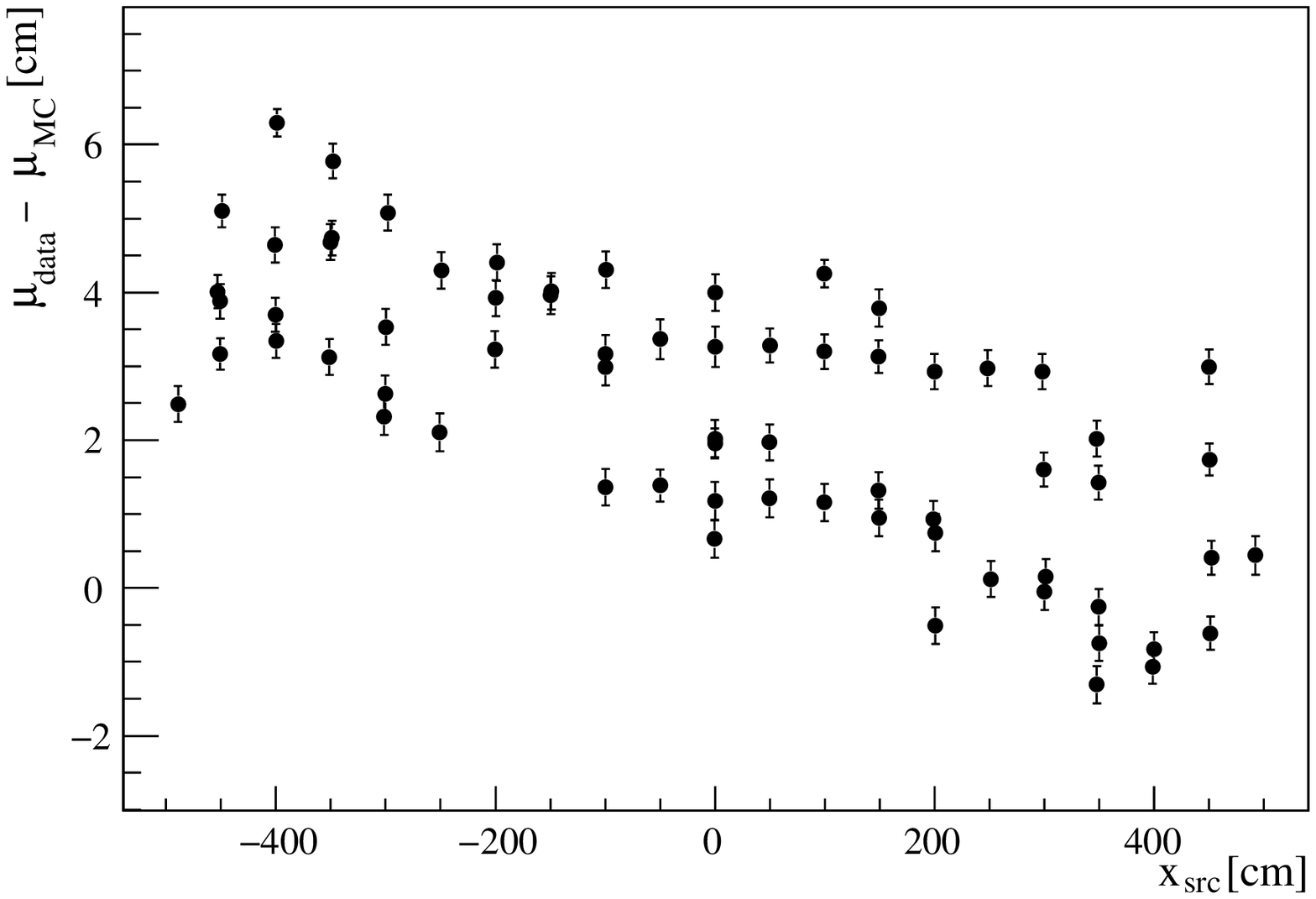}
\includegraphics[width=0.60\textwidth]{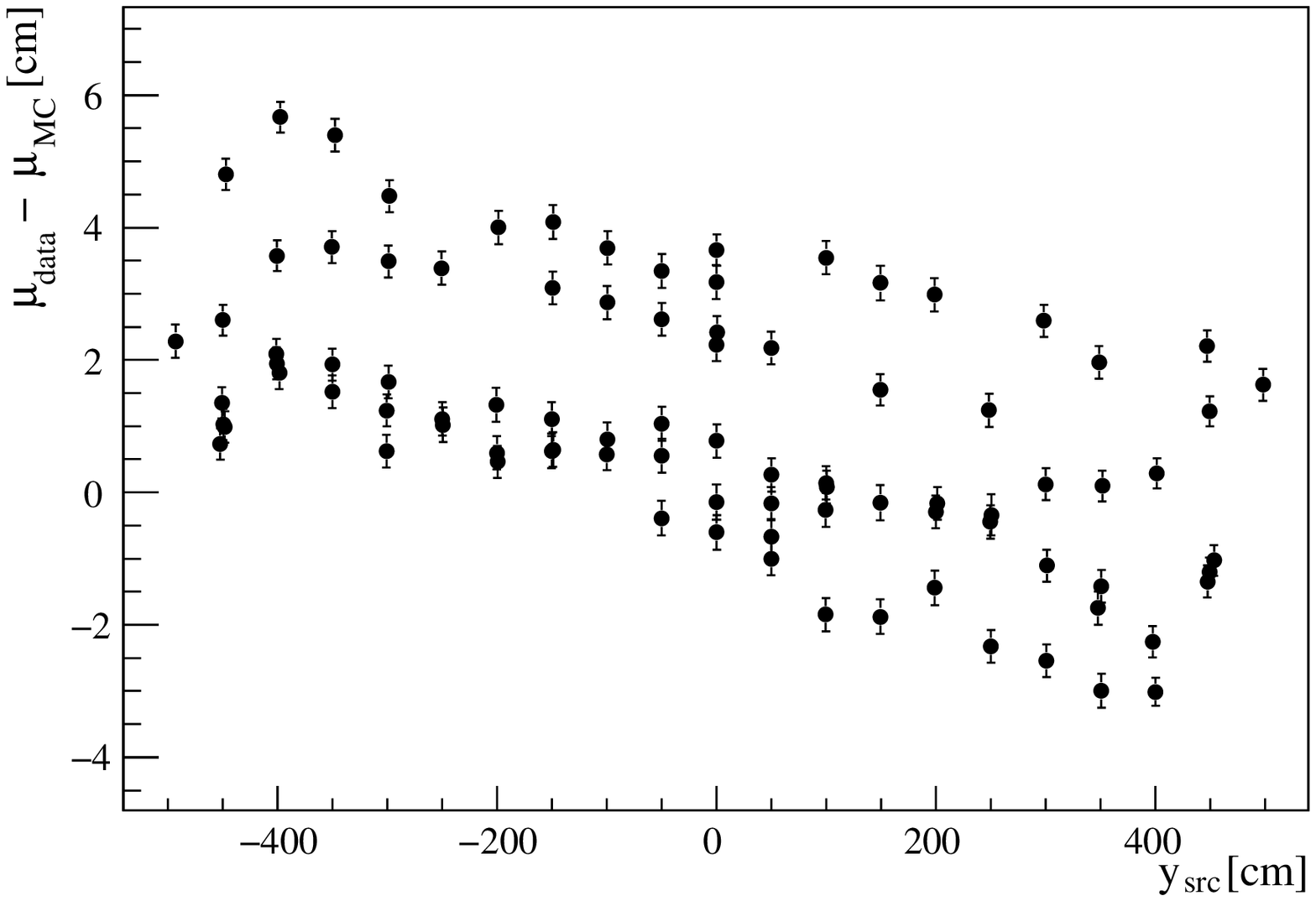}
\includegraphics[width=0.60\textwidth]{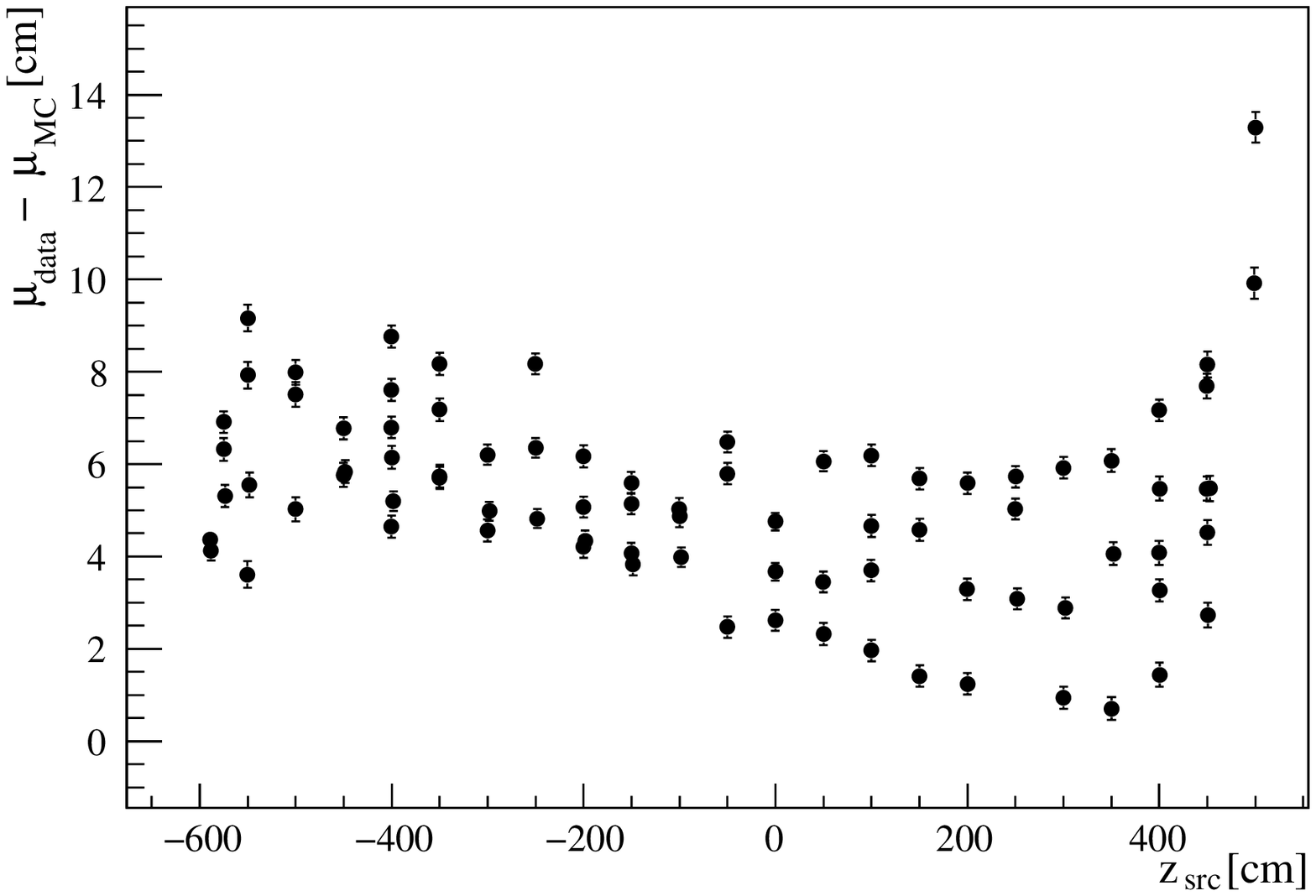}
\caption{Difference between data and MC vertex shifts in the three axis directions as a function of source position for scans along the main axes of the detector.  The top, middle and bottom panels represent the shifts in $x$, $y$, and $z$ coordinates, respectively. }
\label{vacc_dist}
\end{center}
\end{figure}

To evaluate the differences between the true and reconstructed event vertex positions, or the ``vertex shift'', the average reconstructed event position of \NS~\cite{n16} calibration data relative to the source position was compared to that computed from simulated data.  Figure \ref{vacc_dist} shows the difference between the data and Monte Carlo vertex shift as a function of the source position for scans along the main axes of the detector.  It shows a spread of 4~cm in the three directions, and this value was taken as the vertex shift uncertainty.  This uncertainty was found to be correlated to the PMT timing calibration of the detector.  An overall offset of 5~cm was also observed in the $z$ direction.

The systematic uncertainty on ``vertex scaling'', a position-dependent inward or outward shift of reconstructed position, can have a direct effect on the fiducial volume for events detected by the PMT array.  It was measured by determining the range of the slope of a first order polynomial that allowed the inclusion of $68\%$ of the data points in Figure \ref{vacc_dist}.  This uncertainty was believed to be caused by physical factors such as a mismatch of the speed of light in different media and was expected to be the same in all directions.  It was estimated to be $0.9\%$ of the Cartesian coordinates.

\begin{figure}
\begin{center}
\includegraphics[width=0.60\textwidth]{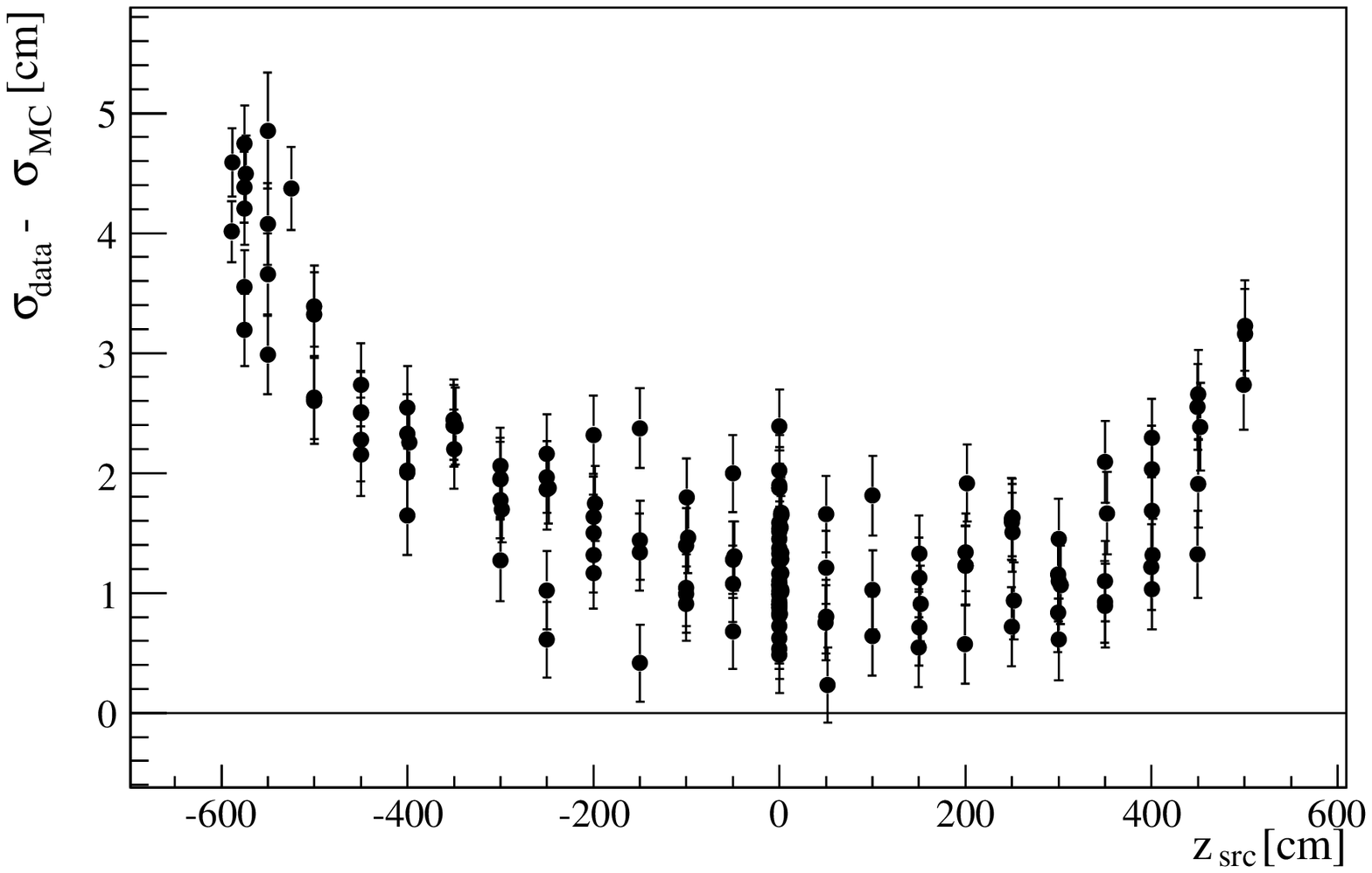}
\includegraphics[width=0.60\textwidth]{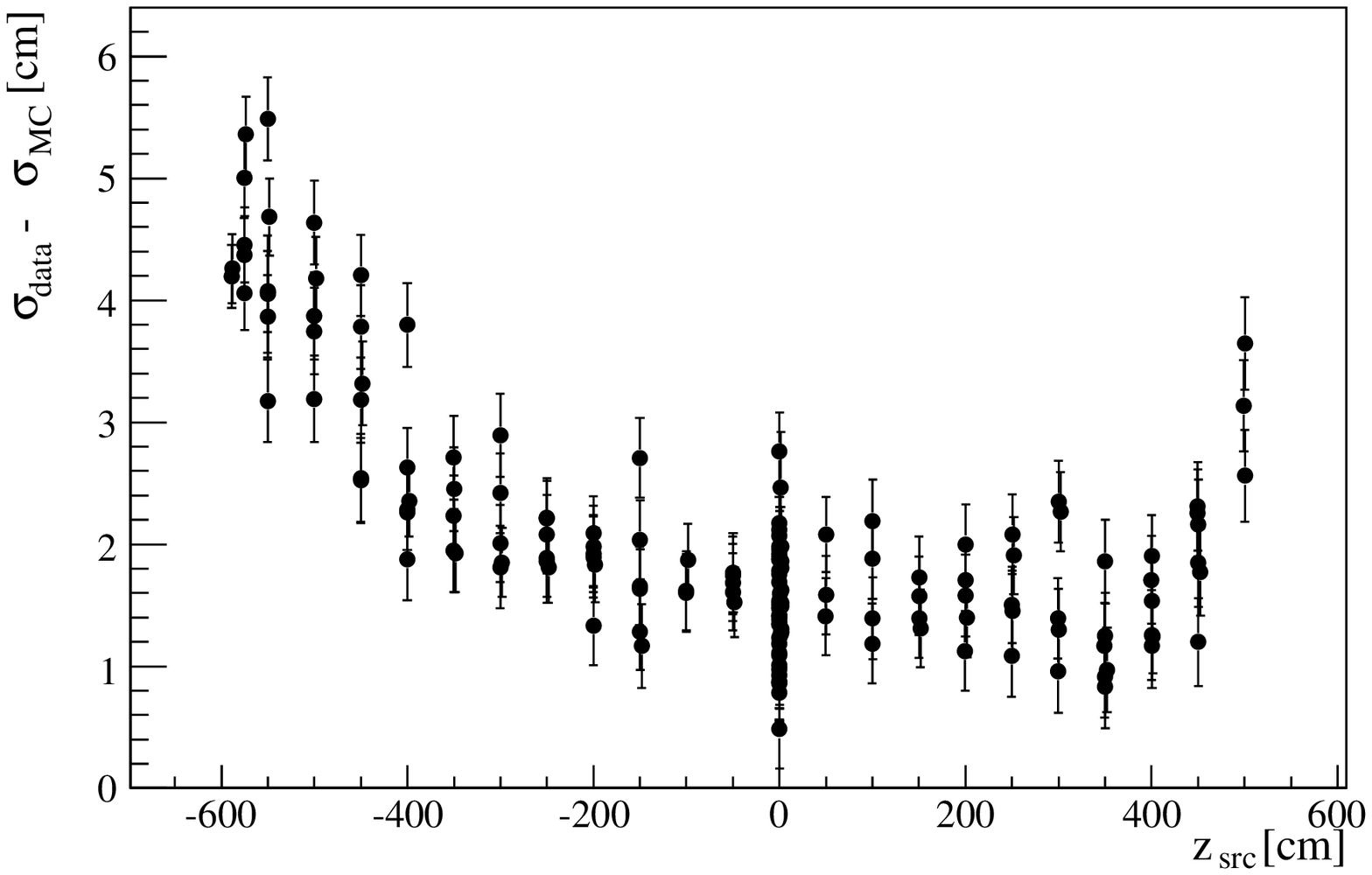}
\includegraphics[width=0.60\textwidth]{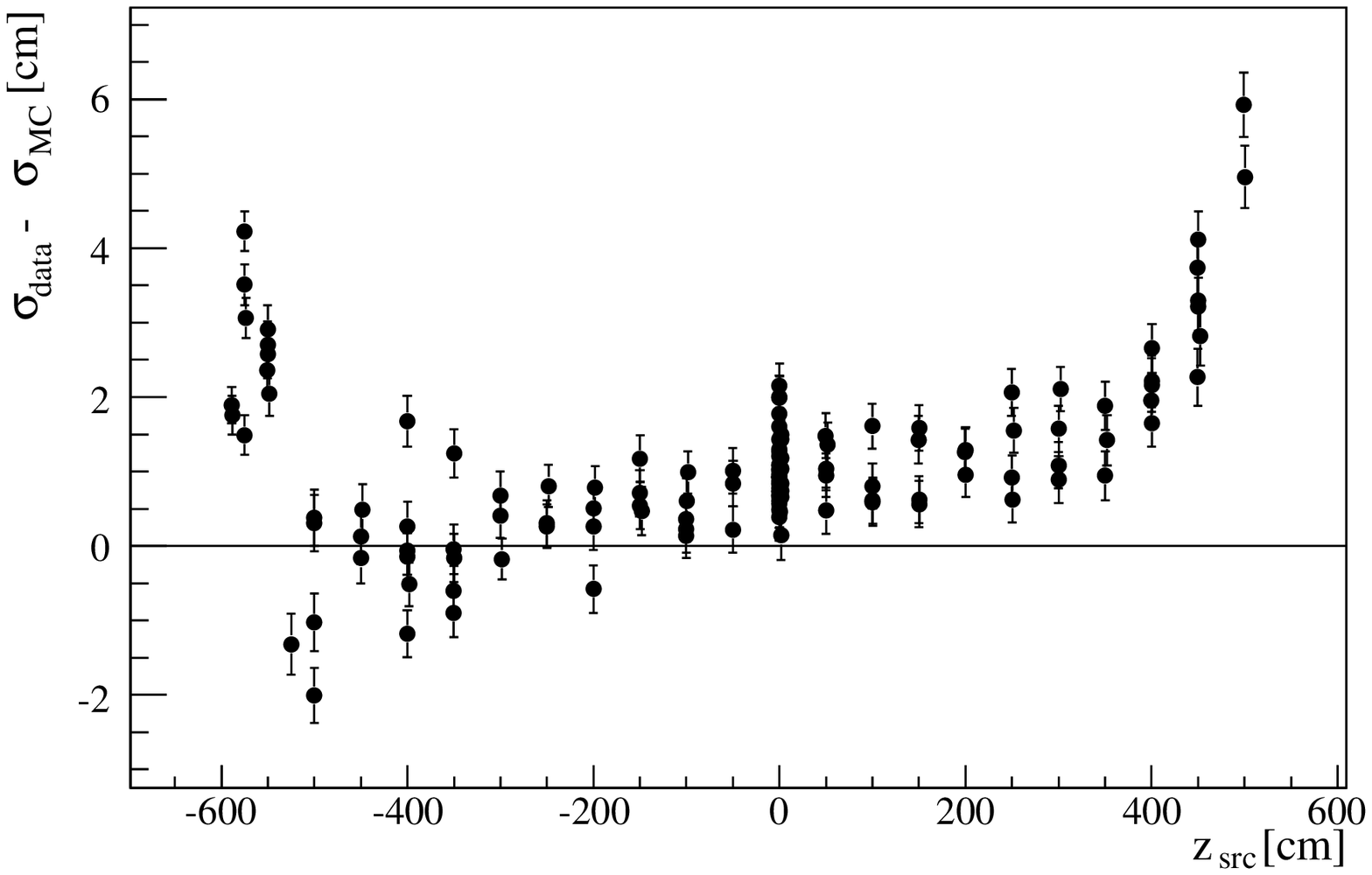}
\caption{Difference between data and MC fitted vertex width in the three axis directions ($x$, $y$ and $z$, from top to bottom) as a function of the position of the source ($z_\textrm{src}$) for \NS\ runs taken along the $z$ axis of the detector.}
\label{vres_z}
\end{center}
\end{figure}

Vertex resolution was another systematic uncertainty that could affect the fiducial volume.  It was assessed by taking the difference between the data and the MC fitted position resolution for all \NS\ calibration data taken inside the acrylic vessel, and propagated by smearing the coordinates of simulated events with a Gaussian random variable such that the width increased by the measured discrepancy.  Figure \ref{vres_z} shows that this uncertainty varies significantly as a function of the $z$ position of the source.
Some smaller fluctuations were also observed for \NS\ calibration scans in the $x-y$ plane and were incorporated in the analysis.  The apparent similarity of the results in $x$ and $y$ directions and the cylindrical symmetry of the detector in the third phase suggested the use of the same parameterization of the systematic effect for these two directions.
The uncertainty on the vertex resolution was expressed as a second order polynomial for $x$ and $y$ directions and as a first order polynomial for the $z$ direction.
Tables \ref{vres_pars} and \ref{vres_cor} present the fitted values for the parameters of these polynomials, along with their associated correlation matrices.

\begin{table}[!ht]
\begin{center}
\caption{Fitted values for the parameters in the polynomial $a_0+a_1z+a_2z^2$ used to evaluate the systematic uncertainty on vertex resolution.}
\label{vres_pars}
\begin{minipage}[t]{5in}
\begin{ruledtabular}
\begin{tabular}{cccc}
Direction & $a_0$ & $a_1$ & $a_2$ \\ 
& [cm]  & [$\times10^{-2}$] & [$\times10^{-5}\ \mathrm{cm}^{-1}$] \\ \hline
$x$, $y$ & 1.19 $\pm$ 0.52 & -0.10 $\pm$ 0.11 & 0.71 $\pm$ 0.36 \\
$z$ & 1.29 $\pm$ 0.51 & 0.21 $\pm$ 0.15 & NA \\
\end{tabular}
\end{ruledtabular}
\end{minipage}
\end{center}
\end{table}

\begin{table}[!ht]
\begin{center}
\caption{Correlation matrices for the fitted parameters of the polynomials used to evaluate the systematic uncertainty on vertex resolution.}
\label{vres_cor}
\begin{minipage}[t]{3.5in}
\begin{ruledtabular}
\begin{tabular}{cccc}
\multicolumn{4}{c}{$x$, $y$ directions} \\ \hline
$\rho$ & $a_0$ & $a_1$ & $a_2$ \\ \hline
$a_0$ & 1.00 & -0.13 & -0.74 \\
$a_1$ & -0.13 & 1.00 & 0.31 \\
$a_2$ & -0.74 & 0.31 & 1.00 \\ 
\end{tabular}
\end{ruledtabular}

\vspace{12pt}

\begin{ruledtabular}
\begin{tabular}{ccc}
\multicolumn{3}{c}{$z$ direction} \\  \hline
$\rho$ & $a_0$ & $a_1$ \\ \hline
$a_0$ & 1.00 & 0.15 \\
$a_1$ & 0.15 & 1.00 \\
\end{tabular}
\end{ruledtabular}
\end{minipage}
\end{center}
\end{table}

\subsubsection{Event direction}

The direction of events in the detector was determined at a later stage, decoupled from position reconstruction.
It relied on a likelihood function composed of a Cherenkov light angular distribution function and a PMT solid angle correction.
To avoid biases caused by scattered light, only PMT hits occurring within a time window of $\pm10$~ns of the prompt light peak were selected.

Since most signals and backgrounds were not correlated with the direction of the Sun's position, the angular resolution uncertainty did not generally have a significant effect.  The direction of electron-scattering (ES) events was however strongly correlated with the incoming neutrino direction and was well modeled by the function
\begin{equation}
P(\cos\theta_\odot)=\alpha_\mathrm{M}\frac{\beta_\mathrm{M}e^{\beta_\mathrm{M}(\cos\theta_\odot-1)}}{1-e^{-2\beta_\mathrm{M}}}+(1-\alpha_\mathrm{M})\frac{\beta_\mathrm{S}e^{\beta_\mathrm{S}(\cos\theta_\odot-1)}}{1-e^{-2\beta_\mathrm{S}}}, \label{ares}
\end{equation}
where $\beta_\mathrm{S}$ is the parameter for the exponential component associated with the main peak and $\beta_\mathrm{M}$ is associated with  the multiple scattering component.  To determine the systematic uncertainty on these parameters, the function in Eqn.~\ref{ares} was fitted using \NS\ calibration data and simulations.
In this analysis, $\cos\theta_\odot$ was replaced by the cosine of the angle between the reconstructed and the true electron direction, the latter being approximated by the fitted vertex position relative to the source, for events reconstructed 120~cm or more away from the source position.  Due to the correlation between these resolution parameters and the complexity associated with the smearing of angular resolution for single events using Eqn.~\ref{ares}, the uncertainty was propagated using the expression
\begin{equation} \label{ctt}
\cos\theta'_\odot=1+(\cos\theta_\odot-1)(1\pm\delta), 
\end{equation}
where $\delta=0.12$ is the relative uncertainty on $\beta_\mathrm{M}$ and $\beta_\mathrm{S}$ parameters.  This parameterization was shown to be a good approximation for ES events.

\subsection{Energy calibration of Cherenkov events}

The fundamental measure of event energy in SNO was the number of Cherenkov photons produced by fast electrons, and the most basic energy observable for Cherenkov events was the number of triggered PMTs $(\nhit)$.  The energy reconstruction algorithm discussed below attempts to determine the number of photons $(\ngamma)$ that would have been produced by an electron, given the reconstructed position $(\rrfit)$ and direction $(\uhfit)$ of the event in the detector, to yield the number of triggered PMTs observed.  An estimate of event kinetic energy $(\teff)$ can then be derived from the one-to-one relationship ($\fecal$) between electron kinetic energy $(\te)$ and the mean of the corresponding distribution of \ngamma~\cite{ft_footnote}. 

The derivation of \fecal\ was performed via Monte Carlo simulation. The only free parameter in the simulation, the average PMT collection efficiency (\epnot) for the PMT array, was determined by comparing the energy scale of \nst\ calibration data and its simulation (see Sec.~\ref{sec:ecal}).

\subsubsection{The energy reconstruction algorithm}

For an initial estimate of an event's equivalent electron kinetic energy \teff, a corresponding estimate of \ngamma\ could be calculated via ${\fecal}^{-1}$.  The number of photons expected to trigger the \ith\ PMT, $N_i$, was calculated as
\begin{equation}
  N_i=\ngamma\frac{\sum_\lambda R_i\left(\lambda,\rv,\uh\right) \frac{1}{\lambda^2}}{\sum_\lambda \frac{1}{\lambda^2}},
  \label{equ:pmtresponse}
\end{equation}
where $R_i$, discussed in detail in Sec.~\ref{sec:pmtri},  is the response of the \ith\ PMT to a photon of wavelength $\lambda$.  The sum over $\lambda$ was done in 10\,nm steps from 220 to 710\,nm, the wavelength range over which the detector was sensitive.  The total number of direct  PMT hits (i.e. not from reflected light) predicted by the energy reconstruction $(\npred)$ is then
\begin{equation}
  \npred=\sum_i^{\npmts} N_i M\left(N_i\right),
\end{equation}
where $M$ is a correction function that accounts for the possibility of multiple photons counting in the hit PMT.  The initial estimate of \ngamma\ was modified by
\begin{equation}
  \ngamma\rightarrow\frac{\neff}{\npred}\ngamma,
\end{equation}
where $\neff$ is the effective number of PMT hits, the number of PMT hits within the prompt light window after corrections for dark noise were made.  This process was iterated until agreement was reached between \npred\ and the noise-corrected number of prompt PMT hits.

\subsubsection{PMT optical response $(R_i)$\label{sec:pmtri}}

The optical response of the \ith\ PMT, $R_i$, in Eqn.~\ref{equ:pmtresponse} was calculated as%
\begin{equation}
  \begin{split}
    R_i&=\epsilon_i(\lambda,\rrfit,\pv_i,\nh)\;
    \Omega_i(\rrfit,\pv_i,\nh)\;
    D(\te|\rrfit,\pv_i,\uhfit) \; \times \\
    &\hspace{1cm} \qquad F(\rrfit,\pv_i)
    \exp\left(-\sum_{m=1}^3 d_m(\rrfit,\pv_i)\alpha_m(\lambda)\right),
  \end{split}
  \label{equ:pmtopticalresponse}
\end{equation}
where $\lambda$ is photon wavelength, $\rrfit$ is the reconstructed position, $\uhfit$ is the reconstructed direction, $\pv_i$ is the vector between the PMT position and $\rrfit$, $\nh$ describes the orientation of the PMT, and $\te$ is the true electron kinetic energy.  The terms in the sum over the media ($m=$~\hw, acrylic, and \lw) are the product of the average optical path length $(d_m)$ and the inverse attenuation length $(\alpha_m)$ measured in Sec.~\ref{sec:opticalparam} above.  The probability of photon transmission through the acrylic vessel $(F)$ was calculated from the Fresnel coefficients, $D$ is the Cherenkov angular distribution, $\Omega_i$ is the solid angle of the PMT as seen from $\rrfit$, and $\epsilon_i$ is the efficiency of photons to trigger a PMT upon their entering the light collecting region of the PMT and reflector assembly.  Each factor in Eqn.~\ref{equ:pmtopticalresponse} required the average optical path of a photon to be calculated from $\rrfit$, $\pv_i$, and the detector geometry.  The calculation was done to the center of the PMT photocathode surface.  $\Omega_i$ was determined from the optical paths to multiple points around the PMT light concentrator assembly.

The PMT efficiency $(\epsilon_i)$ was broken down into the factors
\begin{equation}
  \epsilon_i=\epsilon_{\textrm{PCE}} E(\lambda,\cos\theta_n) \epopt \epelec \epncd,
\end{equation}
where \epnot\ is the aforementioned average PMT collection efficiency, $E$ is the relative response as a function of incidence angle $(\theta_n)$ for a typical PMT, and \epopt\ and \epelec\ are the relative optical and electronic channel efficiencies.  $E$ was determined at multiple wavelengths by a combination of optical calibrations and Monte Carlo simulations (see Sec.~\ref{sec:opticalparam}).  It was normalized to the quantum efficiency for a typical PMT, which was also a function of wavelength, at normal incidence.  In the central region of the detector, the obstruction of photons by the NCD array reduced the average number of direct PMT hits by up to 20\%.  The efficiency factor $(\epncd)$, calculated based on a Monte Carlo simulation of Cherenkov light from single electrons in the \dto, accounted for this effect.  

\subsubsection{Energy calibration\label{sec:ecal}}

Once the detector optical parameters have been determined, the PMT collection efficiency $(\epnot)$ and the energy calibration function $(\fecal)$ were still to be set.  High rate ($\sim200$\,Hz) central \nst\ source calibration runs were compared to Monte Carlo simulations using an initial estimate of $\epnot=0.645$ (as was determined for Phase II).  This value was adjusted to match the means of the \ngamma\ distributions obtained from the source data and from simulations.  The result was a value of \epnot=0.653 for the third phase.  

\begin{figure}
  \includegraphics[width=0.80\textwidth]{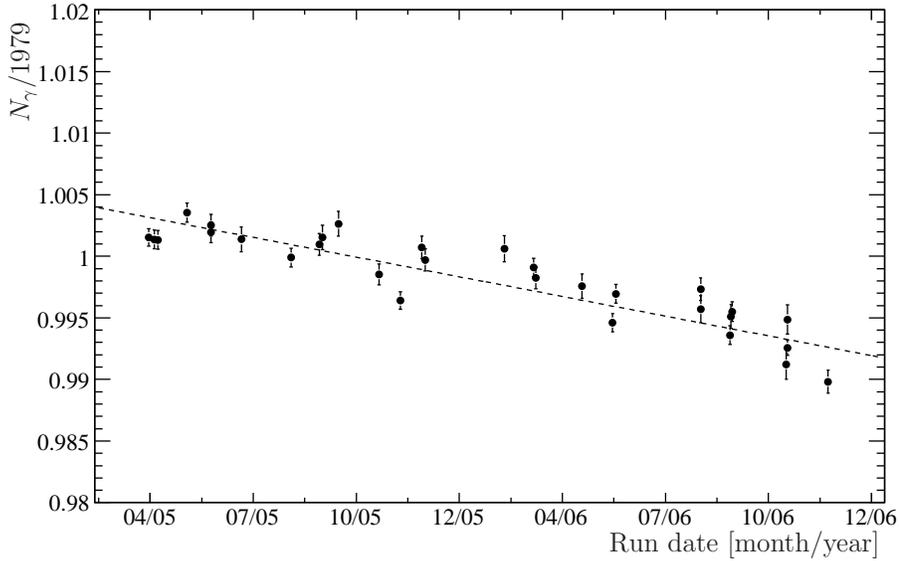}
  \caption{\label{fig:ndrift}Relative photon collection efficiency of selected \nst\ calibration runs as a function of time.  The dashed line represents \ddrift\ which was scaled to unity on September 27, 2005.}
\end{figure}

Figure~\ref{fig:ndrift} shows the relative photon collection efficiency of the detector using the same central \nst\ calibration runs as above.  The slight time variation, not accounted for by energy reconstruction, was described by the function
\begin{equation}
  \ddrift=1.197-1.751\times10^{-5}t,
\end{equation}
where $t$ is the number of days since the reference date December 31, 1974 (UTC).  The function was arbitrarily normalized to unity on September 27, 2005.  As the PMT collection efficiency in the Monte Carlo simulation was tuned to track the time variation observed in Fig.~\ref{fig:ndrift}, a correction $\teff\rightarrow\teff/\ddrift$ was required to be performed on both the reconstructed energy of data and simulated events.

To derive the energy calibration function $\fecal$, a series of mono-energetic electron Monte Carlo simulations, from 2 to 130~MeV, were performed.  The mean of the \ngamma\ distribution for each set of simulated electron events, corrected by \ddrift, was then matched to the known electron kinetic energy $\te$.  \fecal\ consisted of the interpolation between these values.  The average electron energy response can be characterized by a Gaussian function with resolution $\sigma_T =$ $-0.2955 + 0.5031\sqrt{T_e} + 0.0228 T_e$, where $T_e$ is in MeV.  

\begin{figure}
  \includegraphics[width=0.80\textwidth]{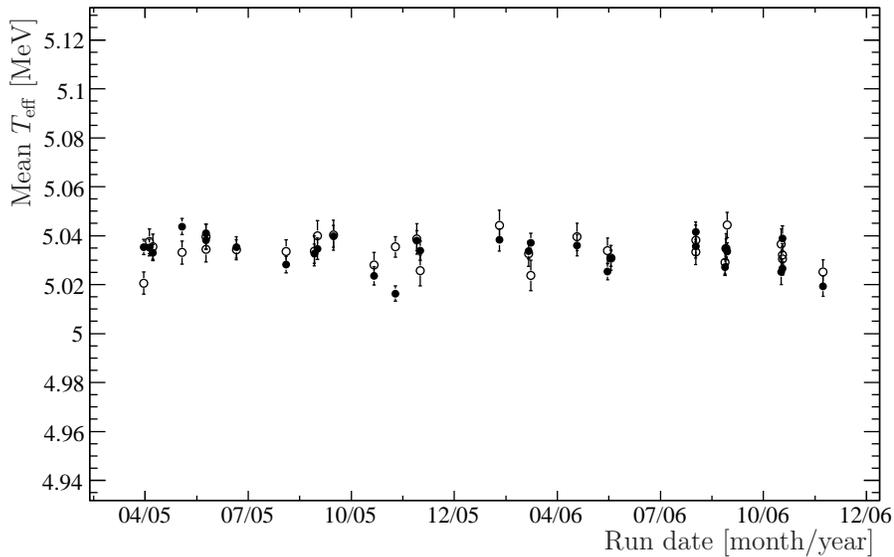}
  \caption{\label{fig:escale}The mean effective electron kinetic energy of selected central \nst\ calibrations (solid points) and Monte Carlo simulation (open points).}
\end{figure}

Figure~\ref{fig:escale} shows the resulting mean \teff\ of selected \nst\ calibration runs (solid points) after application of the drift correction.  The mean \teff\ of the full Monte Carlo simulation with $\epsilon_{\textrm{PCE}}=0.653\,\ddrift$ are also shown (open points).

\subsubsection{Energy systematic uncertainties\label{sec:egysys}}

Two of the most important systematic uncertainties on the measured neutrino fluxes were the energy scale and energy resolution of the PMT array.  These uncertainties were determined by comparing data and MC simulations using the $^{16}$N source.  This source was deployed on nearly a monthly basis and was used to probe not only the center of the detector but also to scan along the $x$, $y$, and $z$ axes. A total of 1053 $^{16}$N runs were used in the energy systematics analysis.  

In order to correctly determine the energy of an event, the energy estimator
must use the number of working PMTs.  The number
of non-working PMTs considered as working by the energy estimator could be
approximated by counting the number of PMTs that fell outside a region of
$5\sigma$ from the average PMT occupancy.  The ratio of this number by the total number of working PMTs for a run was taken as the potential influence of
the uncertainty of the detector state on the estimated energy.  The
uncertainty due to the detector state was determined to be 0.03\%.

The temporal stability of the energy response of the data and Monte
Carlo was evaluated using the $^{16}$N runs taken at the center of the
detector.   A comparison of the mean kinetic energy and resolution distributions in data and simulation was used to
determine the temporal stability systematic uncertainty.  The energy
drift/stability uncertainty on the energy scale was found to be 0.40$\%$ and that on the resolution was determined to be 1.19$\%$.

\begin{figure}
\begin{center}
\includegraphics[width=0.75\textwidth]{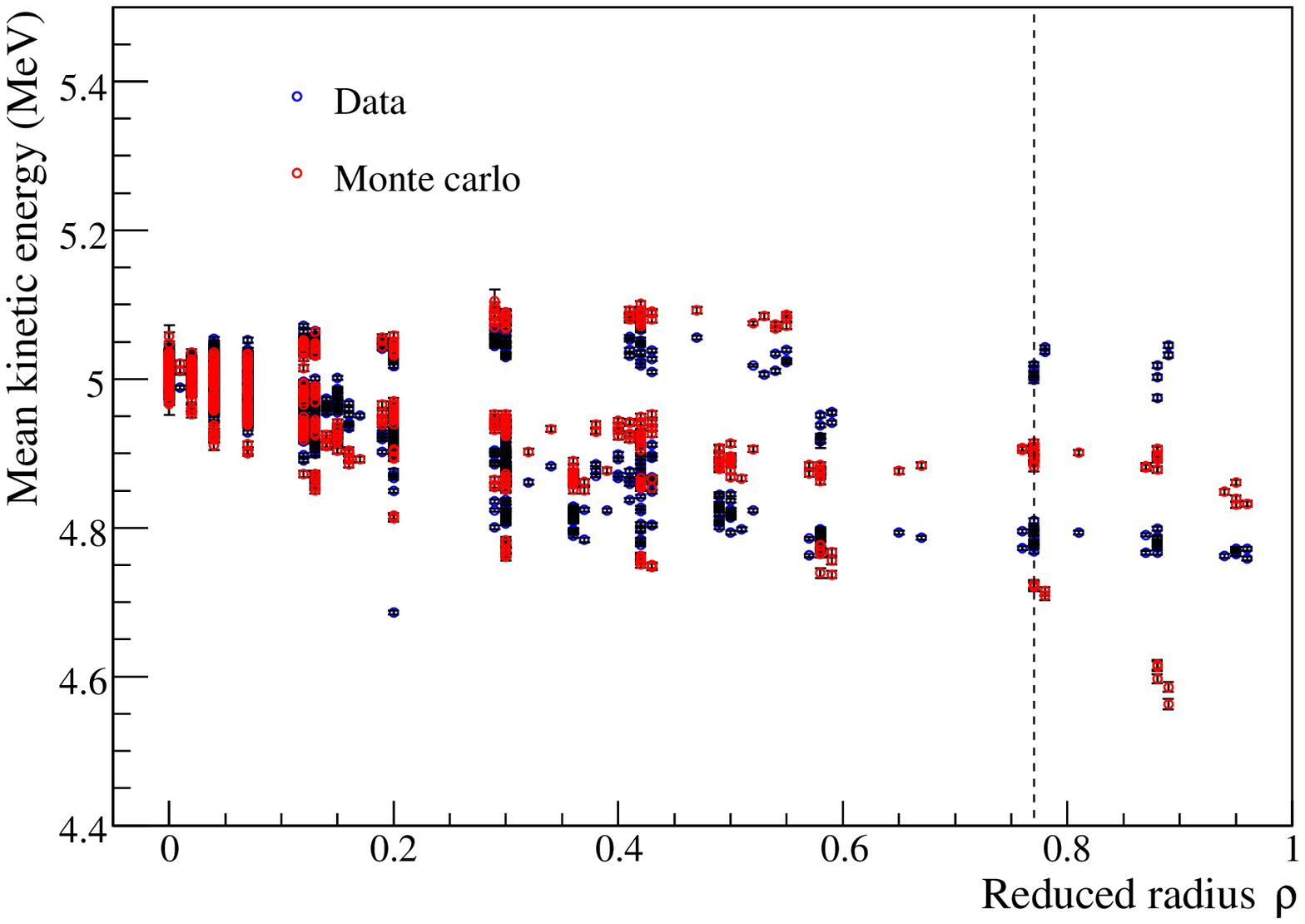}\\
\includegraphics[width=0.75\textwidth]{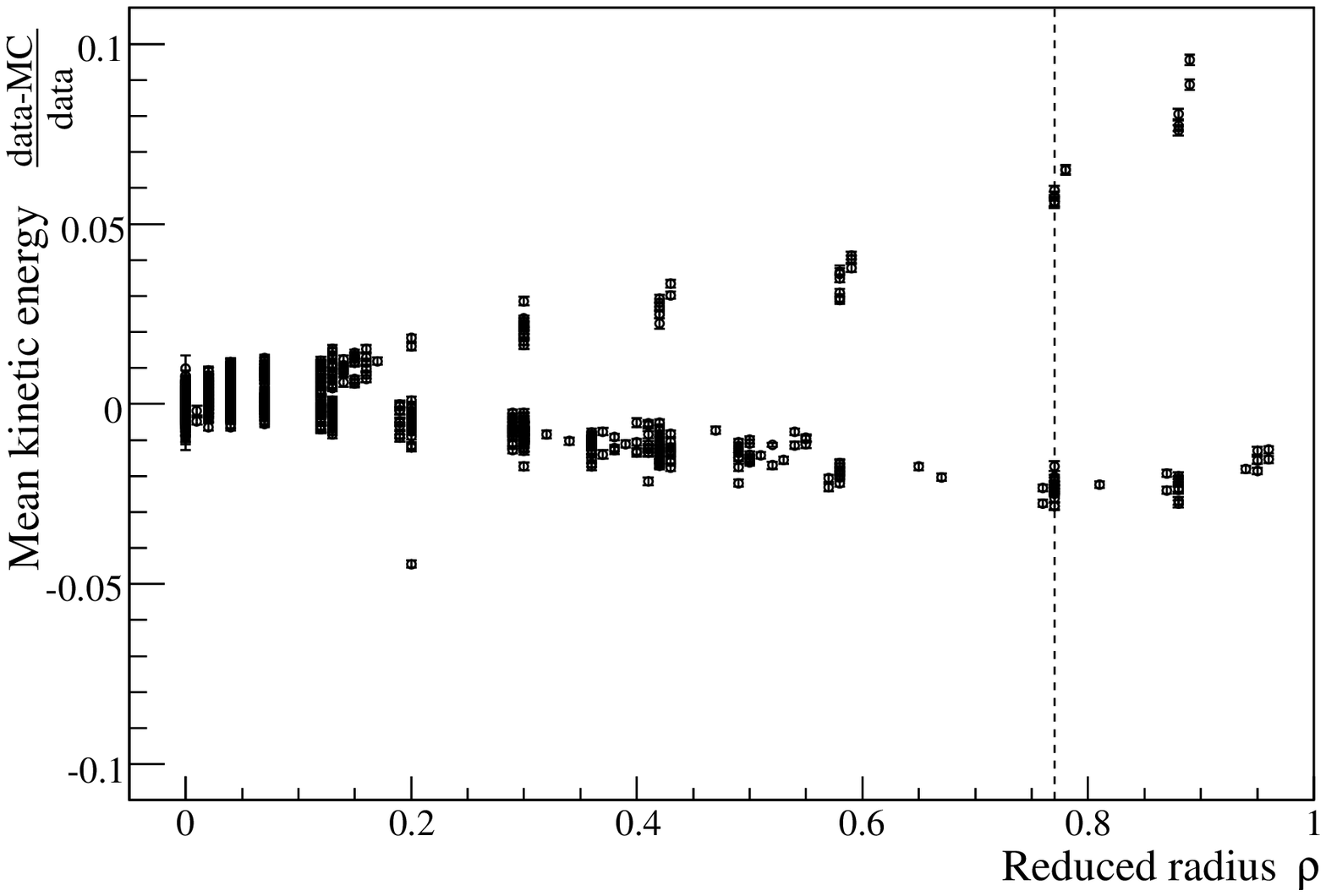}
\caption{\label{fig:energysys}The top plot shows the mean effective kinetic energy $\teff$ versus volume-weighted radius $\rho$  for the $^{16}$N source deployed throughout the detector.  The
bottom plot shows the fractional difference between data and Monte Carlo (the ratio (data-MC)/data) for the  $^{16}$N source
deployed throughout the detector.  The dashed line at $\rho=0.77$ represents the edge of the fiducial volume at 550 cm.  Points at the same value of $\rho$ could have different energy response in data or Monte Carlo because of local point-to-point variation in detector response.   The AV neck and the NCD array accentuated the variations at large $\rho$. }
\end{center}
\end{figure}

One of the  largest contributions to the energy scale and to the
energy resolution uncertainties was due to spatial variations in the detector.  Figure~\ref{fig:energysys} shows a comparison of $\teff$ between data and
MC simulations as a function of the volume-weighted position $\rho$=$\rfit/\rav$, where $\rav=$~600~cm is the radius of the acrylic vessel. 
The point-to-point variations and radial biases of the detector response were
determined by dividing the detector into radial and polar angle
bins assuming an azimuthal symmetry.  The average differences between data and
MC simulations and the variance of the mean kinetic energy and resolution were determined in each of these bins.  The volume-weighted average of these differences
among the bins was then taken as the spatial variation on the energy
scale and resolution.  The uncertainty on the spatial variation of the
energy scale was determined to be 0.64\% and the spatial variation of
the energy resolution was determined to be 1.04\%.

The $^{16}$N source was typically run at a rate on the order of several
hundred hertz whereas neutrino
data were taken with the detector operating at an event rate of an order
of magnitude lower.  In order to evaluate the potential rate
dependence, the $^{16}$N source was periodically run at `low rate'
(several hertz). Comparing the mean kinetic energy  of low-rate runs
taken close in time to high-rate runs,  the
uncertainty on the energy scale related to the rate dependence was determined to be 0.20\%.

A series of dedicated PMT high-voltage scans were performed to
quantify the dependence of the detector response on the PMT gain.  Inspecting the value of the upper half-maximum height of the single photoelectron
charge distribution  allowed for an
estimation of the gain effects on the energy scale uncertainty.  The
ratio of the value of the upper half-maximum height of the single photoelectron
charge distribution at the nominal $^{16}$N
energy compared to this slope allowed for a conservative estimation of
the energy scale systematic due to gain changes.  This uncertainty was determined to be 0.13\%.

A series of dedicated threshold scans were performed to
quantify the dependence of the detector response on the PMT channel  
threshold. Comparing the ratio of the
energy scale to the value of the lower half maximum height of the single
photoelectron charge distribution allowed for an estimation of uncertainty on the energy scale due to 
threshold changes.  This uncertainty was conservatively estimated 
at 0.11\%.

The energy scale could be affected by the change in the timing
position of the prompt light peak.  This is because the energy estimator utilized only
the prompt light within a limited time window of 20~ns.  In order to
determine an uncertainty due to changes in the time residual position
compared to the energy estimator's time window, the mean and width  of the prompt timing peak were determined by
a Gaussian fit on  all the central $^{16}$N runs. The extracted timing
peak means and widths were quite
stable and similar to what were seen during previous phases.  Since
there were no indications that the timing had changed from the last
phase, the uncertainty from the second phase, 0.10\%, was used and was considered conservative.

Incomplete modeling of the $^{16}$N source contributed to the energy scale
uncertainty as Monte Carlo simulations of the source were used to compare to real data.  Effects such as approximations to the source geometry, uncertainties in the $^{16}$N decay branching ratios, and the finite step size in {\sc EGS4}~\cite{EGS4} simulation were studied.  Their combined contribution to the energy scale uncertainty was determined to be 0.65\%.

Table \ref{tab:energy_systematics} summarizes the contributions to the
systematic uncertainty on the energy scale and resolution.  The energy scale
uncertainty contributions were added in quadrature to yield an overall
uncertainty of 1.04\%.  The resolution uncertainty contributions were
determined to be a shift of 1.19\% along with an uncertainty of 1.04\%. 

\begin{table}
\begin{center}
\caption{Summary of PMT array's energy scale and resolution systematic uncertainties.}
\label{tab:energy_systematics}
\begin{minipage}[t]{5in}
\begin{ruledtabular}
\begin{tabular}{l r}
\multicolumn{2}{c} {Scale uncertainty} \\
\hline
Source             &Uncertainty  \\
\hline
Detector state             & 0.03\% \\
Drift/stability (data-MC)  & 0.40\% \\
Spatial variation          & 0.64\%  \\
Gain                       & 0.13\% \\
Threshold                  & 0.11\% \\
$^{16}$N source modeling               & 0.65\% \\
Rate dependence            & 0.20\% \\
Time calibration           & 0.10\% \\   
\hline
Total                     & 1.04\%\\
\hline\hline
\multicolumn{2}{c} {Resolution uncertainty} \\
\hline
Source         &Uncertainty \\ 
\hline
Spatial variation         & 1.04\%\\
\hline\hline
\multicolumn{2}{c} {Resolution shift} \\
\hline
Source         &Shift \\ 
\hline
Drift/stability (data-MC)  & 1.19\%\\
\end{tabular}
\end{ruledtabular}
\end{minipage}
\end{center}
\end{table}

\section{Electronic calibration of NCD array\label{sec:ncdresponse}}

As described in Sec.~\ref{sec:ncddescr} the data stream of the NCD array consisted of events from the shaper-ADC and the MUX-scope subsystems.  The shapers could provide the total charge in an event and the MUX-scope subsystem could digitize the log-amplified waveform of the signal.  The primary goal of the electronic calibration was to measure the parameters of the electronic model, so that the transformations of the counter signals as they propagated through the front-end electronic and data acquisition systems were quantified.  A calibration system was implemented to pulse the preamplifiers.  The output of a programmable waveform generator was attenuated by 30~dB and injected into a pulse distribution system (PDS) board, whose amplified outputs could be sent to selected preamplifiers through computer control.  

\subsection{Linearity}

\begin{figure}[htbp]
\includegraphics[width=0.80\columnwidth]{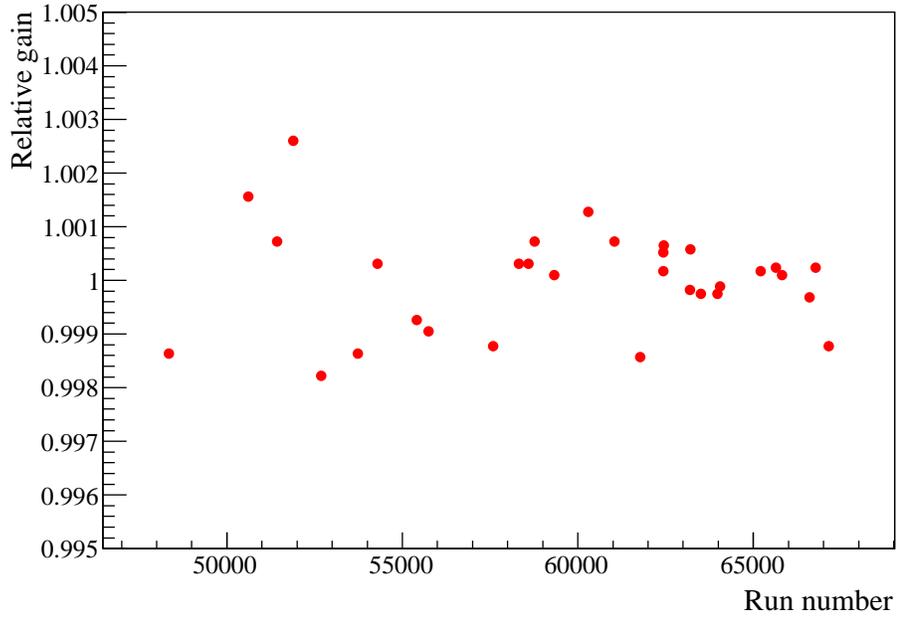}
\includegraphics[width=0.80\columnwidth]{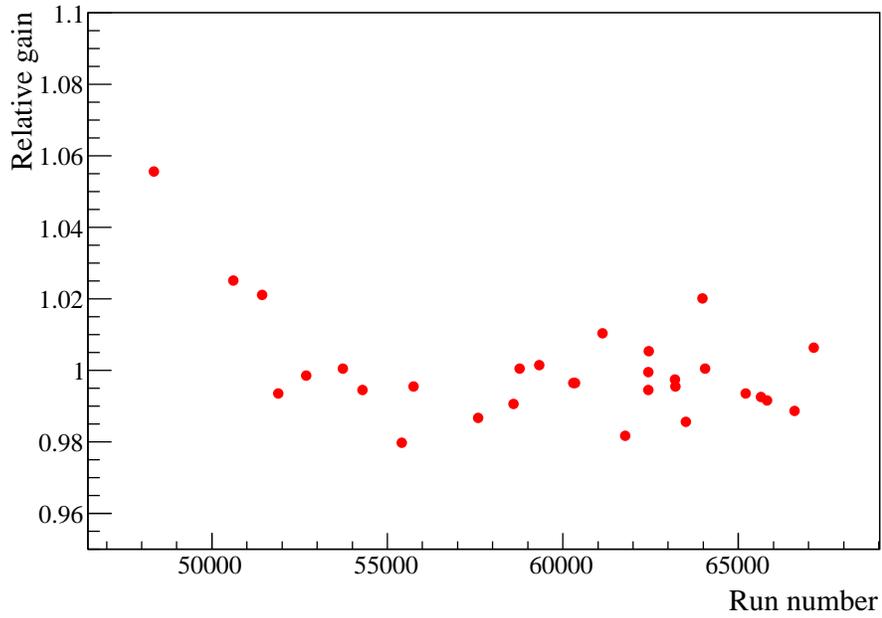}
\caption{\label{fig:ncligain} Temporal variation of the relative gain in the shaper-ADC (top) and MUX (bottom) channels for string N1.  In these plots the gain measured from each monthly extended electronic calibration was normalized to the mean gain.   The data shown here extended from the commissioning (prior to run~50000) to the completion of Phase-III data taking ($\sim$run~67000).}
\end{figure}

The gain and  linearity of the shaper-ADC and MUX-scope channels were calibrated by sending rectangular pulses of known amplitudes to the preamplifiers, one preamplifier at a time.   These calibrations were performed once a week at five different pulse amplitudes.  Extended electronic calibrations with twenty different pulse amplitudes over an expanded range were performed monthly.  Rectangular pulses were used since their start and stop times were easily determined, which facilitated the integration of digitized waveforms in the analysis.  The measured charge of the signal as a function of the calculated input charge was fit to a linear function, which measured the gain and offset of each channel and tested the channels' linear response. Since the digitized waveforms were logarithmically amplified and recorded by the digital oscilloscopes, they were first de-logged (inverting Eqn.~\ref{eq:logamp} below) for the MUX-scope channel linearity analysis. This tested the linearity of the MUX-scope system as well as the measured log-amp parameters.   The shaper-ADC channels were found to be linear to within 0.5\%  across all channels.  The transfer function of the logarithmic amplifier was responsible for an observed non-linearity of up to $\sim$5\%, and a model was developed to account for this behavior.  Figure~\ref{fig:ncligain} shows the temporal variation of the relative gain in the shaper-ADC and MUX channels, measured from the monthly extended calibration runs in Phase III, for string N1.

The shaper-ADC channel corresponding to the mean of neutron signal peak in Am-Be calibrations was used to determine the conversion gain for the measured shaper-ADC charges to event energies in the $^3$He counters.  For the $^4$He counters, which were insensitive to neutrons, the energy peak from $^{210}$Po alpha decays was used.

\subsection{Threshold\label{ncdthrescal}}
The threshold calibration involved injecting offset, single-cycle, sine waves with a constant width and varying amplitudes to all the preamplifiers simultaneously.  The range of these pulser output amplitudes extended above and below each channel's threshold level.  Sine waves were chosen for this measurement because the calibration pulse amplitude and calculation of the total injected charge were more stable with sine waves than with signals that are not smooth, such as a rectangular or triangular pulses. The high-frequency components of non-smooth waveforms could produce transient currents, which could be difficult to calculate accurately for each channel.

\begin{figure}[htbp]
\includegraphics[width=0.80\columnwidth]{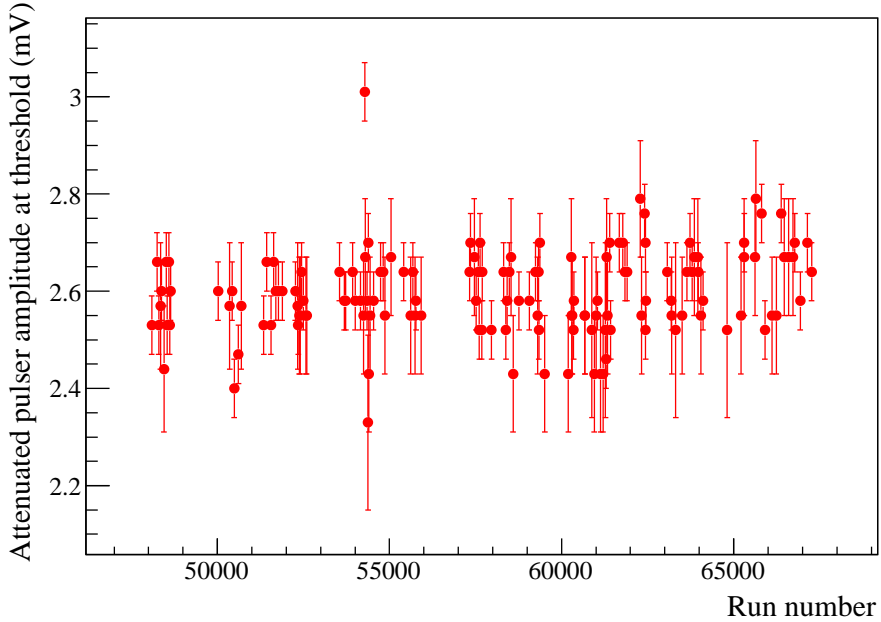}
\caption{\label{fig:ncth} Temporal variation of the shaper-ADC channel threshold for string N1.  In the threshold calibration, pulses from the pulser were attenuated by 30~dB.  The channel threshold was determined by finding the attenuated pulser amplitude at which half of the expected events were observed.  The data shown here extended from the commissioning (prior to run~50000) to the completion of Phase-III data taking ($\sim$run~67000).}
\end{figure}

The threshold levels of the shaper-ADC and MUX channels were determined by finding the pulser amplitude at which half of the expected events were observed.  The algorithm searched over the range of pulser amplitudes, estimating the charge and current thresholds of the shaper-ADC and MUX channels. The thresholds were stable over the course of the experiment except for when they were intentionally changed to account for sporadic electromagnetic pickup or a malfunctioning string.  In the latter case, the thresholds for the channels associated with the malfunctioning string were set to their maximum values to ensure that they were offline.  Figure~\ref{fig:ncth} shows the temporal variation of the shaper-ADC channel threshold for string N1 during Phase III.

\subsection{Log-amp}

The log-amp calibration pulse was an offset, single-cycle 1-$\mu$s-wide sine wave preceded 6~$\mu$s by a narrow rectangular trigger pulse.   These pulses were injected to each preamplifier channel at 3~Hz for a duration of 15~seconds.  A sine wave was selected because its smoothly varying shape and frequency characteristics were similar to the expected counter signals.   If the sine wave were used to trigger the channel, it could be possible that the beginning of the pulse would not be recorded as the time for the sine wave to go from zero to the MUX threshold level might be longer than the electronic delay time.  The width of the trigger pulse was set to a small value in order to reduce the amount of baseline offset produced by its integrated charge before the sine wave arrived at the input.

The logarithmic amplification in the MUX electronic chain was modeled as
\begin{equation}
\label{eq:logamp}
V_{\textrm{log}}(t) = A \cdot \log_{10}\left(1+\frac{V_{\textrm{lin}}(t - \Delta t)}{B}\right) + C_{\textrm{chan}} +V_{\textrm{PreTrig}},
\end{equation}
where $V_{\textrm{log}}$ and $V_{\textrm{lin}}$ are the logarithmic and linear voltages,  $\Delta t$ represents the time delay for each channel in the MUX, and $A$, $B$, $C_{\textrm{chan}}$, and $V_{\textrm{PreTrig}}$ are constants determined by calibrations.   Details of this parameterization of the MUX-scope electronic chain can be found in Ref.~\cite{cox}.

The log-amp calibration analysis involved a $\chi^2$-minimization that estimated the set of five log-amp parameters that best fit a simulated signal to each measured calibration pulse.  A weighted average of the parameter values extracted for each event, along with the uncertainty of the weighted average, was calculated for each NCD electronic channel.  Because some of these parameters were dependent on which of the two oscilloscopes recorded the event, there were two sets of log-amp parameters to be measured for each string. The electronic calibrations measured these parameters and the current threshold level of digitization separately for each string.

\section{Neutron detection efficiency calibration\label{sec:neutron}}

The NC interaction produced a uniform distribution of neutrons in the D$_{2}$O volume.   The primary method for  determining the neutron capture efficiency of the NCD array was to deploy an evenly distributed $^{24}$Na source in this volume~\cite{na24_nim_paper}.  Two such calibrations were performed in October 2005 and in November 2006.

In these ``spike'' calibrations, about one liter of neutron-activated brine containing $^{24}$Na was injected into the D$_{2}$O volume and mixed.  The $^{24}$Na isotope, with a 14.96-hour half-life, decays to $^{24}$Mg, almost always producing a beta decay electron with an end-point of 1.39~MeV and two gamma rays with energies of 1.37~MeV and 2.75~MeV.  The 2.75-MeV gamma ray is capable of photodisintegrating a deuteron, the binding energy of which is 2.225~MeV.  A perfectly mixed spike would produce a uniform distribution of neutrons up to within about 30~cm (one Compton scattering length) of the acrylic vessel, where the neutron intensity would drop off because the 2.75-MeV gamma ray has a significant chance of escaping the D$_{2}$O volume.   A correction factor, based on simulations of neutrons from NC interactions and from photodisintegration due to an evenly distributed $^{24}$Na brine, was applied to compensate for the differences in the neutron energy and neutron radial distribution between the solar neutrino and the spike calibration data.  After this correction, the neutron capture efficiency from the $^{24}$Na calibration was equal to the efficiency for neutrons produced by the NC interaction. However, in calculating the efficiency by this technique and its uncertainty, possible deviations from perfect mixing that might occur near boundaries, such as the walls of the acrylic vessel and the NCD strings, were also considered as described below.

The temporal behavior of the NCD array response was monitored by deploying $^{252}$Cf and AmBe sources in different parts of the \dto\ volume using the source manipulator system~\cite{NIM}.   The data from these point calibrations were also used to calibrate the Monte Carlo code.  The technique that was used to tune the Monte Carlo is discussed in Sec.~\ref{sec:ndiscrete}.

In our previous paper~\cite{ncdprl}, the reported neutron capture efficiency of the NCD array was measured with the $^{24}$Na spikes, and the neutron detection efficiency of the PMT array was calculated from the calibrated Monte Carlo code.  A subsequent analysis of the PMT array's neutron detection efficiency using the $^{24}$Na data is reported in Sec.~\ref{sec:pmtnaspike}.  The results from direct $^{24}$Na calibration and calculations from the tuned Monte Carlo are consistent.

\subsection{Inputs for determining the neutron capture efficiency with a uniformly distributed $^{24}$Na source}

The input elements needed to determine the neutron capture efficiency for the NCD array and the neutron detection efficiency for the PMT array are made explicit in the following formula:
\begin{equation}
\label{eqn:eff_sol}
\epsilon_{\textrm{sol}} = f_{\textrm{non-unif}} \cdot f_{\textrm{edge}} \cdot \epsilon_{\textrm{spike}} \; ,
\end{equation}
\noindent where $\epsilon_{\textrm{sol}}$ is the capture or detection efficiency for neutrons produced by solar neutrinos and $\epsilon_{\textrm{spike}}$ is the efficiency determined from the $^{24}$Na spike calibration.  The two factors multiplying $\epsilon_{\textrm{spike}}$ correct for differences in the distribution of neutrons in solar neutrino and $^{24}$Na calibration data.  The factor $f_{\textrm{edge}}$ accounts for the differences in the neutron energy and neutron radial distribution, while $f_{\textrm{non-unif}}$ is a factor that accounts for the effect of possible non-uniformity of the activated brine in the D$_{2}$O.

The neutron capture efficiency of the NCD array and the neutron detection efficiency of the PMT array measured from $^{24}$Na calibration are given by the ratio of the observed signal rate $R_{\textrm{spike}}$ at a reference time and the $^{24}$Na source strength $A_{^{24}\textrm{Na}}$ at that time:
\begin{equation}
\label{eqn:eff_spike}
\epsilon_{\textrm{spike}} = \frac{R_{\textrm{spike}}}{A_{^{24}\textrm{Na}}}.
\end{equation}

\subsubsection{$A_{^{24}\textrm{Na}}$: The $^{24}$Na source strength measurement\label{subsec:a_na24}}

The strength of the $^{24}$Na source was determined using three different detectors: a germanium detector (\textit{ex situ} measurement), the SNO PMT array (\textit{in situ}), and the NCD array (\textit{in situ}). In the following discussion, the quantity to be calibrated is the rate of neutrons produced by an encapsulated sample of the $^{24}$Na brine placed at the center of the SNO detector. For the calibration with the germanium detector, the total rate of 2.75-MeV gamma rays produced by the $^{24}$Na brine sample was measured and a Monte Carlo simulation program was used to calculate the expected rate of neutrons produced in the heavy water if the $^{24}$Na were positioned at the center of the SNO detector~\cite{james_loach_phd}. For the calibration measurement using the SNO PMT and NCD arrays, a comparison was made between the neutron rates observed in these detector arrays when the $^{24}$Na brine sample was placed at the center of SNO and the rate observed when a well calibrated $^{252}$Cf neutron source was placed at the same location, with a small further correction determined from  simulations to deal with the different neutron spatial and energy distributions of the two sources.

In the germanium detector measurement, a small sample of the activated brine, with a mass measured to better than 0.5\%, was placed on the detector (the liquid was contained in a Marinelli beaker~\cite{na24_nim_paper}) and the rate of the 1.37-MeV and 2.75-MeV gamma rays was measured.  The measured rate was converted to the gamma radiation rate of the sample by correcting for the detector dead time, gamma ray acceptance, and the effect of the detector's dead layer.  Given this measurement and the photodisintegration cross section of the deuteron at 2.75~MeV (with a 2\% uncertainty~\cite{james_loach_phd}), the effective source strength if it were placed at the center of the SNO detector at a reference time could be determined in units of neutrons per second.

Unlike the above measurement, the SNO PMT and NCD arrays were used to measure directly the neutron production rate due to a small activated brine sample in the D$_{2}$O.  A 10 g sample (measured to better than 0.5\%) was placed in a sealed container and deployed to the center of the D$_{2}$O volume using the calibration source manipulator.  The 2.75-MeV gamma rays from the activated brine mostly escaped the container without any interactions; about one in 385 of these gamma rays photodisintegrated a deuteron to produce a free neutron.  Gamma rays that interacted with the container lost energy by Compton scattering; when the effect of this energy loss was included, the average photodisintegration probability decreased to about 1/390 or 1/395, depending on the type of source container used (different ones were used in 2005 and 2006).  The decrease in probability was estimated with a 0.7\% uncertainty using Monte Carlo simulation.

In the \textit{in situ} SNO PMT array measurement, the PMT array detected the Cherenkov radiation produced by the 6.25-MeV gamma ray following the capture of these neutrons on deuterons.  It was necessary to apply event selection cuts on the reconstructed radius and the energy in order to isolate these events from those due to background noise from the beta and gamma rays produced directly by the $^{24}$Na decay.  $\teff$ was required to be between 5 and 9.5~MeV.  This event selection criterion discriminated against radioactive background events, which had an average energy of about 3.0 to 3.5~MeV.  The reconstructed event vertex was required to be 200 to 450~cm away from the source.   This selection effectively removed background gamma-ray events whose range was limited by the Compton scattering length.

The calibration of the neutron rate from the brine sample was obtained by comparing this rate to the rate from the $^{252}$Cf neutron source, whose strength was known with an uncertainty of 0.7\%~\cite{nsp}.  The rate measurement from the $^{252}$Cf source was obtained within a few days of the $^{24}$Na measurement so that the detector condition would be as similar as possible.  Neutrons produced by the  $^{252}$Cf source had a similar, but not identical, radial capture profile in the D$_{2}$O compared to that from the brine.  Whereas the neutrons from the $^{252}$Cf source were produced inside the source container, those from the $^{24}$Na source were produced by photodisintegration in a sphere of radius about 30~cm.  After production, the neutrons typically diffused by about 100~cm, so these initial differences were, to a great extent, mitigated.  Yet, the capture profiles were different enough to introduce significant uncertainty in the $^{24}$Na source strength measurement.  Detector simulation was used to determine the effect of this difference; the capture efficiency of neutrons passing the radial selection cut was found to be about 2\% smaller for the $^{24}$Na source than for the $^{252}$Cf source.  The combined statistical and systematic uncertainties were about as large as this correction.

The \textit{in situ} measurement with the NCD array was performed almost identically as that with the PMT array and data from the same runs were analyzed.  The main difference was the detection of the neutrons with the NCD array instead of with neutron captures by deuterons.  As with the PMT array measurement, the neutron detection rate in the NCD array  was measured with the activated brine in the detector, and this rate was divided by the measured rate of neutrons from the $^{252}$Cf source.  The ratio multiplied by the known $^{252}$Cf source strength gave a good estimate of the brine source strength.  Again, as with the PMT array data, the difference in the radial neutron capture profile required a correction which was obtained using detector simulation.  The neutron capture efficiency for the $^{24}$Na source was about 2\% greater than that for the $^{252}$Cf source.  The uncertainty on this figure was much smaller (about 0.4\%) than for the corresponding one for the PMT measurement because neutrons were captured much closer to the production region so that there was less reliance on the accuracy of modeling neutron diffusion to large radii.

\begin{figure} 
   \centering
   \includegraphics[width=0.70\textwidth]{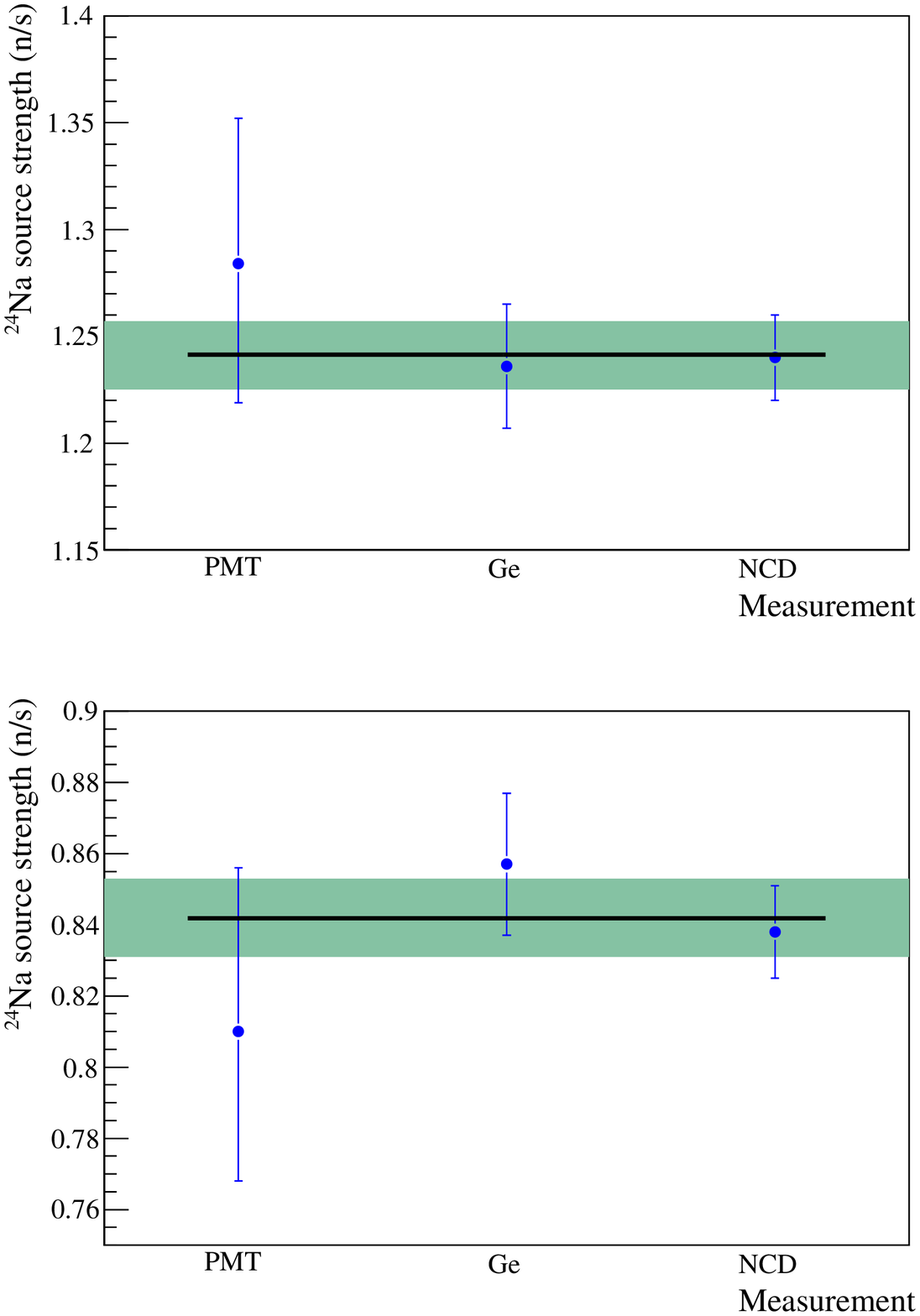} 
   \caption{The results of the brine source strength measurements in 2005 (top) and 2006 (bottom).  The vertical axis in each frame shows the source strength $A_{^{24}\textrm{Na}}$ (in neutrons per second) in the SNO detector, deduced from the three different sample source strength $A_{\textrm{samp}}$ measurements (horizontal axis), at a reference time (see text and Eqn.~\ref{eq:neutronrate} for details).  This reference time ($t=0$) was defined as the beginning of the first run when the brine was judged to be well mixed.  The error bars show combined statistical and systematic uncertainties.  The line through the data points is the result of the best fit, while the band shows the uncertainty of the fit.}
   \label{fig:na24_ss}
\end{figure}

\begin{table}
\begin{center}
\caption{\label{tab:na24_ss_fit}The result of fitting a constant value to the data points in Fig.~\ref{fig:na24_ss}.}
\begin{minipage}[t]{5in}
\begin{ruledtabular}
\begin{tabular}{lccc} 
Data Set & Fit $A_{^{24}\textrm{Na}}$ (n/s) & Percent uncertainty & Fit $\chi^2/\mbox{d.o.f.}$ \\ \hline
2005 & $1.241 \pm 0.016$ & $1.3\%$ & $0.22$ \\ 
2006 & $0.842 \pm 0.011$ & $1.3\%$ & $0.59$ \\ 
\end{tabular}
\end{ruledtabular}
\end{minipage}
\end{center}
\end{table}

The results for the total neutron production rate $A_{^{24}\textrm{Na}}$ from the 2005 and 2006 measurements are shown in Fig.~\ref{fig:na24_ss} and Table~\ref{tab:na24_ss_fit}.  These numbers were derived from the actual measured values of the source strength in the following manner:
\begin{equation}
\label{eq:neutronrate}
A_{^{24}\textrm{Na}} = f_{\textrm{P}} \cdot \frac{m_{\textrm{main}}}{m_{\textrm{samp}}} \cdot e^{-\Delta t / \tau_{\textrm{24Na}}} \cdot A_{\textrm{samp}} \; ,
\end{equation}

\noindent where $A_{\textrm{samp}}$ is the actual measured source strength of the sample of mass $m_{\textrm{samp}}$ (10-30~g), while $A_{^{24}\textrm{Na}}$ is the derived source strength of the main body of brine ($m_{\textrm{main}}=$~500 or 1000~g) that was injected into the SNO detector several days later.  The exponential correction factor $e^{-\Delta t/\tau_{\textrm{24Na}}}$ corrects for the exponential decay of the source strength between the time of measurement and the reference time.  The time offset $\Delta t$ for the run in 2005 was about 4 days, while in 2006, it was 6 days.  The initial source strength was much larger in 2006, so more time was required to allow the source to cool down to levels that the SNO data acquisition system could handle.  The quantity $\tau_{\textrm{24Na}}$ is the $^{24}$Na lifetime, 21.58 hours.  The quantity $f_{\textrm{P}}$ applies only to the \textit{in situ}  measurements, and corrects for the effects (discussed several paragraphs above) due to the radial distribution of neutron production and to a small fraction of gamma rays that scattered off the source container: $f_{\textrm{P}} = 1.0122 \pm 0.0044~(\mbox{stat.}) \pm 0.0053~(\mbox{syst.})$ for the 2005 run, and $f_{\textrm{P}} = 1.0288 \pm 0.0050~(\mbox{stat.}) \pm 0.0053~(\mbox{syst.})$ for the 2006 run, where the statistical uncertainty is due to Monte Carlo statistics.  The value of $f_{\textrm{P}}$ differed between 2005 and 2006 because different source containers were used.

The brine source strength measurements from the three techniques were combined by taking a weighted average.  The weighting of each data point was inversely proportional to the quadratic sum of the statistical and systematic uncertainties.  Thus the PMT array measurement, which had the largest uncertainty, made only a minor contribution to the final result.  The germanium detector and NCD array measurements, having similar uncertainty magnitudes, contributed about equally.  The large uncertainty in the PMT array measurement was, in large part, due to the difficulty in determining the neutron diffusion profile beyond 200~cm from the source, which was necessary to obtain a pure sample of neutron events.

\subsection{Neutron capture efficiency of the NCD array\label{neffncd}}

\subsubsection{$R_{\rm{spike}}$: Neutron capture rate measurement by the NCD array\label{subsec:r_spike}\label{sec:rspike}}

\begin{figure} 
\begin{center}
   \includegraphics[width=0.80\textwidth]{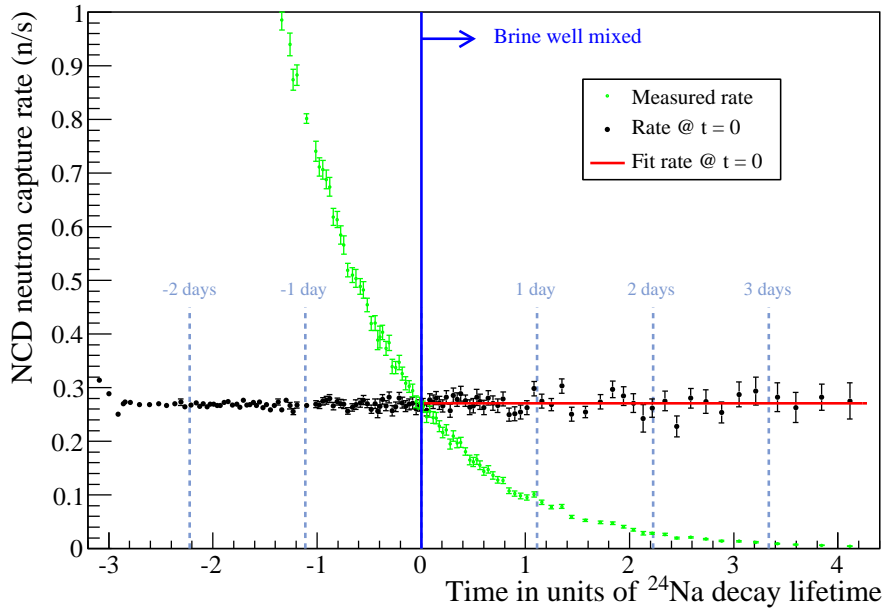} 
\end{center}
   \caption{Neutron capture rate in the NCD array as a function of time, from the $^{24}$Na spike in 2005.  The vertical axis is in units of neutron events per second, while the horizontal axis is time in units of $^{24}$Na mean lifetimes (one lifetime = 21.58 hours).  The exponentially decaying points are the actual measured rate $R(t_{i})$, while the data with the flat trend ($R_{i}(0)$) is the rate with the exponential decay factor corrected for.  The line through the points at time $> 0$ is the best fit to the data.  Time $t=0$ was 4.53 $^{24}$Na mean lifetimes after the spike was added.}
   \label{fig:neut_capt_rate_vs_time_v6_alt3_2005}
\end{figure}

The neutron capture rate in the NCD array as a function of time for the $^{24}$Na spike in 2005 is shown in Fig.~\ref{fig:neut_capt_rate_vs_time_v6_alt3_2005}; the plot for the spike in 2006 is similar.  The horizontal axis can be divided into three regions:

\begin{enumerate}
\item{During the first few hours, the spike was highly non-uniform, and the rate varied in an erratic manner.}
\item{For the two $^{24}$Na lifetimes preceding the reference time $t = 0$, the spike was not quite uniform, as assessed using the distribution of Cherenkov light produced directly by the beta and gamma rays from the decay of $^{24}$Na.  However, the NCD array appeared to be quite insensitive to this moderate non-uniformity of the spike, as the measured rate was indistinguishable from the equilibrium distribution.  During this time period, the rate appeared to decay exponentially as can be seen in the figure.  Time $t=0$ was 4.53 $^{24}$Na mean lifetimes after the spike was added.}
\item{After $t = 0$, the spike was determined to be well-mixed according to the Cherenkov light distribution.  The rate continued to decay exponentially.}
\end{enumerate}

The neutron rate at $t = 0$ could be derived from the measured rate $R(t_{i})$ at the average time $t_{i}$ of the $i^{th}$ run as follows:
\begin{equation}
R_{i}(0) = R(t_{i}) \; e^{t_{i}/\tau_{\textrm{24Na}}}.
\end{equation}
The set of measurements $\left\{ R_{i}(0) \right\}$ was found to be equal to each other within the statistical uncertainty; the result of a fit of a constant value to this set of measurements is shown by the horizontal line through the data in Fig.~\ref{fig:neut_capt_rate_vs_time_v6_alt3_2005}.  The quantity $R_{\textrm{spike}}$ was obtained from this fit.  The results from the runs in 2005 and 2006 are summarized in Table~\ref{tab:r_spike}.

\begin{table}
\begin{center}
\caption{The neutron capture rate by the NCD array at the reference time ($t = 0$), obtained by fitting a constant value to the set of measurements $\left\{ R_{i}(0) \right\}$.  The statistical uncertainty is from the fit, while the systematic uncertainty is from the instability of the rate (see text for discussion).\label{tab:r_spike}}
\begin{minipage}[t]{5in}
\begin{ruledtabular}
\begin{tabular}{lcc} 
Year & Best-fit rate (n/s) & $\chi^2/\mbox{d.o.f.}$ \\ \hline
2005 & $0.2708  \pm 0.0020(\mbox{stat.}) \pm 0.0027(\mbox{syst.})$ & $45.6/48$ \\ 
2006 & $0.1811 \pm 0.0016(\mbox{stat.}) \pm 0.0018(\mbox{syst.})$ & $35.6/38$ \\ 
\end{tabular}
\end{ruledtabular}
\end{minipage}
\end{center}
\end{table}

In addition to the statistical uncertainty, several sources of systematic uncertainty existed for the neutron capture rate in the NCD array.  In general, the rate can be written as follows:
\begin{equation}
R = \frac{N \cdot f_{\textrm{inst}}}{t_{\textrm{data}} \cdot L \cdot \epsilon_{\textrm{comb}}} \; ,
\end{equation}
\noindent where $N$ is the number of detected events, $t_{\textrm{data}}$ is the length of time data were taken, $L$ is the fraction of the time the detector was live (referred to as the ``live fraction''), $\epsilon_{\textrm{comb}}$ is a product of cut and threshold efficiencies, and $f_{\textrm{inst}}$ is a factor used to remove the estimated contribution of instrumental noise.  Studies showed that uncertainties from these input terms were negligible in comparison to a 1\% long-term fluctuation in the neutron detection rate, as assessed using standard point $^{252}$Cf and Am-Be sources.  Since the spike calibration runs were essentially two snapshots of the detector performance, the long-term fluctuation implied that the rate could have been different by $\pm 1\%$ if the calibration were performed at any other time.  For this reason, we assigned a systematic uncertainty of 1\% to $R_{spike}$.

\subsubsection{$\epsilon_{\rm{spike}}$: Combining the rate and source strength measurements}

The capture efficiency by the NCD array of neutrons produced by the activated brine that was injected and mixed in the SNO detector is defined as the ratio of the number of neutrons captured in the NCD string live volume to the total number of neutrons produced.  This ratio can be obtained experimentally from the two measured quantities $A_{^{24}\textrm{Na}}$ (Sec.~\ref{subsec:a_na24}) and $R_{spike}$ (Sec.~\ref{subsec:r_spike}): $\epsilon_{\textrm{spike}} = R_{\textrm{spike}}/A_{^{24}\textrm{Na}}$.  The central value of $\epsilon_{\textrm{spike}}$ can be obtained by simply dividing the numbers in Table~\ref{tab:r_spike} by those in Table~\ref{tab:na24_ss_fit}.  The error propagation was performed with care because the numerator and denominator depend at least partly on measurements in the NCD array.  The systematic uncertainty of the rate in the NCD array was dominated by the long-term fluctuation of 1\%.  If the time between the source strength measurement and the spike rate measurement could be considered short, then the instability systematic should cancel out and the 1\% systematic uncertainty in the numerator and denominator could be ignored.  On the other hand, if this time period was not sufficiently short, then these uncertainties should be combined in quadrature.  Although there was strong evidence for stability within a day or two of running, no data exist to demonstrate stability over a 4 to 6-day period, as was the case for the present analysis.  Thus we decided to combine the uncertainties in quadrature.  The result for $\epsilon_{\textrm{spike}}$ is shown in Table~\ref{tab:eff_spike}.

\begin{table}
\begin{center}
\caption{The NCD array's capture efficiency for neutrons produced by $^{24}$Na brine injected and well-mixed in the SNO detector.  The combined statistical and systematic uncertainty is shown here.\label{tab:eff_spike}}
\begin{minipage}[t]{5in}
\begin{ruledtabular}
\begin{tabular}{ccc} 
Year & $\epsilon_{\textrm{spike}}$ & Percent uncertainty \\ \hline
2005 & $0.2182 \pm 0.0046$ & 2.1\% \\ 
2006 & $0.2151 \pm 0.0043$ & 2.0\% \\ 
\end{tabular}
\end{ruledtabular}
\end{minipage}
\end{center}
\end{table}

\subsubsection{$f_{\rm{edge}}$: Correction factor for neutron density near the acrylic vessel}

Although the activated brine produced a neutron distribution in D$_{2}$O that was similar to that produced by solar neutrinos, the neutron density near the acrylic vessel was different.  For $^{24}$Na decays occurring within about 30 cm (one Compton scattering length) of the acrylic vessel, the probability of the 2.75-MeV gamma ray to escape from the D$_{2}$O volume was significant, so the neutron density dropped quickly as a function of the distance to the acrylic vessel wall.  The neutrons produced by the 2.75-MeV gamma rays in $^{24}$Na decays started with an energy of 260~keV and then were moderated by scattering, whereas those from the NC interaction started with a range of energies. ÊThis difference in the initial neutron energy could affect the neutron density at the edge of the vessel. ÊNeutrons produced near the vessel wall were significantly less likely to be detected by the NCD array than those produced elsewhere.  Thus the volume-averaged capture efficiency of the NCD array for neutrons from the spike was somewhat larger than that for neutrons from solar neutrinos. 

This difference near the acrylic vessel could be determined accurately by simulations, and was accounted for with the correction factor $f_{\textrm{edge}}$:
\begin{equation}
f_{\textrm{edge}} = 0.9702 \pm 0.0078.
\end{equation}

\subsubsection{$f_{\rm{non-unif}}$: Correction factor for source non-uniformity}

Another potential source of difference in the neutron distribution between that from the activated brine and from solar neutrinos was imperfect mixing of the brine.  Although it was not possible to directly measure the salinity as a function of position in the D$_{2}$O, there were a number of indications that the brine was well mixed and that any residual non-uniformity would not have a large impact on the measurements of the neutron capture efficiency. As can be seen in Fig.~\ref{fig:neut_capt_rate_vs_time_v6_alt3_2005}, the NCD array's neutron detection rate stabilized well after about a day of mixing, whereas the signal from the gamma rays  emitted by the $^{24}$Na did not stabilize completely for about  three more days~\cite{na24_nim_paper}. This indicated that the neutron capture rate of the NCD array was not sensitive to the remaining inhomogeneities for several days prior to the time $t=0$ used for the start of the analysis of the $^{24}$Na data and therefore could certainly be considered stable after $t=0$. Whereas the stability in the rate only indicated that the brine distribution reached some equilibrium configuration, the only reasonable regions where it might conceivably be non-uniform were near the inside of the acrylic vessel wall or near the NCD strings where boundary layers might have different salinity. Therefore these regions were considered carefully and used to establish systematic uncertainties on the uniformity of the brine.   

A possible way of measuring the brine distribution was by detecting the Cherenkov radiation from the Compton scattering of gamma rays and betas from the decay of $^{24}$Na with the SNO PMT array and comparing with the expectations for a uniform distribution.  This method, however, lacked precision because the calculations of the detector response were not very accurate for the low energies deposited by these decay products.  When the Cherenkov light data were compared with Monte Carlo simulation of perfectly uniform brine, variations at the level of about $\pm 10\%$ were seen, which were consistent with our ability to model the detector at such low energies.  Similar variations were seen in the studies of solar neutrino data at such energies.

\begin{figure} 
\begin{center}
   \includegraphics[width=0.85\textwidth]{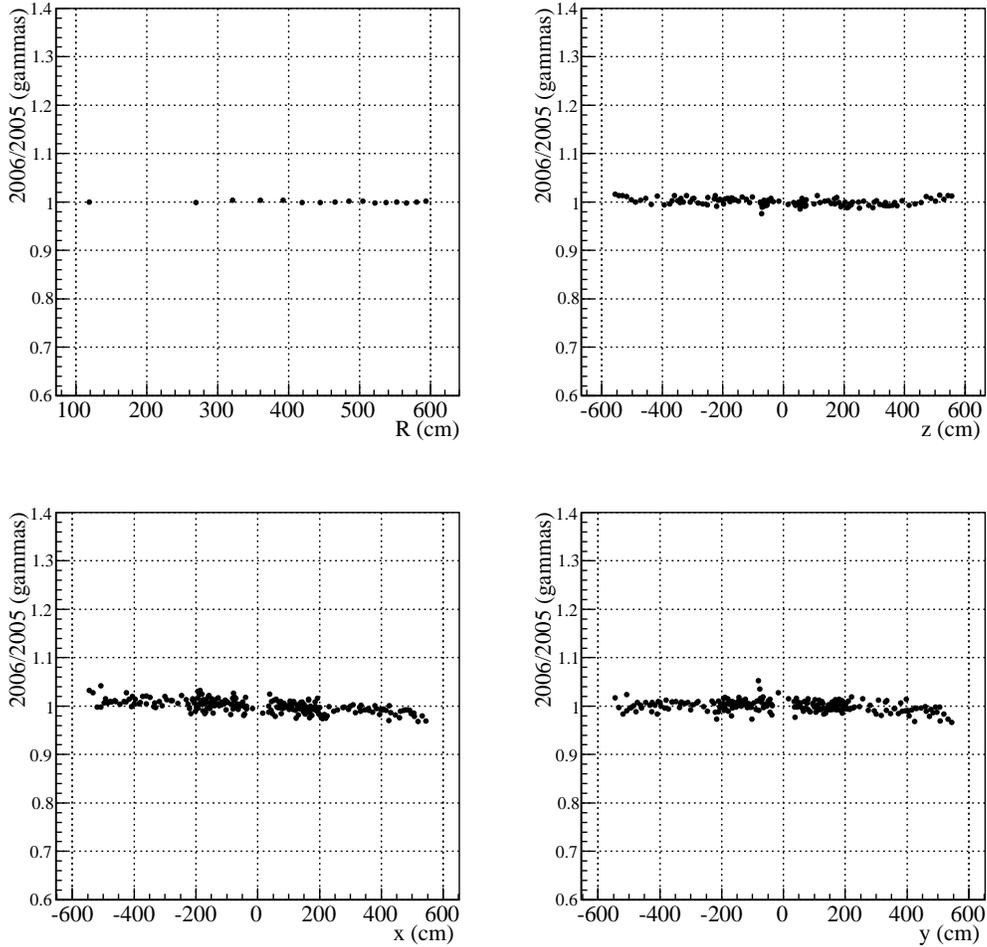} 
\end{center}
   \caption{Ratio of the brine concentration as observed by reconstructed Cherenkov events in 2006 to that in 2005, as a function of detector coordinates. Top left: $R = $~reconstructed radial distance from the center of SNO; top right: $z$, bottom left: $x$; bottom right: $y$.}
   \label{fig:nonuniformity_2005_vs_2006}
\end{figure}

Another piece of evidence suggesting that the brine was well-mixed in the central regions away from the AV or the NCD array came from the comparison of the distribution of Cherenkov events from $^{24}$Na gamma rays  observed in the 2005 and 2006 data.  The injection and mixing methods for the two years were very different~\cite{na24_nim_paper}.  In 2005, the brine was injected at several positions along the central vertical axis and then the water circulation was turned on.  In this configuration D$_{2}$O was pulled out from the bottom of the detector and returned at the top.  In 2006, a more sophisticated method involving flow reversal and temperature inversion was employed.  Using this method, an extensive eddy current was set up by causing the entire bulk of the D$_{2}$O to rotate in one direction, then reversing the direction of the flow.  In both years, the data showed that the brine distribution reached a stable equilibrium.  If one examines the Cherenkov light data from each year alone, one cannot say with much confidence whether or not the equilibrium configuration was uniform because of the limitations associated with modeling the low-energy response.  However, because the brine was mixed so differently, it is implausible that a non-uniform equilibrium configuration in 2005 could be the same as that in 2006. Figure~\ref{fig:nonuniformity_2005_vs_2006} shows, moreover, that the spatial distributions of the Cherenkov events in the brine were very similar. This comparison does not depend on the Monte Carlo simulations but is simply a study of the Cherenkov light data observed in the two cases. A plausible explanation for the similarity is that the brine was well-mixed in the central regions away from the NCD array and AV for both calibration sessions.

Based on the above argument, we assumed that the spike was well-mixed, so the correction factor $f_{\textrm{non-unif}}$ has a central value of 1.0.  The uncertainty on this value was obtained by considering the areas where stable inhomogeneities could possibly be established in the detector, namely near the NCD strings and near the wall of the acrylic vessel. 

In the vicinity of the NCD strings, calculations based on the laminar flow rates measured from the velocities observed from the Cherenkov light events indicated that the boundary layer could be on the order of 5~cm.  A 0.5\%  effect of such a boundary layer on neutron capture efficiency was estimated from redistributing the salinity from within the layer uniformly in the \dto\ target.

Careful studies of the fall-off of reconstructed gamma-ray signals from the $^{24}$Na brine in the vicinity of the acrylic vessel, coupled with uncertainties in the knowledge of the optical properties of the detector in this region, led to an upper limit for a boundary layer thickness of 3.6~cm.  Estimates based on flow rates in this vicinity gave smaller values for this boundary layer thickness.  The effect of such a boundary layer on neutron capture efficiency was considered by redistributing the  salinity from within the layer uniformly in the \dto\ target.  This 1.5\% effect was then combined in quadrature with the uncertainty from the NCD string region to give a full uncertainty 1.6\% on the efficiency.  From this conservative analysis of the systematic uncertainties, we obtained:
\begin{equation}
f_{\textrm{non-unif}} = 1.000 ^{+0.016}_{-0.016}.
\end{equation}

\subsubsection{$\epsilon_{\rm{sol}}$:  Capture efficiency for neutrons produced by solar neutrinos}

The NCD array's capture efficiency for neutrons produced by solar neutrinos was obtained from the product of $\epsilon_{\textrm{spike}}$, $f_{\textrm{edge}}$, and $f_{\textrm{non-unif}}$ (Eqn.~\ref{eqn:eff_sol}).  Before performing this multiplication, however, the value and uncertainties for $\epsilon_{\textrm{spike}}$ from the runs in 2005 and 2006 were combined, taking into account uncertainty components that were correlated between the two years.  The final result is:
\begin{equation}
\epsilon_{\textrm{sol}} = 0.211 \pm 0.005.
\end{equation}

\noindent This is in good agreement with the result from detector simulation, $0.210 \pm 0.003$ (see Sec.~\ref{sec:ndiscrete}), and with our previously published value of $0.211\pm 0.007$~\cite{ncdprl}.  In this paper, we have improved on the determination of the neutron capture efficiency by the NCD array and have further examined the assumptions that were made in the first Phase-III results reported in Ref.~\cite{ncdprl}.  This resulted in a small reduction in the systematic uncertainty in the present analysis.   This small improvement would have negligible effect on the measured NC flux, and was not incorporated in the solar neutrino flux analysis reported in this paper.

\subsection{Neutron detection efficiency of the NCD array\label{sec:ncddeteff}}

Several corrections must be applied to the capture efficiency, described in the previous section, in order to determine the NCD array's detection efficiency of NC neutrons in the solar neutrino analysis.  These corrections, averaged over the duration of data-taking in Phase III, included the mean live fraction of the MUX ($l_{\textrm{MUX}}$) and the digitizing oscilloscope ($l_{\textrm{scope}}$), the average MUX threshold efficiency ($\epsilon_{\textrm{MUX}}$), signal acceptance in the shaper energy window ($\langle \epsilon_{\textrm{shaper}}\rangle$), and the acceptance of data reduction cuts ($\langle \epsilon_{\textrm{cut}} \rangle$). The overall correction $C$ is the product of these individual factors:
\begin{equation}
C = l_{\textrm{MUX}} \cdot l_{\textrm{scope}} \cdot \langle \epsilon_{\textrm{MUX}} \rangle \cdot \langle \epsilon_{\textrm{shaper}} \rangle \cdot \langle \epsilon_{\textrm{cut}} \rangle
\end{equation}
Table~\ref{ncdcorr} provides a summary of these factors.

The MUX live fraction in a data run was determined by comparing the readings on two live time scalars: one that determined the total run time, and the other that was stopped when the MUX system was unable to take in events.   A pulser was installed to inject pulses at random times to provide additional validation.  The mean MUX live fraction $l_{\textrm{MUX}}$ was the run-time-weighted average of the solar neutrino run measurements.  The scope live fraction as a function of event rate was determined using the number of observed partial MUX events. The mean scope live fraction, $l_{\textrm{scope}}$, was then established by numerically integrating the neutrino run time, weighted by the rate-dependent scope live fraction. This calculation of $l_{\textrm{scope}}$ was verified using the random pulser system.  The methodology for measuring the MUX threshold efficiency has been described in Sec.~\ref{ncdthrescal}, and the results from regular calibration runs were averaged to provide an estimate of $\langle \epsilon_{\textrm{MUX}} \rangle$.   NCD array events with shaper energy $E_{\textrm{NCD}}> 0.4$~MeV were selected for the solar neutrino flux measurement described in Sec.~\ref{sec:sigex}.  The fraction of shaper events above this energy threshold was evaluated for each NCD counter using AmBe calibration data.  The mean acceptance $\langle \epsilon_{\textrm{shaper}} \rangle$ for NC neutron events was then calculated by averaging these individual counter estimates, weighted by the expected fraction of NC neutrons that each counter would capture.  The last correction factor is the neutron signal acceptance of the data reduction cuts, which are described in Sec.~\ref{sec:dataselect} and Appendix~\ref{sec:apdxa}.  The overall correction factor is
\begin{equation}
C = 0.862 \pm 0.004.
\end{equation}

\begin{table}
\caption{\label{ncdcorr} Summary of correction factors in the determination of the neutron detection efficiency.  The NCD array's detection efficiency of NC neutrons, with a shaper energy threshold $E_{\textrm{NCD}} > 400$~keV, is the product of the neutron capture efficiency (Sec.~\ref{neffncd}) and the combined factor $C$.  The factors shown in this table are the average values for solar neutrino data presented in this paper. }
\begin{minipage}[t]{6in}
\begin{ruledtabular}
\begin{tabular}{lc}
Correction factor  &  Value \\ \hline
MUX live fraction ($l_{\textrm{MUX}}$) & $0.9980 \pm 0.0001$ \\
Scope live fraction ($l_{\textrm{scope}}$) & $0.957 \pm 0.004$ \\
Average MUX threshold efficiency ($\langle \epsilon_{\textrm{MUX}} \rangle$) & $0.99491 \pm 0.00031$ \\
Average shaper energy window acceptance ($\langle \epsilon_{\textrm{shaper}} \rangle$) & $0.91170 \pm 0.00014$ \\
Data reduction cut acceptance ($\langle \epsilon_{\textrm{cut}} \rangle$) & $0.99521 \pm 0.00011$  \\ \hline
Combined ($C$) & $0.862 \pm 0.004$ \\ 
\end{tabular}
\end{ruledtabular}
\end{minipage}
\end{table}

\subsection{Neutron detection efficiency of the PMT array\label{sec:pmtnaspike}}

Neutrons can also be captured by deuterons with the emission of a 6.25-MeV gamma ray that could be detected by the PMT array.  The efficiency for this detection channel was much smaller than that for the NCD array because of the large difference in the thermal neutron capture cross section between the deuteron and $^3$He.  In this section, we present the analysis of the PMT array's neutron detection efficiency.  The basic analysis approach, i.e. the evaluation of individual terms in Eqn.~\ref{eqn:eff_sol}, was nearly  identical to that for the NCD array, discussed in Sec.~\ref{neffncd} above.  

The source non-uniformity factor $f_{\textrm{non-unif}}$ was evaluated in the same manner as the case for the NCD array.  We assumed that the $^{24}$Na spike was well mixed, but allowed for possible inhomogeneity near the acrylic vessel and the NCD strings.  Based on simulation studies, these effects could lead to a combined uncertainty of 0.8\% on the neutron detection efficiency for the PMT array, that is,
\begin{equation}
f_{\textrm{non-unif}} = 1.000 \pm 0.0080.
\end{equation}

The difference in the detection efficiency for NC neutrons and the $^{24}$Na photodisintegration neutrons, characterized by $f_{\textrm{edge}}$, was also different for the PMT array and the NCD array.  This is because the radial dependence of the neutron capture efficiency was different for the two detection mechanisms.  This difference was determined to be 0.92\% by simulation, or a $f_{\textrm{edge}}$ value of:
\begin{equation}
f_{\textrm{edge}} = 0.9789 \pm 0.0090.
\end{equation}

For the measurement of $R_{\textrm{spike}}$ in Eqn.~\ref{eqn:eff_spike}, a maximum likelihood analysis was performed to statistically separate the neutron signal from the $^{24}$Na beta and gamma-ray backgrounds.   Cherenkov light events were selected using the energy and fiducial volume cuts for the solar neutrino analysis, i.e. $\teff\geq6$~MeV and $\rfit\leq 550$~cm.  A fit of the $\teff$ spectrum from the combined data of the $^{24}$Na runs in 2005 and 2006 to a combination of neutron signal and $^{24}$Na beta-gamma background was performed.  The number of neutrons was found to be $n_n=8205.3\pm121.9$ and the number of background events was found to be $n_\gamma=1261.7\pm88.8$ with a correlation of $-0.61$.  Energy and reconstruction related systematic uncertainties in the solar neutrino analysis were propagated in this analysis, and their combined effect was found to be $^{+1.71}_{-1.38}$\%.  After symmetrizing this systematic uncertainty and combining it with the statistical uncertainty in the $\teff$ spectral fit in quadrature, $n_n$ was determined to be $8205.3\pm 185.4$.  With the neutron rates given in Table~\ref{tab:na24_ss_fit} and the time span of the calibration runs,  $\epsilon_{\textrm{spike}}$ could be evaluated, and the result for $\epsilon_{\textrm{sol}}$ is 
\begin{equation}
\epsilon_{\textrm{sol}} = 0.0502 \pm 0.0014.
\end{equation}

This is in agreement within uncertainties with the MC calculations ($\epsilon_{\textrm{sol}} = 0.0485 \pm 0.0006$) in Sec.~\ref{sec:ndiscrete}, which was used in our previously published results~\cite{ncdprl}.  This small difference has a negligible effect on the solar neutrino flux results.

\subsection{Neutron calibration with discrete sources\label{sec:ndiscrete}}

Calibration data were used to tune the Monte Carlo simulation code, which was then used to predict the neutron detection efficiencies for both the PMT and NCD arrays. The uncertainties were calculated by propagating those on the tuning parameter and on other Monte Carlo input parameters.  

The calibration data used to tune the Monte Carlo were acquired using AmBe sources, which were periodically deployed  around the \dto\ target during the data taking period. In each of the calibrations data were taken with the source positioned at a series of well-defined, repeatable locations. The tuning parameter was the hydrogen concentration in the heavy water; its measured value, from Fourier transform infrared spectroscopy, was $(9.8\pm 0.5)\times 10^{-4}$~atom of hydrogen per atom of deuterium.

The method for tuning the hydrogen concentration was to take averaged source data for a given source location and calculate the relative detection efficiency of rings of counters around the position. The relative capture efficiency of a ring of counters close to the source to one further away was sensitive to the neutron diffusion length and therefore to the concentration of hydrogen, which has a large neutron capture cross section. These ratios were independent of source strength and, because they were of rings of counters, they were relatively insensitive to the exact source position. Simulations were run with a range of hydrogen concentrations and source locations, and a maximum likelihood fit was used to extract the most probable hydrogen concentration. The measured hydrogen concentration was used as a constraint in the fit.

The central values for the neutron detection efficiencies in the PMT and NCD arrays were calculated using a simulation run at the best-fit hydrogen concentration of $(9.45^{+0.50}_{-1.05})\times 10^{-4}$~atom of hydrogen per atom of deuterium, and the uncertainty propagated by rerunning the simulation with the hydrogen concentration set to its upper and lower bounds.  Additional sources of uncertainty in the Monte Carlo, such as parameters relating to the modeling of the counter and AV geometry, were studied separately. The only significant uncertainty came from the modeling of the shape of the NCD counter live regions, which was estimated to impart a 1.0\% uncertainty on the NCD neutron detection efficiency.

The final prediction for the neutron capture efficiency of the NCD array was $0.210 \pm 0.003$. The prediction for the neutron detection efficiency of the PMT array was $0.0485 \pm 0.0006$.

\section{Backgrounds\label{sec:backgrounds}}

Several sources of radioactive backgrounds were present in the 
PMT array and NCD array data.  The majority were associated with naturally occurring
$^{238}$U and $^{232}$Th, cosmogenic activity and atmospheric
neutrino interactions in the detector.  The impact of these backgrounds on the
neutrino analysis was reduced by optimization of analysis
cuts.  Those that remained were included in the fits for the
background and the solar neutrino signals.  A summary of background contributions is given in
Table~\ref{nppd}. 

In this section, we will discuss the identification and measurement of the neutron and Cherenkov light
backgrounds in the solar neutrino measurement.   Alpha decays from the construction materials were the largest source of backgrounds in the neutron signal region in the NCD array.  This background source was difficult to calibrate due to variation in the spatial distribution and in the composition of trace radioactivity in different counters.  The treatment of alpha decay backgrounds will be presented in the next section (Sec.~\ref{sec:pulsesim}).

\begingroup
\begin{table}
\squeezetable
\caption{\label{nppd} Summary of backgrounds in the PMT and NCD arrays.}
\begin{ruledtabular}
\begin{tabular}{lcc}
Source & PMT events & NCD events \\ \hline
Neutrons generated inside D$_2$O: & & \\
$^{2}$H photodisintegration [U,Th in D$_2$O]      & $7.6 \pm 1.2$ & $28.7\pm 4.7$\\
$^{2}$H photodisintegration [U,Th in NCD bulk]    & $4.3^{+1.6}_{-2.1}$ &$25.8^{+9.6}_{-12.3}$\\
$^{2}$H photodisintegration [U,Th in Hotspots]    & $17.7\pm 1.8$ & $64.4\pm 6.4$\\
$^{2}$H photodisintegration [U,Th in NCD Cables]  & $1.1 \pm 1.0$ & $8.0 \pm 5.2$ \\
n from spontaneous fission [U]                             & $0.2 \pm 0.1$ & $0.3 \pm 0.1$\\
$^{2}$H($\alpha, \alpha n$)$^{1}$H [Th, $^{222}$Rn] & $0.2 \pm 0.1$ & $0.4\pm 0.1$\\
$^{17,18}$O($\alpha , n$)$^{20,21}$Ne [Th]       & $0.3 \pm 0.1$ & $1.8 \pm 0.4$\\
Atmospheric $\nu$                       & $24.1 \pm 4.6$ & $13.6\pm 2.7$\\
Cosmogenic muons                               & $0.009 \pm 0.002$ & $0.04\pm 0.004$\\
Reactor and terrestrial neutrinos              & $0.3 \pm 0.1$ & $1.4 \pm 0.2$\\
CNO solar $\nu$                                      & $0.05 \pm 0.05$&$0.2\pm 0.2$ \\
\hline
Total internal neutrons & $55.8^{+5.6}_{-5.4}$ & $144.7^{+13.4}_{-15.5}$\\
\hline \hline
Neutrons generated from AV and \hto\ radioactivity: & & \\
$^{2}$H photodisintegration [U,Th in H$_2$O]      & $2.2 ^{+0.8}_{-0.7} $ & $7.1 ^{+5.4}_{-5.2}$\\
($\alpha , n$) in AV        & $18.3 ^{+10.2}_{-7.3} $ & $33.8 ^{+19.9}_{-17.1} $\\ \hline 
 Total external-source neutrons       & $20.6 ^{+10.2}_{-7.3} $ & $40.9 ^{+20.6}_{-17.9}$\\
\hline \hline
Cherenkov events from radioactivity inside the D$_2$O: & &\\
beta-gamma decays (U,Th)  &$0.70^{+0.37}_{-0.38}$ & N/A\\
Decays of spallation products in D$_2$O: 
$^{16}$N following muons       &$0.61 \pm 0.61$ & N/A\\
\hline \hline
Cherenkov backgrounds produced outside D$_2$O: & &\\
 beta-gamma decays (U,Th) in AV, H$_2$O, PMTs & $5.1^{+9.7}_{-2.9}$ & N/A \\
Isotropic acrylic vessel events & $<0.3$ (68\% C.L.)& N/A\\
\end{tabular}
\end{ruledtabular}
\end{table}
\endgroup

\subsection{Photodisintegration backgrounds}

Gamma rays with energy greater than 2.225 MeV can break apart a
deuterium nucleus releasing a free neutron, which was indistinguishable
from one produced by a NC interaction.  Such gamma rays are
emitted by beta-gamma decays of $^{208}$Tl and $^{214}$Bi from
the $^{232}$Th and $^{238}$U chains respectively.  An accurate
measurement of these radioisotopes was crucial for the 
determination of the total $^{8}$B neutrino flux.  Concentrations of
$3.8 \times 10^{-15}$ gTh/gD$_2$O and $30 \times 10^{-15}$ gU/gD$_2$O
are each equivalent to the production of one neutron per day via
photodisintegration. Two independent approaches were developed to
measure these backgrounds.  These are broadly classified as
\textit{ex situ} and \textit{in situ} techniques. 

\subsubsection{\textit{Ex situ} determination of radioactivity in \dto}

Three \textit{ex situ} methods were developed to assay parent isotopes
of $^{208}$Tl and $^{214}$Bi in the D$_2$O and H$_2$O regions of the
detector.  Common to all three techniques was extraction and filtering of a
known amount of water from the detector and external counting of the resultant
sample.  Two methods extracted $^{224}$Ra and $^{226}$Ra, one using
beads coated with manganese oxide (MnO$_x$) \cite{MNOX} and the other
using filters loaded with hydrous titanium oxide (HTiO)
\cite{HTIO04,HTIO08}.  For MnO$_{x}$ and HTiO assays, up to 500~tonnes
of water passed through the loaded columns over a 4 to 5-day period.  In
the MnO$_{x}$ technique, Ra isotopes were identified by alpha spectroscopy of 
Rn daughters.  In the HTiO method, Ra isotopes were stripped from the
filters, concentrated and identified using beta-alpha coincidence
counting of the daughter nuclides.

The equilibrium between $^{238}$U and $^{214}$Bi was broken by the
ingress of $^{222}$Rn (halflife = 3.82 d), primarily from the laboratory
air and emanation from construction materials.  The amount of
$^{222}$Rn in the water was measured by degassing, cryogenically
concentrating the dissolved gases and counting the sample using a
ZnS(Ag) scintillator \cite{RN}.  Each Rn assay processed approximately
5 tonnes of water in a 5 hour period. 

During the third phase of SNO, 20 MnO$_{x}$ and 16 HTiO assays were conducted at
regular intervals in the heavy water region.  The results from each
independent assay method were in good agreement. The activity measured by
each assay was a combination of activity from the D$_2$O and
water systems piping.  The variation of Th~(Ra) activity in the water
and piping was modeled as a function of time, taking into account other sources of Th~(Ra) in the flow path.  The resultant concentration was
the live-time weighted combined HTiO and MnO$_{x}$ activity, which was 
$0.58 \pm 0.35 \times 10^{-15}$ gTh/gD$_2$O.  The quoted uncertainty
was combined from the systematic and statistical uncertainties.

A total of 66 Rn assays were performed in the D$_2$O region at regular
intervals throughout the third phase. To calculate the mean $^{222}$Rn
concentration, the individual assay results were time and volume weighted.
The equivalent mean $^{238}$U concentration was  $5.10 \pm 1.80 \times
10^{-15}$ gU/gD$_2$O, where the total uncertainty was combined from the systematic and statistical uncertainties.  

\begin{figure}
\begin{center}
\includegraphics[width=0.65\columnwidth]{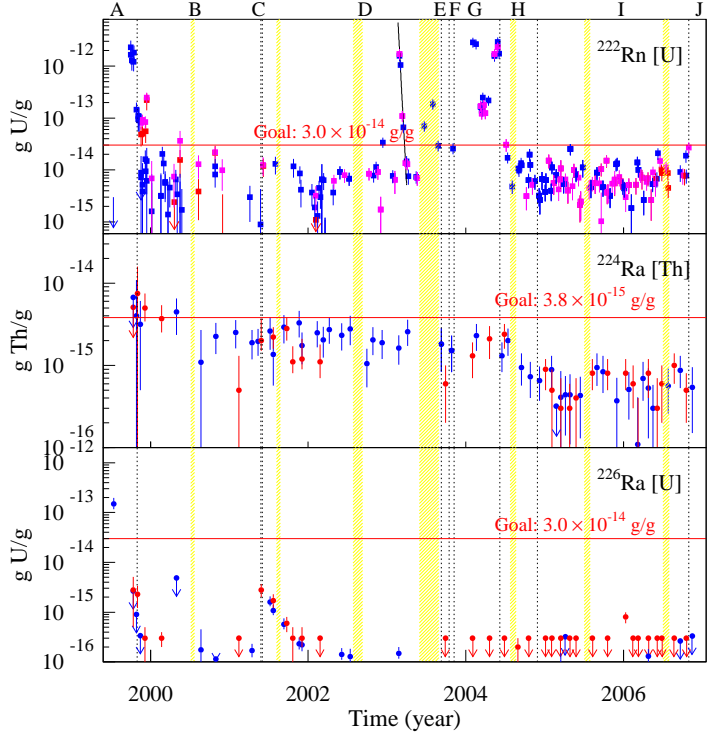} \\
\caption{\label{fig:exsitudto} 
\dto\ radioactivity measurements by \textit{ex situ} assays are shown for $^{222}$Rn (top),
$^{224}$Ra (middle) and $^{226}$Ra (bottom). The dashed lines are the bounds for different SNO detector configurations (from periods A to J as indicated): A - Phase-I (unadulterated \dto) commissioning; B - Phase-I operation; C - salination; D - Phase-II (salt) operation; E - desalination; F - preparation for NCD array installation; G - NCD array deployment; H - NCD array commissioning; I - Phase-III (NCD array) operation; and J - removal of the NCD array.  Periods highlighted in yellow were times with reduced access to the underground laboratory.  For the radon data, color represents different sampling points: red is at the top (near the chimney-sphere interface), purple 1/3 down and blue at the bottom of the AV.   The radon level was well below target for essentially the whole duration of the experiment, except for the high level at the beginning of the experiment, a large calibration spike in Phase-II, and during the deployment of the NCD array (when the cover gas protection was temporarily turned off).  For the radium data, color indicates the technique used: red is for HTiO and blue for MnO$_x$.  Radium assays sampled the heavy water either from the top or the bottom of the AV. Again the initial higher concentrations quickly went below target.  It should be noted that the data shown in this plot were not the only input to determining the radiopurity of the \dto\ target. }
\end{center}
\end{figure}

Figure~\ref{fig:exsitudto} is a summary of the \dto\ assay results since the beginning of the SNO experiment.   During SNO's Phase-III operation, an aggressive program of system purification combined with minimum recirculation of the heavy water led to a factor of five reduction in thorium.  The concentration of $^{224}$Ra was routinely measured at 0.1 atom/tonne. 

\subsubsection{\textit{In situ} determination of radioactivity in \dto\ and NCD housing}

The \textit{in situ} technique measured the $^{232}$Th and $^{238}$U content of the
water and NCD array housings directly from the Cherenkov light data \cite{HMOK}.  
In the energy window $4.0 < \teff < 4.5$~MeV,  the selected events were
dominated by $^{214}$Bi and $^{208}$Tl, from the $^{238}$U and
$^{232}$Th chains respectively.  The observed Cherenkov light was dominated by the direct
beta decay of $^{214}$Bi to the ground state of $^{214}$Po with an end
point of 3.27 MeV. $^{208}$Tl decays almost always emitted a 2.614-MeV
gamma ray, accompanied by one or more lower energy gamma ray and a beta with
an end point of up to 1.8 MeV.  $^{208}$Tl events produced a more
isotropic Cherenkov light distribution when compared with
$^{214}$Bi events and it was this difference in light isotropy that was used,
in addition to differences in the radial distributions, to separate
background components.   

The light isotropy parameter was $\beta_{14}\equiv\beta_1+4\beta_4$, where
\begin{equation}
\beta_l = \frac{2}{N(N-1)}\sum_{i=1}^{N-1} \sum_{j=i+1}^N
P_l(\cos\theta_{ij}).
\end{equation}
In this expression $P_l$ is the Legendre polynomial of order $l$,
$\theta_{ij}$ is the angle between triggered PMTs $i$ and $j$ relative
to the reconstructed event vertex, and $N$ is the total number of
triggered PMTs in the event.  Details of $\beta_{14}$ can be found in Ref.~\cite{nsp}.

\begin{figure}
\begin{center}
\includegraphics[width=0.7\textwidth]{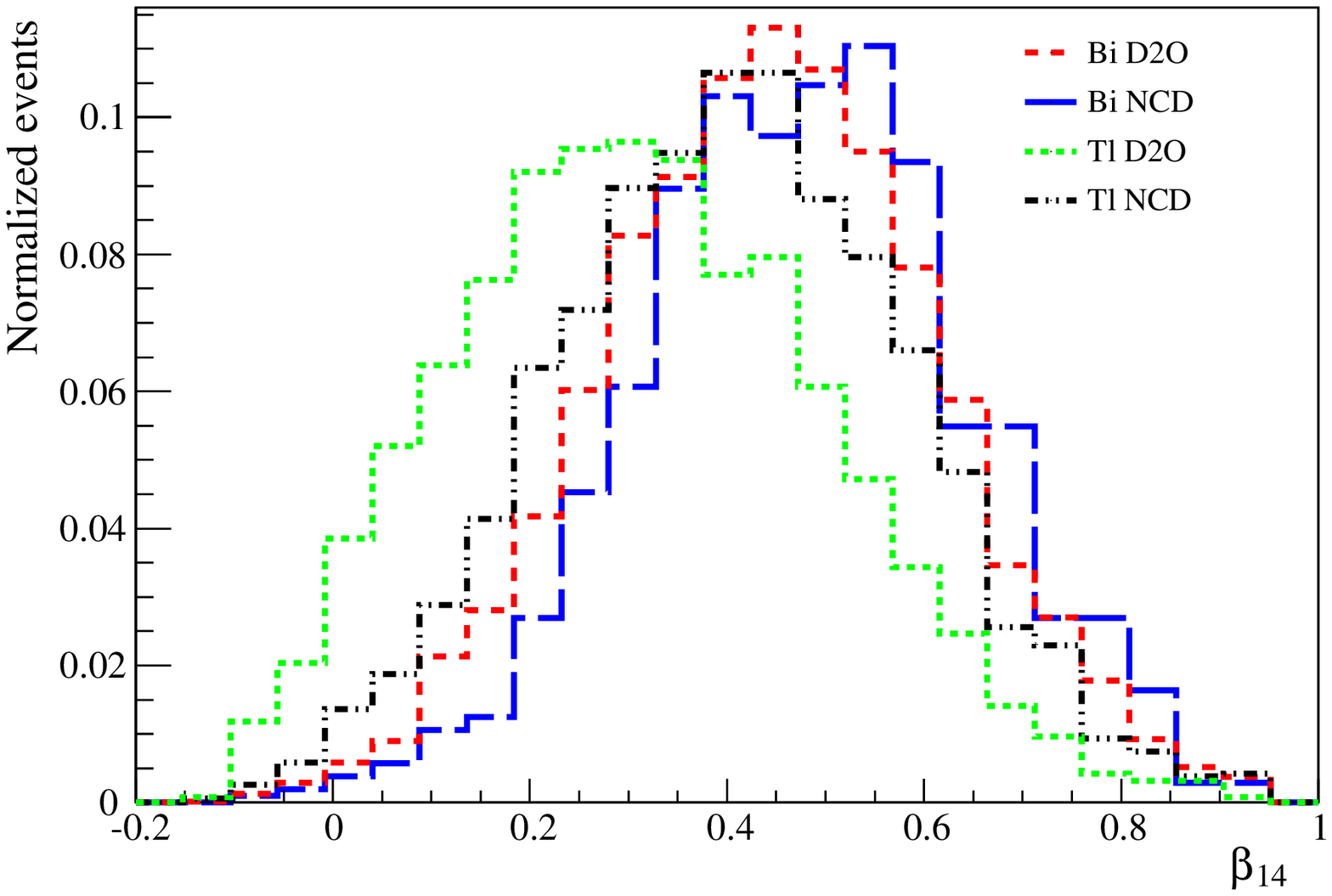} \\
\includegraphics[width=0.7\textwidth]{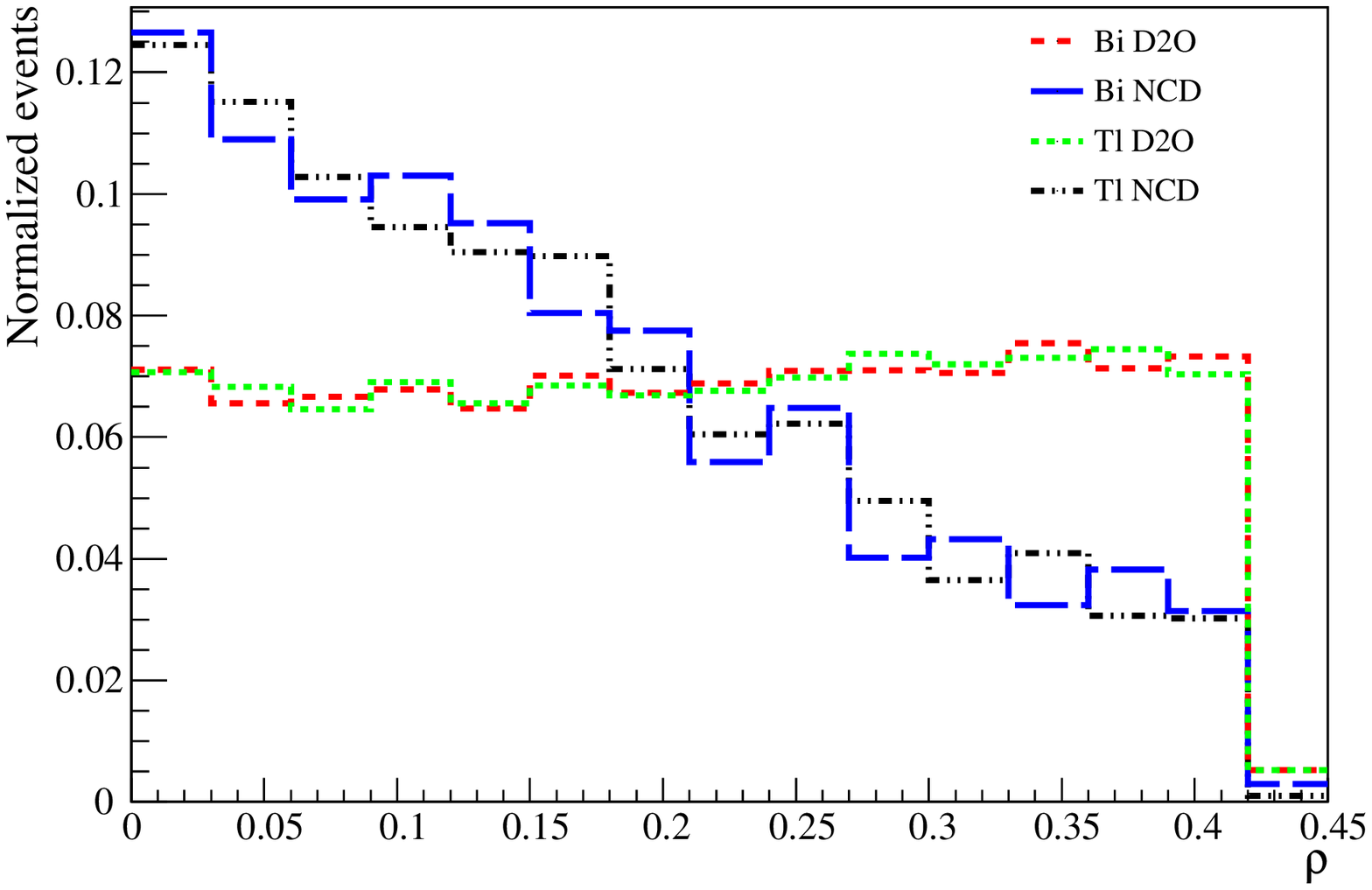} \\
\caption{\label{fig:insitu_pdf} Probability distribution functions of $\beta_{14}$ (top) and $\rho$ (bottom) for different background sources in the \textit{in situ} measurement of \dto\ and NCD array backgrounds. }
\end{center}
\end{figure}

Radioactive decays originating from the NCD array had an exponential radial profile
while those from the D$_2$O had an approximately flat radial profile.  The radial
profiles were statistically indistinguishable for different
radioisotopes that originated from the same location.  Therefore, to
distinguish between $^{208}$Tl and $^{214}$Bi, differences in event
isotropy were used.  By analyzing events that reconstructed with $\rfit
< 450$~cm they can be classified as $^{208}$Tl and $^{214}$Bi in the
D$_2$O or NCD strings.  The \textit{in situ} method provided
continuous monitoring of backgrounds in the neutrino data set and a
direct measurement of $^{214}$Bi and $^{208}$Tl, both of which could 
cause photodisintegration, without making any assumptions about
equilibrium in the decay chain.  The $\beta_{14}$ and the volume-weighted radial position, $\rho$, distributions for the different background sources in this \textit{in situ} measurement are shown in Fig.~\ref{fig:insitu_pdf}.

There were four main radioactive signals in the D$_2$O region: 
uranium- and thorium-chain activities in the \dto\ and in
the NCD strings.  Assuming these were the dominant contributions in
the \textit{in situ} analysis signal region, a two-dimensional (radial and $\beta_{14}$) maximum likelihood 
fit was made to the data.  The normalized PDF associated with each background was constructed from Monte Carlo
simulations of $^{214}$Bi and $^{208}$Tl from the D$_2$O and NCD strings. The simulated events were selected using the same 
cuts used on the data.  Cherenkov light produced by $^{208}$Tl decays in the NCD string bodies 
was dominated by the 2.614-MeV gamma rays.  There
was very little contribution from betas as nearly all of them were stopped
in the nickel of the NCD counter housing.  Thus,  $^{208}$Tl decay events in the NCD string bodies were less isotropic and had a higher average value of $\beta_{14}$, when compared with events from $^{208}$Tl decays in the D$_2$O.  The mean value of $\beta_{14}$ for these
events was very similar to that of $^{214}$Bi decays in D$_2$O.  Bismuth
decays in the NCD array produced even less light than $^{208}$Tl decays.  In addition to betas being stopped in the nickel bodies, gamma rays from $^{214}$Bi decays have lower energies than those in $^{208}$Tl decays.  These $^{214}$Bi decays were very similar in
isotropy to $^{214}$Bi D$_2$O events.  Therefore, $^{214}$Bi D$_2$O, $^{214}$Bi
and $^{208}$Tl ``NCD bulk'' isotropy distributions were very
similar, but were significantly different from that of $^{208}$Tl
D$_2$O.  

Combining radial and
isotropy information, the $^{208}$Tl D$_2$O, $^{214}$Bi D$_2$O and NCD bulk events could 
be separated.  It was not possible to distinguish
between $^{208}$Tl and $^{214}$Bi events originating from the NCD
bulk.  The ratio of U to Th in the NCD bulk obtained by
coincidence studies of NCD array's alpha data~\cite{LCS} was used to separate the NCD
events obtained by the 2D maximum likelihood fit into $^{208}$Tl and
$^{214}$Bi.  In the $^{232}$Th chain, $^{220}$Rn alpha decays to $^{216}$Po which
decays by alpha emission.  The signature
for this coincidence was a 6.288-MeV alpha followed by a 6.778-MeV
alpha with a half-life of 0.15 seconds. In the $^{238}$U chain the
coincidence was between $^{222}$Rn and $^{218}$Po, the signature being
a 5.49-MeV alpha followed 3.10~minutes later by a 6.02-MeV alpha.  The
results from the alpha coincidence measurements
were $2.8^{+0.6}_{-0.8} \times 10^{-12}$ gU/gNi and $5.7^{+1.0}_{-0.9}
\times 10^{-12}$ gTh/gNi.

For comparison with \textit{ex situ} assay and alpha coincidence
measurements, the number of $^{214}$Bi and $^{208}$Tl events were
converted into equivalent amounts of $^{238}$U and $^{232}$Th by
assuming secular equilibrium and using Monte Carlo simulations.  The
equivalent concentrations, integrated over the solar neutrino data set, 
were found to be $6.63^{+1.05}_{-1.22} \times 10^{-15}$ gU/gD$_2$O,
$0.88^{+0.27}_{-0.27} \times 10^{-15}$ gTh/gD$_2$O, 
$1.81^{+0.80}_{-1.12} \times 10^{-12}$ gU/gNi and $3.43^{+1.49}_{-2.11}
\times 10^{-12}$gTh/gNi.  The \textit{in situ} results of the NCD bulk were in good agreement with those obtained from the
alpha coincidence analysis.

Results from the \textit{in situ} and \textit{ex situ} analyses of the
D$_2$O were found to be consistent.  As the two methods and their
systematic uncertainties were independent, the best measurement of the
equivalent concentrations of $^{232}$Th in
the D$_2$O was obtained by taking the
weighted mean of the \textit{in situ} and combined HTiO and MnO$_x$
results. The weighted mean of the \textit{in situ} and \textit{ex situ} Rn
results were used to obtain the best measurement of the $^{238}$U
content in the D$_2$O. The weighted mean concentrations were $6.14 \pm
1.01 \times 10^{-15}$ gU/gD$_2$O and $0.77 \pm 0.21 \times 10^{-15}$
gTh/gD$_2$O.  Only the \textit{in situ}
results for the NCD array bulk concentrations were used because the alpha
coincidence method could not provide data for all NCD counters and it could
not sample the whole array.  

\subsubsection{Radioactive hotspots on NCD strings}
Two areas of increased activity (hotspots) were identified in strings K5 and K2. The \textit{in situ} method
identified an excess of events close to each of these strings, but it could
not prove conclusively that these events were caused by radioactivity.
The isotropy distribution of the events associated with K5 was more
isotropic than that of the NCD bulk, implying either 
radioactivity or scintillant on the surface of the NCD strings.  If the
hotspot was radioactivity, an excess of neutrons should have been
captured by the contaminated string. However, K5 had a gain drift problem (Sec.~\ref{sec:dataset}) and the number of neutrons captured by this string could not be quantified.  No excess alphas were observed in the data from K2
suggesting that the contamination was embedded in the dead region of
the counter, a conclusion that was supported by the \textit{in situ} analysis. 
An extensive experimental program was developed to measure the radioactive content of these hotspots, and more details can be found in Ref.~\cite{HSNIM}.

The lower-chain hotspot activities expressed in terms of equivalent masses of $^{232}$Th and $^{238}$U are summarized in Table~\ref{hsres}. ÊFor the analysis presented here, the neutron rates have been calculated using the weighted average of the \textit{in situ} and \textit{ex situ} data. The updated analysis presented in Ref.~\cite{HSNIM} yields changes that are negligible relative to the uncertainties in the final result.

\begingroup
\begin{table}
\squeezetable
\caption{\label{hsres} 
Equivalent masses of uranium and thorium for lower-chain activities in the K5 and K2 hotspots. ÊMore detailed and updated results may be found in Ref.~\cite{HSNIM}.}
\begin{minipage}[t]{5in}
\begin{ruledtabular}
\begin{tabular}{l c c c}
 & String & $^{232}$Th ($\mu$g) & $^{238}$U ($\mu$g) \\ \hline
Total \textit{ex situ}& K5          &$1.28 \pm 0.14$     & $0.10 ^{+0.05}_{-0.05}$\\
Total \textit{in situ}& K5          &$1.48 ^{+0.24}_{-0.27}$  & $0.77 ^{+0.19}_{-0.23}$\\
Total \textit{ex situ}& K2          &$1.43 \pm 0.17$     & $< 0.40$\\
Total \textit{in situ}& K2          &$< 0.93$ & $\equiv 0$ \\
\end{tabular}
\end{ruledtabular}
\end{minipage}
\end{table}
\endgroup

\subsection{Other neutron backgrounds}

\subsubsection{Internal-source neutrons}

In addition to photodisintegration backgrounds, there were 
other neutron backgrounds that were generated in the \dto.  These included 
contributions from ($\alpha$,$n$) reactions on nuclei, spontaneous
fission from $^{238}$U, cosmic-ray spallation and anti-neutrinos from
nuclear reactor and atmospheric neutrinos.

Neutrons can be produced by alpha reactions on $^{2}$H, $^{17}$O and
$^{18}$O.  The most significant contribution to this background arose
from the 5.3-MeV alpha produced by $^{210}$Po decay.  In the third 
phase, Po isotopes on the external surface of the NCD array were of 
particular concern.  Taking the average alpha activity of 18 samples, a total surface area of 2.40~m$^2$ that was counted using a multi-wire proportional counter, yielded a neutron production rate of $1.32\pm 0.28\times 10^{-2}$~neutron/day generated from the entire array, 64.72~m$^2$, given that $6.4\times 10^{-8}$ neutron were produced per alpha. This rate resulted in $5.1\pm 1.1$~neutrons produced during Phase III.

The contribution from spontaneous fission of $^{238}$U was determined
from the results of \textit{ex situ} HTiO assays which placed limits on
the concentration of $^{238}$U in the detector.   

The {\sc NUANCE}~\cite{nuance} neutrino Monte Carlo simulation package was used in the
calculation of neutron backgrounds produced by atmospheric neutrino
interactions.   These atmospheric neutrino interactions were often associated with a burst of events in the detector.  
After applying time-correlation cuts that removed event bursts and other data reductions cuts to these simulated events, the expected number of observed neutrons from atmospheric neutrino interactions was determined to be 13.6 $\pm$2.7 for the NCD array and 24.1$\pm$4.6 for the PMT array.   The dominant systematic uncertainties associated with these estimates were those in the neutrino interaction cross section and atmospheric neutrino flux.

The muon flux incident on the SNO detector was measured to be
$3.31\pm 0.09 \times 10^{-10}$~cm$^{-2}$~s$^{-1}$ \cite{snoatm,CK}.   The
possibilities that the muon tag could be missed and that neutrons could be produced in the surrounding rock, led to an estimate of their total rate of $0.18\pm 0.02$ neutron per year, a negligible background.

Anti-neutrinos produced by nuclear reactors afar could also create neutrons
in the D$_2$O.  The magnitude of this background was calculated
assuming an average reactor anti-neutrino spectrum and an average
power output of all commercial reactors within 500 km of the SNO
detector.  Oscillations were taken into account in the calculation, and the estimate was 1.4$\pm$0.2 neutrons per year.  
An estimate was also made of the number of neutrons produced by
anti-neutrinos from radioactive decays within the Earth and this was
found to be 0.02 neutron per year.  

The SNO detector was also sensitive to CNO neutrinos from the Sun.  It was
estimated that 1 neutron per year would be produced by this
process.  A signal of $<0.2$~event was expected in the NCD array.

\subsubsection{External-source neutrons}

Radioactive backgrounds in the acrylic vessel and the surrounding \hto\ could bring forth photodisintegration neutrons in the \dto\ target.  Neutrons could also be produced via ($\alpha, n$) reactions in the AV.  The total neutron backgrounds due to these external sources were found to be 20.6$^{+10.2}_{-7.3}$~events for the PMT array data and 40.9$^{+20.6}_{-17.9}$~events for the NCD array data.

The \textit{in situ} and \textit{ex situ} techniques were applied to the H$_2$O region between the AV and PSUP.  In total, 29 MnO$_{x}$ and 25
HTiO  \hto\ assays were conducted during the third phase.  The results from the HTiO and MnO$_{x}$ assays were in good agreement.
The results were corrected for the neutrino live time and the weighted average was calculated to produce a single \textit{ex situ} $^{232}$Th-chain measurement for
the H$_2$O region. The activity was found to be $26.9 \pm 12.3 \times 10^{-15}$ gTh/gH$_2$O.  Due to hardware issues that were discovered towards the 
end of the phase, Rn assays, although performed, were not used in this analysis.

The \textit{in situ} analysis window for the H$_2$O region was $4.0 < \teff< 4.5$ MeV and $650 < \rfit < 680$ cm.  The equivalent $^{238}$U
and $^{232}$Th concentrations were determined using an isotropy fit to the data.  The background levels determined by the \textit{in situ}
analysis were $30.0^{+9.2}_{-19.4} \times 10^{-15}$ gTh/gH$_2$O and $35.0^{+9.9}_{-5.4} \times 10^{-14}$ gU/gH$_2$O.  The photodisintegration neutron background from Th activity in the \hto\ was determined from a weighted mean of the \textit{in situ} and \textit{ex situ} assay results, while that from Rn was determined exclusively from the \textit{in situ} results.  The photodisintegration neutron backgrounds due to Th and U in the \hto\ region were found to be $2.2^{+0.8}_{-0.7}$~events for the PMT array data and $7.1^{+5.4}_{-5.2}$~events for the NCD array data.

During its construction the acrylic vessel was exposed to Rn in the underground laboratory air.  The subsequent Rn daughters became embedded in the acrylic and could initiate ($\alpha , n$) reactions on $^{13}$C, $^{17}$O and $^{18}$O.  The activity on the surface of the AV was directly counted using silicon counters.  Results from measurements performed at the end of the third phase were in agreement with those performed at the end of the second phase.  Thus the rate of these external-source neutrons from the vessel was taken to be the same as for Phase II~\cite{nsp}.  Adding the photodisintegration neutron backgrounds due to intrinsic Th and U in the acrylic, which were determined from \textit{ex situ} assays~\cite{longd2o}, the total external-source neutron backgrounds from the AV were found to be $18.3 ^{+10.2}_{-7.3} $~counts for the PMT array data and  $33.8 ^{+19.9}_{-17.1}$~counts for the NCD array data.

\subsection{Other Cherenkov light backgrounds}

A 20-second veto following a tagged muon event removed the majority of radioactivity that followed.  The residual background from the decay of cosmogenic $^{16}$N was estimated at 0.61$\pm$0.61~event in the PMT array data.  Other Cherenkov light background events inside and outside the fiducial volume were estimated using calibration source data, measured activities, Monte Carlo calculations, and controlled injections of Rn~\cite{na24_nim_paper} into the detector.  These backgrounds were found to be small above the analysis energy threshold and within the fiducial volume, and were included as an additional uncertainty on the flux measurements.  Isotropic acrylic vessel background (IAVB) events were identified in previous phases~\cite{nsp}.  It was estimated that $<0.3$~IAVB event (68\% CL) remained in the PMT array data after data reduction cuts.

\section{Simulation of pulses in the NCD array\label{sec:pulsesim}}

The largest source of backgrounds in the neutron signal region in the NCD counters was alpha decays from the construction materials of the array.  This was a very difficult background to calibrate, as any alpha particles from external calibration sources would not have sufficient energy to penetrate the counter wall.  The spatial distribution and the composition of trace radioactivity in the counters also varied from counter to counter;  therefore,  background samples from the $^4$He counters were not sufficient to fully characterize this all-important background to the neutron signal.  An extensive Monte Carlo, discussed in this section, was developed to simulate ionization pulses in the NCD counters, and was used in defining the alpha background spectral shape for the solar neutrino analysis.  Further details can be found in Refs.~\cite{ncdsim,tseung_thesis,oblath_thesis}.

\subsection{Physics model}

The NCD counter simulation created ionization tracks for protons, tritons, alphas, and betas in the NCD counter gas.  Alpha energy loss in the nickel wall was also calculated if necessary.  Track formation for betas was handled by EGS4~\cite{EGS4}.  Proton, triton, and alpha tracks were all calculated using the same procedure as follows.  The track was divided into $N$ (typically 5,000-20,000) 1-$\mu$m long segments such that each segment could be approximated as a point charge.  The total current resulting from the whole track at time $t$ was the sum of the individual currents from each track segment, $i$.  The current induced on the anode wire from each segment was mainly a result of a positive charge, $q_i = e n_{i}$, drifting towards the cathode~\cite{Wilkinson}:
\begin{equation}\label{eq:summedcurrent}
I_{\mathrm{track}}(t) = \sum_{i=1}^{N} G_i n_{i} \frac{q_i}{2\,\ln(b/a)}\frac{1}{t-t_0+\tau},
\end{equation}
where $G_i$ is the gas gain, $n_i$ is the number of electron-ion pairs created in segment $i$, $a$=25~$\mu$m is the anode radius, $b$=2.54~cm is the NCD-counter inner radius, $t_0$ is the ``start time'' for the current from the $i^{\rm th}$ segment, 
and $\tau$ is the ion-drift time constant.  The number of ion pairs depended on the stopping power, $\frac{\mathrm{d}E}{\mathrm{d}x}$; the mean energy required to produce an electron-ion pair in the gas, $W$; and the segment length $l$, such that $n_i = \frac{\mathrm{d}E}{\mathrm{d}x} \frac{l}{W}$.

The description of an ionization track involved knowing where each segment was located and how much energy had been deposited there.  Multiple scattering of the ionizing particle was simulated with the Ziegler-Biersack-Littmarck method~\cite{ZBL}.  The results of that simulation were in excellent agreement with the full TRIM Monte Carlo calculation~\cite{TRIM}.  Values of $\frac{\mathrm{d}E}{\mathrm{d}x}$ were determined with stopping-power tables from TRIM for protons, tritons, and alphas.

The average energy required to produce an electron-ion pair in the NCD counter gas was measured using neutron sources and undeployed NCD counters.  Integrating over many current pulses with average energy $\overline{E}$, the ratio $G/W$ is proportional to the total current in a proportional counter, $I$~\cite{Knoll}:
\begin{equation}
\frac{G}{W} = \frac{I}{\eta e \overline{E}},
\end{equation}
where $\eta$ is the rate of neutron captures and $e$ is the electron charge.  $\overline{E}$ was determined with the NCD counter Monte Carlo to be $(701\pm{}7)$~keV.  $W$ is a characteristic of the NCD counter gas.  It is approximately energy-independent, and is approximately equal for protons, tritons, and alphas~\cite{icrureport31}.  We measured $W$ by operating the counter in the ``ion saturation'' mode (200-800~V), and $G/W$ by operating the counter at the standard voltage (1950~V).  In the first case, we found $W = 34.1 \pm 12.4$~eV, and in the latter case, $W = 34\pm 5$~eV, which was used in the Monte Carlo.

A low-energy electron transport simulation 
was developed to evaluate the mean drift times, $t_d$,
of electrons in the NCD counter gas mixture as a function of radial distance from the anode wire $r$. 
The results were in good agreement with GARFIELD \cite{GARFIELD} predictions,  measurements by Kopp \textit{et al} \cite{Kopp} and further verifications made by inspecting specific types of alpha pulses. The $t_d(r)$  curve used in the simulation was:
\begin{equation}\label{equation:driftcurve} 
t_d=121.3r+493.9r^2-36.71r^3+3.898r^4
\end{equation}
with $t_d$ in~ns and $r$ in~cm.

Electron diffusion resulted in a radially-dependent smearing 
effect on all pulses, and dominated the time resolution. 
A smearing factor $\sigma_\textrm{D}$ was  
tabulated as a function of $r$ and applied in pulse calculations. 
$\sigma_\textrm{D}$ and $t_d$ were linearly related: 
\begin{equation}\sigma_\textrm{D}(t_d)=0.0124\, t_d + 0.559\end{equation}

The mean NCD counter gas gain $\bar G$, as a function of voltage, was well described 
by the Diethorn formula \cite{Knoll}.  However, ion shielding from the charge multiplication can significantly change the gas gain.  A two-parameter model was developed to account quantitatively for this space-charge effect.  The change in gas gain, $\delta G$, resulting from a change in wire charge density, $\delta\lambda(\bar r)$, due to the ions formed near the anode at voltage $V$ is 
\begin{equation}\label{equation:dG}
 \delta G 
    \propto \bar G\ln(\bar G)\frac{\ln(b/a)}{2\pi\epsilon_{\naught} V}
 \left[1+\frac{1}{\ln(r_{\textrm{av}}/a)}\right]\delta\lambda(\bar r),
\end{equation}
where $\epsilon_{\naught}$ is the permittivity of free space and $r_{\textrm{av}}= 58 \pm10$~$\mu$m is the mean avalanche radius.  $\delta\lambda$ can be obtained by dividing the induced charge by a characteristic shower width in the spatial dimension parallel to the anode wire, $\cal W$.  The other parameter that needs to be optimized is the constant of proportionality in this equation.  Electrons originating from some segment of a track are affected by the density changes $\delta \lambda_j$ due to ions formed in previous electron cascades. Each of these ion clusters moves slowly towards the cathode while the primary electrons are being collected. In the presence of many ion clusters, the total change in the anode charge density at time $t$,  experienced by electrons from the $i^{th}$ track segment is therefore:
\begin{equation} \label{equation:dlamb} 
\delta\lambda_i = \frac{e}{\cal{W}}\sum_{j=1}^{i-1}
    \frac{\ln(b/\bar r_j(t))}{\ln(b/a)}G_j n_j + 
    \frac{e}{\cal{W}}\frac{\ln(b/\bar r)}{\ln(b/a)}n_i,
\end{equation}
where $n_j$ is the number of ion pairs formed in the $j^{th}$ segment.  $j$ loops over all previous ion clusters, which have moved to different radii $\bar r_j(t)$ at time $t$. $\bar r_j(t)$ is solved by integrating the relation $\frac{\mathrm{d}r_j}{\mathrm{d}t}=\mu_i{\cal E}$:
\begin{equation} 
\bar r_j(t)^2=\frac{2\mu_iVt}{\mathrm{ln}(b/a)}+r_{av}^2,
\end{equation}
where $\mu_i$ is the ion mobility and $\cal E$ is the cylindrical electric field.

The mean gas gain of the $i^{th}$ track segment is $\bar G_i=\bar G-\delta G_i$. The actual $G_i$ applied to the $i^{th}$ segment is sampled from an exponential distribution with mean $\bar G_i$.

The smaller ion mobility, relative to that of the electrons, results in the long tail that is characteristic of pulses from ionization in the NCD counters.  The evolution of a current pulse in a cylindrical proportional counter is described by Eqn.~\ref{eq:summedcurrent}.  The ion time constant, $\tau$, is inversely proportional to the ion mobility, $\mu$:
\begin{equation}
\tau = \frac{a^2 p \ln(b/a)}{2 \mu V},
\end{equation}
where $p$ is the gas pressure.

We measured $\tau$ using neutron calibration data.  Ionization tracks that are parallel to the anode wire have a relatively simple underlying structure; the primary ionization electrons all reach the anode at approximately the same time, with some spread due to straggling.  The ion-tail time constant was extracted by selecting the narrowest neutron pulses from calibration data sets and fitting each pulse with a Gaussian convolved with the ion tail, a reflection and the electronic model.  This model fitted the peaks well enough to allow for a characterization of the ion tail.  We found $\tau = 5.50 \pm 0.14$~ns.  This time constant corresponds to an ion mobility of $\mu = (1.082 \pm 0.027) \times 10^{-8}$~cm$^2$~ns$^{-1}$~V$^{-1}$.

\subsection{Simulation of NCD array electronics}

Propagation of the pulse along the NCD string was simulated with a lossy transmission line model.  Half of the pulse was propagated down the string, through the delay line, and back to the point-of-origin of the pulse.  The delay-line attenuation was also simulated as a lossy transmission line.  Both halves of the pulse (reflected and direct) were then transmitted up to the top of the NCD string.  The model parameters were based on SPICE simulations~\cite{spice} and \textit{ex situ} measurements.

Propagation in the NCD counter cable was simulated with a low-pass filter (RC $\approx$ 3~ns).  There was a small reflection (reflection coefficient = 15\%) at the preamp input due to the slight impedance mismatch between the preamp input and the cable.  This portion of the pulse traveled to the bottom of the NCD string and reflected back upwards.

The preamplifier was simulated with a gain (27,500 V/A), a low-pass filter (RC $\approx$ 22~ns) and a high-pass filter (RC = 58000~ns).  All RC constants in the electronic model were measured by fitting the model to \textit{ex situ} injected pulses.

The frequency response of the multiplexer system before the logarithmic amplifier was simulated with a low-pass filter (RC $\approx$ 13.5~ns).  The constants used to parameterize the logarithmic amplification were the same constants used to ``de-log'' real data pulses, which were determined by regular \textit{in situ} calibrations during data taking.  The circuit elements after the logarithmic amplification were simulated with the final low-pass filter (RC $\approx$ 16.7~ns).  The pulse array values were rounded off to the nearest integer to replicate the digitization.

Noise was added to the pulses as the final stage in the simulation.  It was added to the multiplexer and shaper branches of the electronics independently.  For the multiplexer branch, the frequency spectrum of the noise on the current pulses was measured for each channel using the baseline portions of injected calibration data.

The shaper-ADC branch of the electronics was simulated by a sliding-window integral of the preamplified pulse.  This number was then converted to units of ADC counts by doing an inverse linearity calibration.  The calibration constants used in this ``uncalibration'' were the same constants that were used to calibrate the data.  Noise was added to the shaper value with a Gaussian-distributed random number.  The mean and standard deviation of the noise for each channel were determined uniquely for each of the 40 NCD strings; the typical RMS noise (in units of ADC values) was 2.0, with a variance of 0.7 across the array.

The multiplexer and shaper systems included independent triggers that used the true threshold values.  Event triggers were determined by checking each pulse amplitude or shaper value against the appropriate threshold. The dead times of the two systems were then taken into account within each Monte Carlo event.  The NCD string signals were integrated with the PMT trigger simulation by inserting each signal into the time-ordered array of PMT signals.  As the simulation scanned over the combined PMT and NCD array  signals, any individual NCD string signal was sufficient to cause a global trigger of the detector.

Certain parts of the overall simulation were relatively slow due to the loops over the large ($N = 17,000$) arrays containing the simulated pulses.  As a result we implemented a fast alternative to the full simulation.  The ionization track was simulated to determine the timing of the event and the energy deposited in the gas.  That energy was converted directly to an approximate shaper-ADC measurement and was smeared with a Gaussian to roughly account for the missing physics and electronic noise.  This option allowed simulations for preliminary comparisons with the data because they did not require pulse-shape calculations.

\subsection{Verification and systematic uncertainties}
The ratio of the number of bulk uranium and thorium alpha events to that from cathode-surface polonium in the NCD strings could not be fully represented in the $^4$He string data due to string-to-string variation in these backgrounds.  Simulations were used to calculate the alpha energy spectrum PDFs instead.  Therefore, it was important to accurately simulate these PDFs, and to assess their systematic uncertainties for use in the region of interest for the solar neutrino analysis.

We optimized and validated the NCD array signal simulation by comparing with several types of data, including neutron source calibrations, high-energy alpha events, and $^4$He-string alpha data.  This procedure was designed to ensure that the simulations accurately reproduced the data, without in any way tuning on a data set that contained the neutron signal, which was determined by signal extraction (Sec.~\ref{sec:sigex}).

\begin{figure}
\begin{center}
\includegraphics[width=0.8\textwidth]{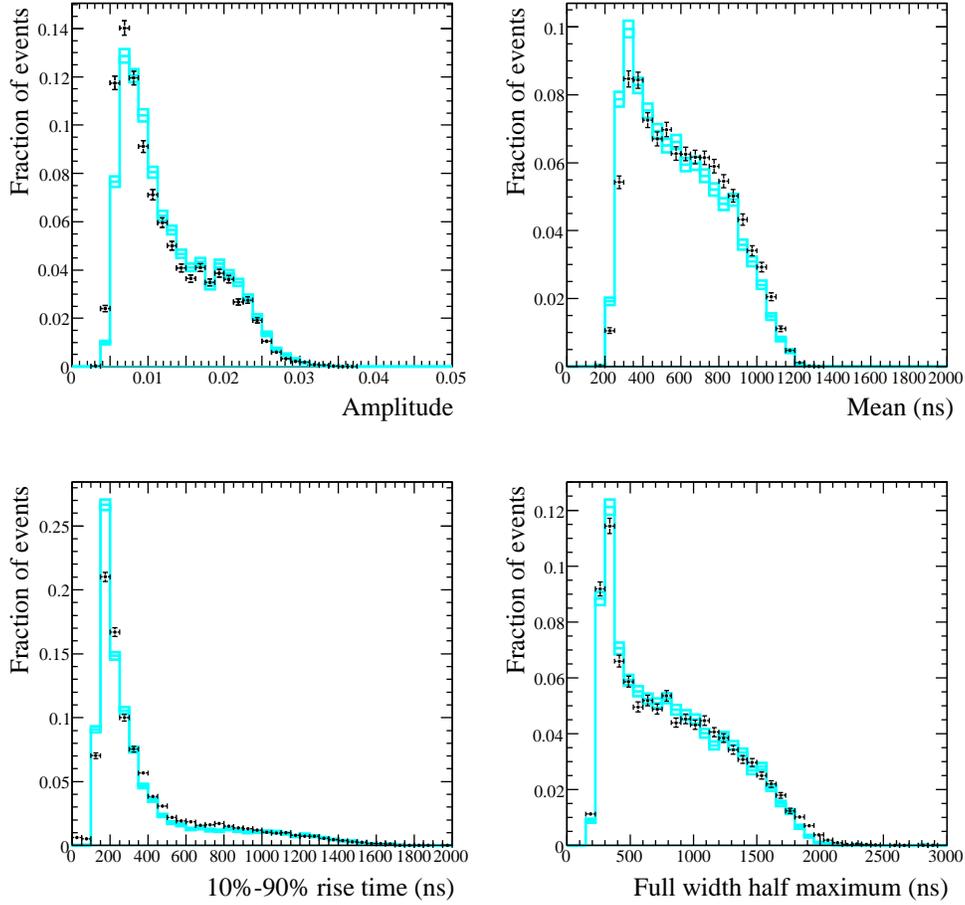} \\
\caption{\label{fig:ncdmcvar} Comparison of pulse shape variables in $^{24}$Na neutron calibration
data (data points) and the NCD array Monte Carlo (histogram) in the neutron
energy window, 0.4 to 1.2 MeV, with statistical uncertainties only. From top left to
bottom right: fraction of events as a function of pulse amplitude, time-axis mean
of the pulse, 10\%-90\% rise time, and full width at half maximum. All distributions
are normalized to unit area.}
\end{center}
\end{figure}

The comparison of the simulation with neutron calibration data tested nearly all aspects of the simulation physics model.  We compared $^{24}$Na neutron calibration data with simulations for a number of pulse characteristics, such as the mean, width, skewness, kurtosis, amplitude, and integral, and timing variables, including the rise time, integral rise time, and full-width at half-maximum.  These comparisons were used to estimate parameter values and uncertainties for electron and ion motion in the NCD counter gas, as well as the space charge model.  Figure~\ref{fig:ncdmcvar} shows a comparison of some of these pulse-shape variables between real and simulated $^{24}$Na neutron data.

\begin{figure}
\begin{center}
\includegraphics[width=0.4\textwidth]{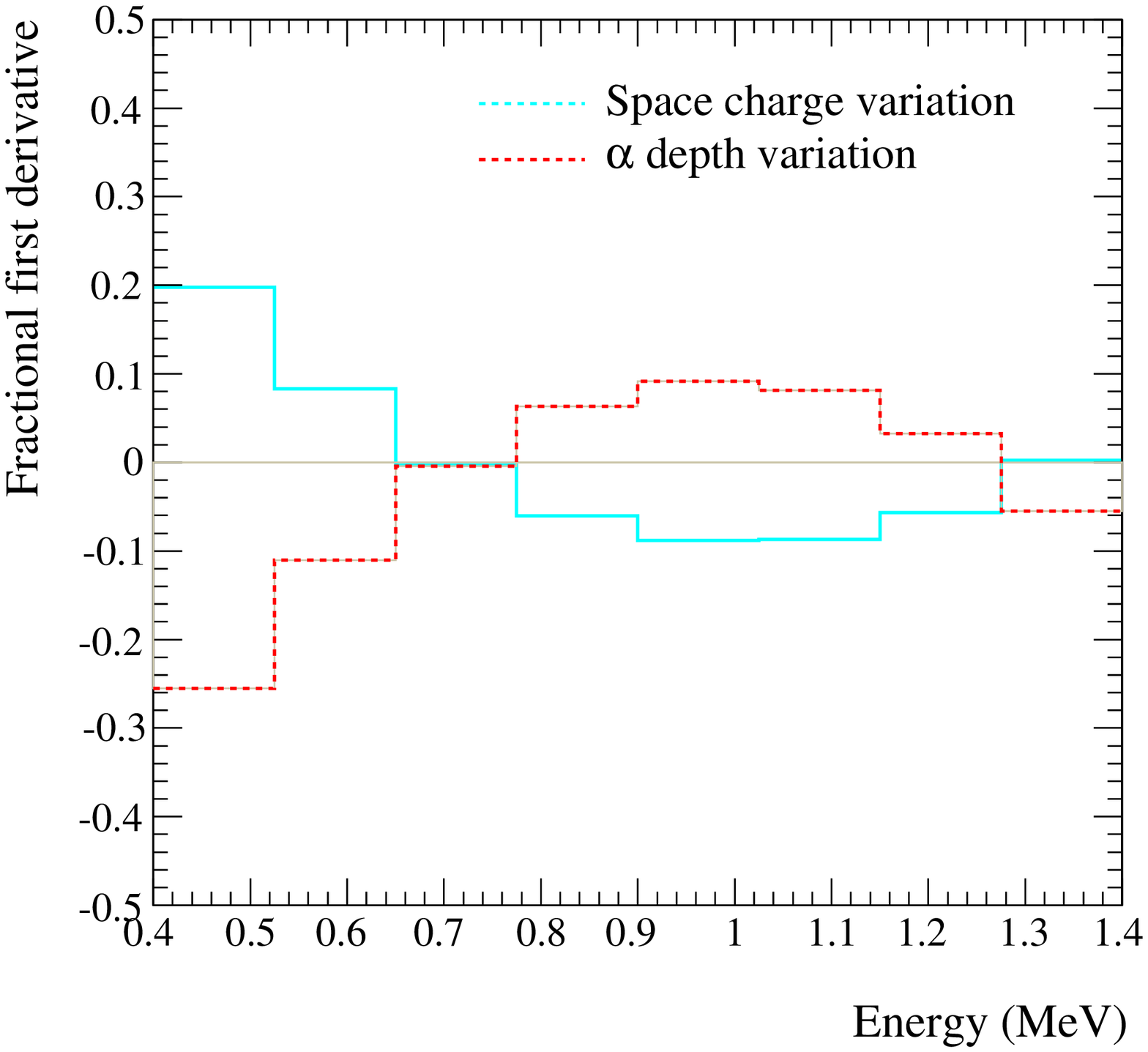} \\
\includegraphics[width=0.4\textwidth]{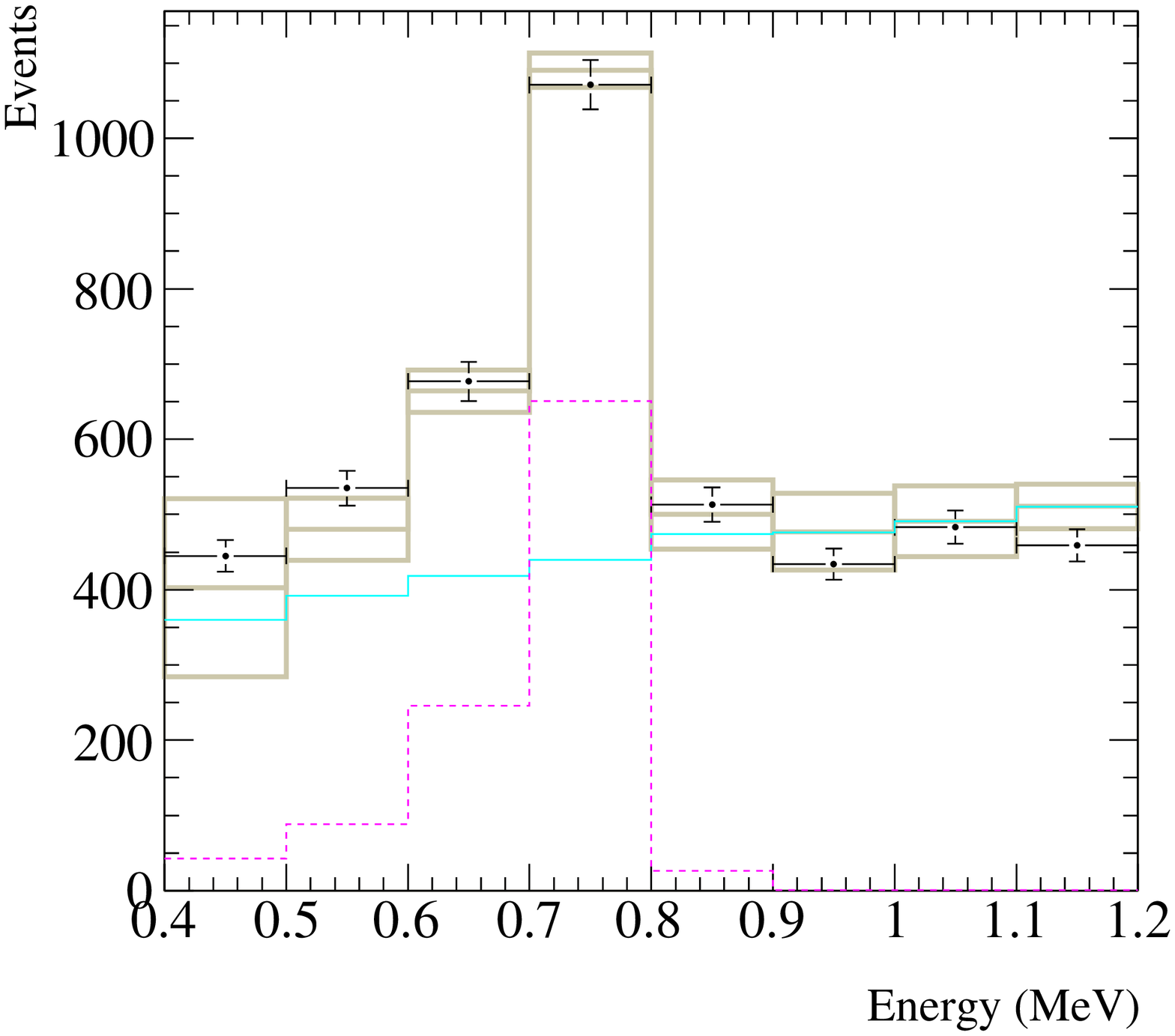} \\
\includegraphics[width=0.4\textwidth]{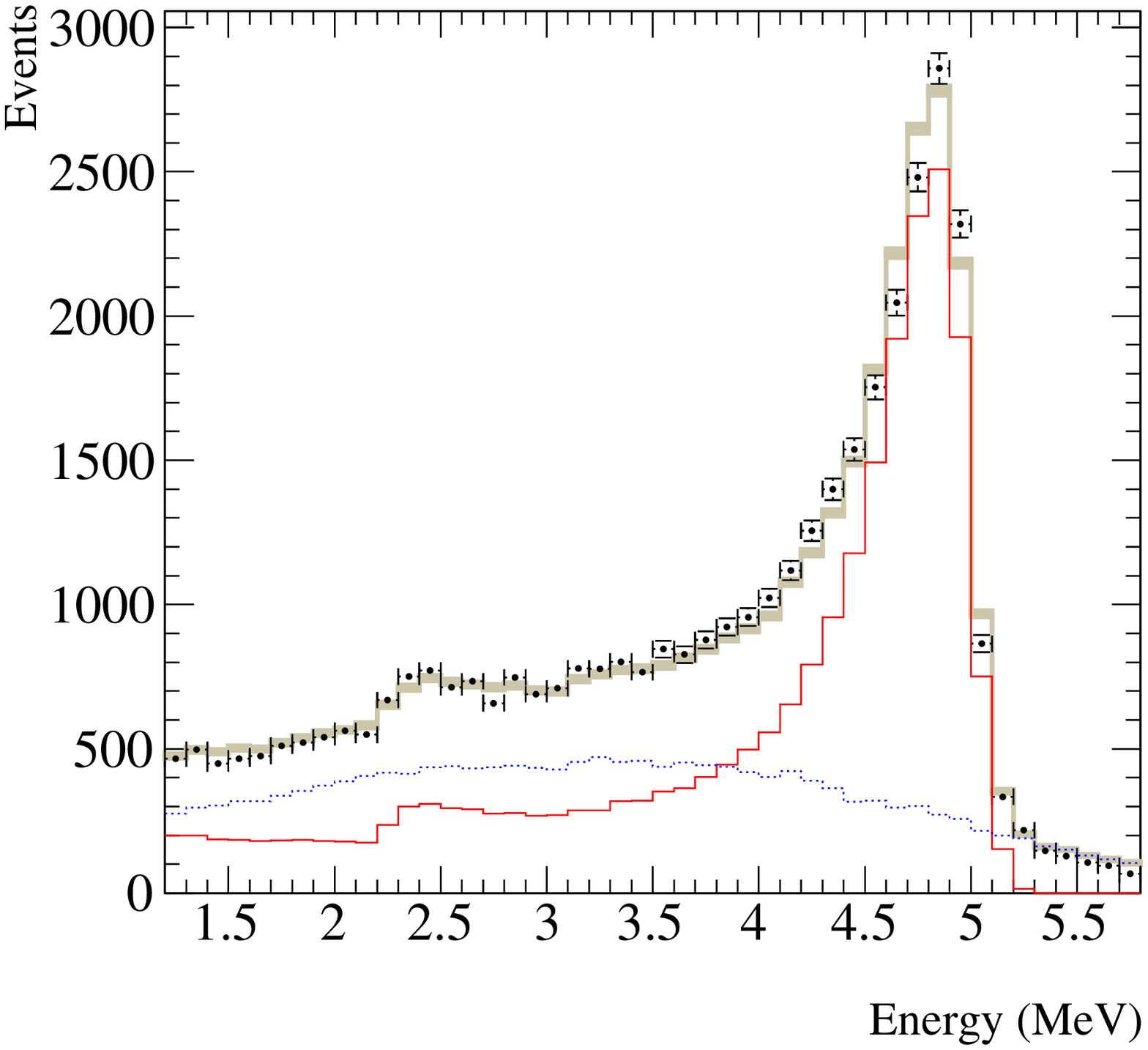}
\caption{\label{fig:energy_systematics} Top: fractional first derivatives vs. event energy (MeV) for two systematic variations (space charge and alpha depth) of the simulation.  Center: number of events versus energy in the NCD array data analysis energy window, for data (black, with statistical uncertainties), neutron template from calibration data (purple, dashed), alpha cocktail template from simulation (cyan), and total predicted number of events (grey, with systematic uncertainties) in the ``open'' data set (see Sec.~\ref{sec:sigex}).  Bottom: energy spectrum above the NCD array data analysis energy window, for data (black, with statistical uncertainties), alpha ``cocktail'' (grey, with systematic uncertainties), polonium (red), and bulk (blue dashed) simulation.  The ``cocktail'' is a collection of simulated alpha pulses with the appropriate mixture of NCD string cathode-surface polonium and bulk uranium and thorium alpha events. }
\end{center}
\end{figure}

We estimated the fraction of surface polonium and bulk alpha events on each string by fitting the energy distribution above the neutron energy region, shown in the bottom plot in Fig.~\ref{fig:energy_systematics}.  After the fit, we calculated an event weight, which was a function of string number and alpha type (polonium, uranium, or thorium) describing the best-fit fraction of alphas on each string due to each source.  In general, polonium comprised $\sim$60\% of the alpha signal; however, there were $\pm20\%$ variations between strings.  The best-fit alpha fractions and the MC energy scale correction were applied on an event-by-event basis to produce the alpha energy spectrum PDF.  

The systematic uncertainties included the depth of alpha-emitting contaminations (``alpha depth'') within the NCD counter walls, the efficiency of data reduction cuts, the space-charge model parameters, the electron-drift curve, the ion mobility, and the surface polonium to NCD bulk activity fraction on each string. These effects reflected uncertainties in the parameters of the NCD array simulation physics
model.  Systematic uncertainties were assessed by generating a large set of variation Monte Carlo samples, each with one input parameter varied by  $1\sigma$ with respect to its default value.  The size of the variation of the NCD array Monte Carlo systematic uncertainties was estimated from \textit{ex situ} data, off-line measurements of the NCD array signal processing electronic response, and in situ constraints from the NCD array data.  The most significant sources of systematic uncertainty were due to variations of the simulated alpha depth within the NCD counter walls and the data reduction cuts.

We calculated fractional first derivatives to describe the changes in the Monte Carlo alpha energy spectrum allowed by the systematic uncertainties~\cite{ncdsim}.  For each of the systematic uncertainties, the first derivatives for each bin of the energy distribution were calculated by taking the difference between histograms containing the standard Monte Carlo prediction for the shape of the energy distribution, and the variational Monte Carlo energy-distribution shapes. Then, the total systematic uncertainty in each
energy bin was assessed by summing the eight contributions in quadrature.  The functional dependence of those derivatives on energy is shown in the top plot of Fig.~\ref{fig:energy_systematics}.

The final simulated alpha energy spectrum in the solar neutrino analysis window for the NCD array is shown in the center plot of Fig.~\ref{fig:energy_systematics}, with systematic uncertainties.  This alpha spectrum was used as the alpha background PDF, together with the neutron signal template from $^{24}$Na neutron calibration data, to determine the total number of neutron events.

\section{Neutrino signal decomposition\label{sec:sigex}}

This section describes the techniques used in the SNO Phase-III neutrino flux measurement (referred to as signal extraction).  Three different signal extraction methods were developed.  These extended log-likelihood techniques were designed to perform a joint analysis of the data from the PMT and the NCD arrays.  The nuisance parameters (systematic uncertainties), weighted by external constraints determined from calibrations and simulations, were allowed to vary in the fit of the neutrino signals.  This ``floating'' technique enabled the determination of the  correlations between the observed signals and the nuisance parameters.  In the three methods, the energy spectrum of NCD array events was fit with the Monte Carlo alpha background distribution described earlier, the neutron spectrum determined from $^{24}$Na calibration, expected neutron backgrounds, and two instrumental background event distributions.  Events from the PMT array were fitted in the reconstructed effective kinetic energy $\teff$, the cosine of the event direction relative to the vector from the Sun $\cos\theta_\odot$, and the volume-weighted radius $\rho$.

There were in general two classes of systematic uncertainties:  the first had a direct impact on the shapes of the probability density functions (PDF), such as energy scale or angular resolution for the PMT array data; the second are uncertainties on parameters that did  not affect the PDF shapes, such as detection efficiencies.  The three signal extraction techniques differed in the implementation of how the nuisance parameters were floated.  

The first method was an extension to the signal extraction techniques~\cite{james_loach_phd} used in previous analyses~\cite{longd2o,nsp}.  Systematic uncertainties of the second kind were easily floated in this method.  But floating the first kind was much more challenging as it would require PDFs to be rebuilt between evaluation of the likelihood function in the minimization process.  As a result, only a few significant systematic uncertainties were allowed to float in this method due to computational limitations.  Those systematic uncertainties that were not floated were estimated by repeating the fit with the parameters varied by their positive and negative $1\sigma$ deviations.  

In the second method the nuisance parameters were not varied during the fit.  Instead the signal extraction fit to the data was done many times with the necessary PDFs built from a set of nuisance parameters, whose values were drawn randomly according to their distributions.  After this ensemble of fits, the flux result from each trial was re-weighted in proportion to its likelihood.  This sampling method circumvented the need to rebuild PDFs during the numerically intensive likelihood maximization process.  

Markov-Chain Monte Carlo (MCMC) was the third method used in signal extraction.  The flux results from this method were published in the original letter on this phase of SNO~\cite{ncdprl}, and the details in its implementation are presented in the following.  These discussions include a description of how the MCMC parameter estimation works, the extended log-likelihood function used to obtain parameter estimates, and finally the results of fits to the full third-phase data set.

It is noted that a blind analysis procedure was used.  The data set used during the development of the signal extraction procedures excluded a hidden fraction of the final data set and included an admixture of neutron events from cosmic-ray muon interactions.  The blindness constraints were removed after all analysis procedures, parameters and backgrounds were finalized.  The finalization of the three signal extraction procedures before removing these constraints required an agreement of their flux results for the ``open'' data set  to within the expected statistical spread.   A comparison of the results from the three analysis methods after the blindness conditions had been removed revealed two issues, which were understood prior to the publication of the letter~\cite{ncdprl}.  These issues will be discussed later in this section.

\subsection{Markov-Chain Monte-Carlo parameter estimation}

In SNO's previous phases, the negative log-likelihood (NLL) function was simply minimized with respect to all parameters to get the best-fit value.   Minimizing the NLL was very challenging with so many fit parameters, and the likelihood space near the minimum could be uneven.  This means that common function minimization packages, such as MINUIT~\cite{Minuit}, could run into numerical convergence problems.  The use of MCMC circumvented this slow convergence problem by interpreting NLL as the negative logarithm of a joint probability distribution $P$ for all of the free parameters, i.e.  $P=\exp(\log(L))$.  The nuisance parameters were integrated  to determine the posterior distributions for the fluxes.  The origins of this procedure go back to Bayesian probability theory, and in fact our approach could be considered to be a Bayesian analysis with uniform priors assumed for the fluxes.   Both the speed of convergence and the insensitivity to unevenness in the NLL space mean that the MCMC method is better suited to handling large numbers of nuisance parameters.

The basic idea of MCMC is to take a random walk through the parameter space, where each step is taken with a probability given by the likelihood of the new step ($L_{\textrm{prop}}$) compared to the previous step ($L_{\textrm{curr}}$).  The probability of accepting the proposed new parameters is min(1,$L_{\textrm{prop}}/L_{\textrm{curr}}$).  If the step is accepted the parameter values are updated, else the MCMC repeats the current point in the chain and generates a new proposal for the parameters.  By the Metropolis-Hastings theorem~\cite{metro,hast}, the resulting distribution of parameters from the chain will have a frequency distribution given by $L$.  The choice of an appropriate proposal kernel is critical to this method.  If the width of the proposal distribution were too wide, steps are rarely taken; and if the width of the proposal distribution were too narrow, then the MCMC would not sample enough of the parameter space to find the best-fit region, and instead could fall into a local minimum.  In the MCMC signal extraction method, the step proposal distribution was a Gaussian of mean zero and width that was $\sim 1/3$ of the expected statistical uncertainty or the constraint uncertainty.  This choice was checked by the convergence and distributions of different MCMC chains that were started at different random starting points.

\subsection{Observables, signals and fit ranges}

\subsubsection{PMT array data}

For the PMT array data, 33 signals and backgrounds were included in the fit.   An energy-unconstrained fit was done, meaning that CC and the ES fluxes were fitted for each $\teff$ bin.  The energy binning used for this unconstrained fit was 0.5-MeV bins between $\teff$ of 6~MeV and 12~MeV, and a single bin from 12~MeV to 20~MeV, totaling 13 bins per signal.  The other signal and backgrounds in the fit included the NC signal; photodisintegration neutron backgrounds due to radioactivity in the \dto\ target, the bulk nickel in the NCD strings, and the hotspots on the strings K2 and K5; photodisintegration and ($\alpha,n$) neutron backgrounds in the acrylic vessel (which also included the backgrounds from the cables of the NCD array); and neutrons from atmospheric neutrino interactions.

The observables used in the fit to the PMT data were each event's reconstructed $\teff$, $\rho$, and $\cos\theta_\odot$.  Cherenkov light candidate events were selected with the criteria $\rho<\left(\frac{550}{600} \right)^2=0.77025$, $-1\leq\cos{\theta_\odot}\leq1$, and $6.0$~MeV$< \teff<20.0$~MeV.  The signal and background PDFs for the PMT array data were three dimensional in $\rho$, $\cos{\theta_\odot}$, and $\teff$.  The exceptions were the photodisintegration backgrounds due to K2, K5 and the bulk nickel, which were factorized (i.e. $P(\teff)P(\cos\theta_\odot)P(\rho)$) in order to avoid any problems associated with low statistics in their simulations.  

\subsubsection{NCD array data}

The NCD array signals and backgrounds in the fits were the NC signal and various sources of neutron backgrounds.  These backgrounds include photodisintegration due to radioactivity in the \dto\ target, nickel housing of the NCD array, and hotspots on strings K2 and K5; external source neutrons due to radioactivity in the acrylic vessel and the \hto\ shield;  neutrons from atmospheric neutrino interactions and cosmogenic muons; and instrumental backgrounds with characteristics of those seen in strings J3 and N4.

The only observable used in signal extraction was the summed energy spectrum of the shaper-ADC in the NCD array readout (``shaper energy'', $E_{\textrm{NCD}}$), restricted to the range of  $0.4$~MeV$<E_{\textrm{NCD}}<1.4$~MeV.  The shaper energy spectrum from a uniformly distributed $^{24}$Na source was used as the NC PDF.  The alpha background PDFs were derived from the simulation as discussed in Sec.~\ref{sec:pulsesim}.

\subsection{The extended log-likelihood function}

In the extended log-likelihood function, two different approaches were used to handle systematic uncertainties.  To include systematic uncertainties associated with the PMT array data and uncertainties associated with the neutron PDFs in the NCD array data, the respective PDFs were rebuilt from an array of event observables on each evaluation of the log-likelihood function.  For the systematic uncertainties associated with the alpha background PDFs in the NCD array data, the shapes of those PDFs were modified by a multiplicative function.

Twenty six systematic uncertainties were included in the fit.  The systematic uncertainties due to neutron detection efficiency were applied to the neutral-current PDFs separately for the NCD and PMT data streams.   The twelve systematic uncertainties associated with the PDF shapes for the PMT array were related to its energy scale and resolution, and the event vertex reconstruction algorithm's spatial accuracy, spatial resolution and angular resolution.  This last entity was applied to the ES signal only.  The twelve systematic uncertainties associated with the PDF shapes for the NCD array were related to the energy scale ($a_{1}^{E_{\textrm{NCD}}}$) and resolution  ($b_0^{E_{\textrm{NCD}}}$) of the counters, data reduction cut efficiency on alpha backgrounds, spatial distributions and intensity of different alpha background activities, energy spectrum of instrumental backgrounds, and parameters in the Monte Carlo that could affect the observed signal, such as ion mobility.  Details of the parameterization of these systematic parameters can be found in Appendix~\ref{sec:apdxb}.

A constraint term for each of the 26 systematic uncertainties was added to the log-likelihood function. The means and standard deviations of these constraint terms came from calibration measurements or Monte-Carlo studies and are listed in Table~\ref{table:mcalphasys} in Sec.~\ref{sec:fitresults}.  

The NLL function to be minimized was the sum of a NLL for the PMT array data $-\log{L_{\textrm{PMT}}}$ and for the NCD array data ($-\log{L_{\textrm{NCD}}}$):
\begin{equation}
-\log{L} = -\log{L_{\textrm{PMT}}} -\log{L_{\textrm{NCD}}}.
\end{equation}

The following subsections describe how these two NLLs were handled in this analysis.

\subsubsection{PMT array --- NLL with systematic uncertainties\label{sec:pmtnllsys}}

The systematic uncertainties arose from differences in the simulations and reconstructed data.  The PDFs associated with the PMT array ($P_{\textrm{mc}}$) were rebuilt using the scaled and smeared value of $\teff$, $\rho$ and $\cos\theta_\odot$.  The goal was to allow the data to tell us how the scales and resolutions differed between data and simulation within the constraints from calibration data.  We therefore modeled the differences as a possible re-mapping of the observables for the simulated events.  We then fitted for the re-mapped parameters to determine the allowable range of this re-mapping while still matching the data observables.
 
The negative log-likelihood for the PMT array data can therefore be written as:
\begin{equation}\label{eqn:loglpmt}
\begin{split}
 -\log{L_{\textrm{PMT}}} &= \sum_i^{M^{\textrm{PMT}}} f_i^{\textrm{PMT}} S_i^{\textrm{PMT}} \phi_i - \\ 
     & \hspace{0.75cm}
 \sum_j^{N^{\textrm{PMT}}} \log \left( \sum_i^{M^{\textrm{PMT}}} f_i^{\textrm{PMT}} S_i^{\textrm{PMT}} 
   \phi_i P_{\textrm{mc}}^i (\rho_j, \cos\theta_{\odot,j}, T_{{\rm eff},j} ) \right)  -\log{L^{\textrm{PMT}}_{\rm constraints}},
   \end{split}
\end{equation}	
where there were $M^{PMT}$ event classes (such as neutrino signals from different interactions and Cherenkov light backgrounds)  and  $N^{PMT}$ PMT events.  For the $i^{\text{th}}$ signal, the factor to convert from flux to the number of events is $f_i$, the fitted flux is  $\phi_i$, the fiducial volume correction factor is $S_i$, and the remapped normalized probability density function is $P_{\textrm{MC}}^i$.  The constraint terms are:
\begin{equation}\label{eqn:loglpmtconstr}
\begin{split}
-\log L^{\textrm{PMT}}_{\textrm{constraints}} &= \frac{1}{2}\sum_i{ (\frac{p_i-\overline{p_i}}{\sigma_{p_i}})^2 } + \\ 
                          &  \hspace{0.75cm} \frac{1}{2}\sum_{i} \sum_{j} (b_i-\overline{b_i}) \, (b_j-\overline{b_j}) \, (V_{b}^{-1})_{ij} 
\end{split}
\end{equation}
where $p_i$ are all of the PMT systematic parameters other than two sets of correlated position resolution parameters denoted as $b_i$ in the second term.  $V_{b}$ is the covariance matrix for these latter parameters.  The parameterization of the nuisance parameters can be found in Appendix~\ref{sec:apdxb}.

\subsubsection{NCD array --- NLL with systematic uncertainties\label{sec:ncdnllsys}}

For the NCD array data the probability distribution functions $Q$ were one dimensional functions of the shaper energy ($E_{\textrm{NCD}}$).  For each evaluation of
the likelihood, where the neutron PDF systematics were changed, the PDFs were rebuilt using $E_{\rm remap}$:
\begin{equation}
E_{\rm remap} = a_1^{E_{\textrm{NCD}}} E_{\textrm{NCD}} \times (1 + N(0, b_0^{E_{\textrm{NCD}}}))
\label{eqn:eadcremap}
\end{equation}
where $N(\mu_{\naught},\sigma)$ is a Gaussian distribution with mean $\mu_{\naught}$ and width $\sigma$.  As mentioned previously, the simulated alpha background PDF was rebuilt by multiplying the unmodified PDF  ($Q^{\alpha}$)  by a re-weighting factor ($\alpha_i$) and a multiplicative function in shaper energy $s_i(E_{\textrm{NCD}})$:
\begin{equation}
Q_{\textrm{mc}}^{\alpha}(E_{\textrm{NCD}}) = Q^{\alpha}(E_{\textrm{NCD}}) (1+\sum_{i=1}^{8}\alpha_i \, s_i(E_{\textrm{NCD}})).
\end{equation}
The eight reweighting functions $s_i$ included systematic effects in the depth and intensity variations of $^{210}$Po and other natural alpha emitters in the NCD string body, ion mobility, avalanche width and gradient offset, drift time variation and data reduction cut efficiency.  In addition two instrumental background PDFs, based on instrumental background events observed in strings J3 and N4 (which were excluded from the neutrino candidate data set), were parameterized as skewed Gaussian distributions.  A summary of the parameterizations of these systematic uncertainties can be found in Appendix~\ref{sec:apdxb}.

The negative log-likelihood for the NCD array data can therefore be written as:
\begin{equation}
\begin{split}
 -log L_{\textrm{NCD}} &= \sum_i^{M^{\textrm{NCD}}} f_i^{\textrm{NCD}} S_i^{\textrm{NCD}} \phi_i - \\
  & \hspace{0.75cm} \sum_j^{N^{\textrm{NCD}}} \log \left( \sum_i^{M^{\textrm{NCD}}} f_i^{\textrm{NCD}} S_i^{\textrm{NCD}} 
   \phi_i Q_{\textrm{MC}}^i(E_{\textrm{NCD}, j}) \right) -\log L^{\textrm{NCD}}_{\textrm{constraints}},
\label{eqn:loglncd}
\end{split}
\end{equation}
where there are $M^{\textrm{NCD}}$ event classes and $N^{\textrm{NCD}}$ events in the NCD array data.
For the  $i^{\text{th}}$ event class, the flux-to-event conversion factor is
$f_i^{\textrm{NCD}}$, the fitted flux is $\phi_i$, the fiducial volume correction factor is
$S_i$, and the remapped normalized probability density function is
$Q_{\textrm{MC}}^i$.  The constraint terms are:
\begin{equation}
-\log{L^{\textrm{NCD}}_{\rm constraints}} = 
\frac{1}{2}\sum_i{ (\frac{p_i-\overline{p_i}}{\sigma_{p_i}})^2 }, 
\label{eqn:loglncdconstr}
\end{equation}
where $p_i$ are all of the systematic parameters for the NCD array data.

In Eqns.~\ref{eqn:loglpmt} and~\ref{eqn:loglncd} the factors $S_i$ were used to propagate the fiducial volume uncertainty.  They could change as the systematic parameters were changed, and were calculated each time the PDFs were rebuilt.  The other set of conversion factors $f_i$ included live time and efficiencies that were used to convert the number of events in a fiducial volume into a flux above threshold.  A full description of these conversion factors can be found in Ref.~\cite{longd2o}.

\subsubsection{Neutron background and other input constraints}

The neutron backgrounds in the PMT array data were determined from the neutron-background fits in the NCD array data.  For $N_i^{\textrm{NCD}}$ events fitted for a certain neutron background in the NCD array data, the number of background events from the same source is  $f_i^{\textrm{PMT}} N_i^{\textrm{NCD}}$, where $f_i^{\textrm{PMT}}$ is the ratio of the number expected in the NCD array data to that in the PMT array data.  These conversion factors are summarized in Table~\ref{table:values}.

\begin{table}
\begin{center}
\caption{Multiplicative conversion factors for determining the number of neutron background events in the PMT array data from the number in the NCD array data.  Photodisintegration is denoted as ``PD'' in this table.}
 \label{table:values}
 \begin{minipage}[t]{4in}
\begin{ruledtabular}
\begin{tabular}{ c  c c }
Factor      & Description & Value  \\ \hline
$f_{\textrm{ex}}^{\textrm{PMT}}$   &  AV, PD            & 0.5037    \\
$f_{\textrm{d2opd}}^{\textrm{PMT}}$&  D$_2$O, PD       & 0.2677  \\
$f_{\textrm{ncdpd}}^{\textrm{PMT}}$&  NCD, PD    & 0.1667    \\
$f_{\textrm{k2pd}}^{\textrm{PMT}}$ &  K2, PD      & 0.2854    \\
$f_{\textrm{k5pd}}^{\textrm{PMT}}$ &  K5 , PD           & 0.2650    \\
$f_{\textrm{cab}}^{\textrm{PMT}}$  &  NCD Cables, PD      & 0.1407   \\
$f_{\textrm{atmos}}^{\textrm{PMT}}$&  Atmospheric $\nu$   & 1.8134   \\
\end{tabular}
\end{ruledtabular}
\end{minipage}
\end{center}
\end{table}

Several small Cherenkov light backgrounds required adjustments of the CC and ES fluxes.  These backgrounds were Cherenkov events from beta-gamma decays in the \dto\ and in the external regions (AV, \hto, and PMT support geodesic structure), and isotropic light background from the AV.  To account for these small backgrounds at each step in the chain, their contributions were randomly drawn, assuming a Gaussian distribution with means and widths given in Table~\ref{nppd}.  These contributions were then split between the CC and ES channels, with the latter assuming 10\% of the total.  This fraction was applied because the ES peak occupies a corresponding fraction in the $\cos\theta_\odot$ distribution for neutrino signal.

\subsection{Results of fits to the full Phase-III data set\label{sec:fitresults}} 

Fits without systematic uncertainty evaluation were obtained by running a single fit Markov-chain with 320,000~steps where the first 20,000~steps were rejected to ensure convergence.  The fits that allow systematic uncertainty evaluation were obtained by running 92 independent Markov-chains with 6,500 steps each.  Each fit was started with different starting parameters near the best-fit point varied by a Gaussian distribution with width given by the estimated uncertainty on the parameter.  The first 3,500~steps in this fit were rejected in order to minimize the effect of initial values on the posterior inference, and the remaining 3,000~steps of each fit were put into a histogram for each parameter.  A total of 276,000 steps were used to estimate the parameter uncertainties.

To ensure the robustness of the MCMC signal extraction, an ensemble test with all systematic parameters floated was performed.  Each of the mock data sets were assembled with PDFs that were regenerated with different randomly sampled systematic parameter values.   To run this ensemble test of 100 runs with all systematic parameters included, each with a MCMC chain length of 22,000, a substantial amount of computational power was required.  As a result, only the final fit configuration of the full Phase-III data set was tested.  The fits to the real data were performed with far more steps to ensure that the uncertainty estimates were more robust than those in the ensemble test.  This ensemble test showed an acceptable level of bias and pull in the fit parameters, and established the reliability of this signal extraction method.

\begingroup
\begin{table}
\begin{center}
\caption{Results of fit to the full Phase-III data set.  The fitted number of signal and background counts, along with the integral neutrino flux values are shown.   In the data set, the total number of events in the PMT array and NCD array data sets were 2381 and 7302 respectively.  The $^8$B spectrum from Ref.~\cite{wint} was used in deriving the equivalent neutrino fluxes from the fitted number of CC, ES and NC events.  All the fluxes are in units of $\times 10^{6}$/cm$^2$/s.  }
 \label{table:mcalphanocor}
 \begin{minipage}[t]{3.5in}
\begin{ruledtabular}
\begin{tabular}{ l l}  
\multicolumn{2}{c}{Fitted counts --- PMT array} \\ 
\hline
CC         &    1867$^{+ 91}_{-101}$   \\
ES         &     171$^{+ 24}_{- 22}$   \\
NC   &    267$^{+ 24}_{- 22}$   \\
Backgrounds   &    77$^{+ 12}_{- 10}$   \\
Total    &    2382$^{+ 98}_{-107}$   \\
\hline	
\multicolumn{2}{c}{Fitted counts --- NCD array} \\ 
\hline	  					                     
NC  & 983$^{+ 77}_{- 76}$   \\
Neutron backgrounds   & 185$^{+ 25}_{- 22}$   \\
Alpha backgrounds    & 5555$^{+196}_{-167}$   \\
Instrumental backgrounds       &   571$^{+162}_{-175}$   \\
Total       &  7295$^{+ 82}_{- 83}$   \\ \hline	
\multicolumn{2}{c}{Integral flux} \\ \hline                     
$\phi_{\textrm{CC}}$      & 1.67$^{+ 0.08}_{-0.09}$  \\
$\phi_{\textrm{ES}}$      & 1.77$^{+0.26}_{-0.23}$  \\
$\phi_{\textrm{NC}}$      &  5.54$^{+0.48}_{-0.46}$  \\
\end{tabular}
\end{ruledtabular}
\end{minipage}
\end{center}
\end{table}
\endgroup

\begin{table}
\caption[SNO-III statistical correlation coefficients for the CC, ES and NC fluxes.]{Statistical correlation coefficients for the CC, ES and NC fluxes in the full SNO Phase-III data set.}
\begin{center}
\begin{minipage}[t]{3in}
\begin{ruledtabular}
\begin{tabular}{cccc}
Signal& CC & ES & NC\\
\hline
CC&1.000 &0.2376&-0.1923\\
ES & 0.2376  &1.000&0.0171\\
NC&-0.1923 & 0.0171&1.000\\
\end{tabular}
\end{ruledtabular}
\end{minipage}
\end{center}
\label{table:fluxcor}
\end{table}%

In the signal extraction fit, the spectral distributions of the ES and CC events were not constrained to the $^{8}$B shape, but were extracted from the data.   In Table~\ref{table:mcalphanocor}, the number of events in different signal and background classes determined from this ``energy-unconstrained'' fit are tabulated.  The equivalent neutrino fluxes, derived from the fitted number of CC, ES and NC events under the assumption of the $^8$B neutrino spectrum in Ref.~\cite{wint},  were determined to be (in units of $10^6$/cm$^2$/s)~\cite{hep_footnote,es_crosssection}:
\begin{eqnarray}
\phi_{\text{CC}} &=& 1.67^{+0.08}_{-0.09} \nonumber \\
\phi_{\text{ES}} &=& 1.77^{+0.26}_{-0.23} \\
\phi_{\text{NC}} &=& 5.54^{+0.48}_{-0.46} \nonumber 
\end{eqnarray}
where the uncertainties are the total uncertainties obtained from the posterior distributions.  Their correlations are tabulated in Table~\ref{table:fluxcor}.  The ratio of the ${}^{8}$B neutrino flux measured with the CC and NC reactions is
\begin{equation}
\frac{\phi_{\text{CC}}}{\phi_{\text{NC}}}  =  \snoccncratio. \\
\end{equation}
In Table~\ref{table:ccesphi}, the CC and ES electron differential energy spectra from the energy-unconstrained fit are tabulated.

\begingroup
\begin{table}
\begin{center}
\caption{The CC and ES electron differential energy spectrum.   The fluxes in each of the 13~$\teff$ bins are in units of $10^{4}$/cm$^2$/s.  The uncertainties shown are total uncertainties with correlations between all systematic uncertainties described in the text included.}
 \label{table:ccesphi}
 \begin{minipage}[t]{3in}
 \begin{ruledtabular}
\begin{tabular}{ l c c }
$\teff$ (MeV) & CC & ES \\ \hline
6.0 $-$ 6.5  & 19.0$^{+ 2.3}_{- 2.2}$  & 33.4$^{+ 10.6}_{- 8.8}$ \\
6.5 $-$ 7.0  & 23.7$^{+ 2.0}_{- 1.9}$  & 10.5$^{+ 9.2}_{- 8.1}$ \\
7.0 $-$ 7.5  & 21.2$^{+ 2.0}_{- 1.7}$  & 33.2$^{+ 9.3}_{- 7.8}$ \\
7.5 $-$ 8.0  & 18.9$^{+ 1.7}_{- 1.6}$  & 28.1$^{+ 8.4}_{- 7.7}$ \\
8.0 $-$ 8.5  & 17.2$^{+ 1.6}_{- 1.4}$  & 12.4$^{+ 6.7}_{- 5.5}$ \\
8.5 $-$ 9.0  & 14.2$^{+ 1.4}_{- 1.3}$  & 16.3$^{+ 7.3}_{- 5.3}$ \\
9.0 $-$ 9.5  & 13.3$^{+ 1.4}_{- 1.2}$  & 17.8$^{+ 6.9}_{- 5.1}$ \\
9.5 $-$ 10.0  & 10.0$^{+ 1.3}_{- 1.0}$  & 9.1$^{+ 5.8}_{- 3.9}$ \\
10.0 $-$ 10.5  & 10.3$^{+ 1.2}_{- 1.1}$  & 0.2$^{+ 4.8}_{- 0.4}$ \\
10.5 $-$ 11.0  & 6.6$^{+ 1.0}_{- 0.9}$  & 1.2$^{+ 3.3}_{- 6.3}$ \\
11.0 $-$ 11.5  & 4.1$^{+ 0.7}_{- 0.6}$  & 3.16$^{+ 3.2}_{- 2.5}$ \\
11.5 $-$ 12.0  & 3.3$^{+ 0.6}_{- 0.5}$  & 2.32$^{+ 3.1}_{- 2.4}$ \\
12.0 $-$ 20.0  & 5.3$^{+ 1.1}_{- 0.9}$  & 9.00$^{+ 4.8}_{- 3.1}$ \\ 
\end{tabular} 
\end{ruledtabular}
\end{minipage}
\end{center}
\end{table}
\endgroup

\begin{figure}[htbp]
\includegraphics[width=0.40\columnwidth]{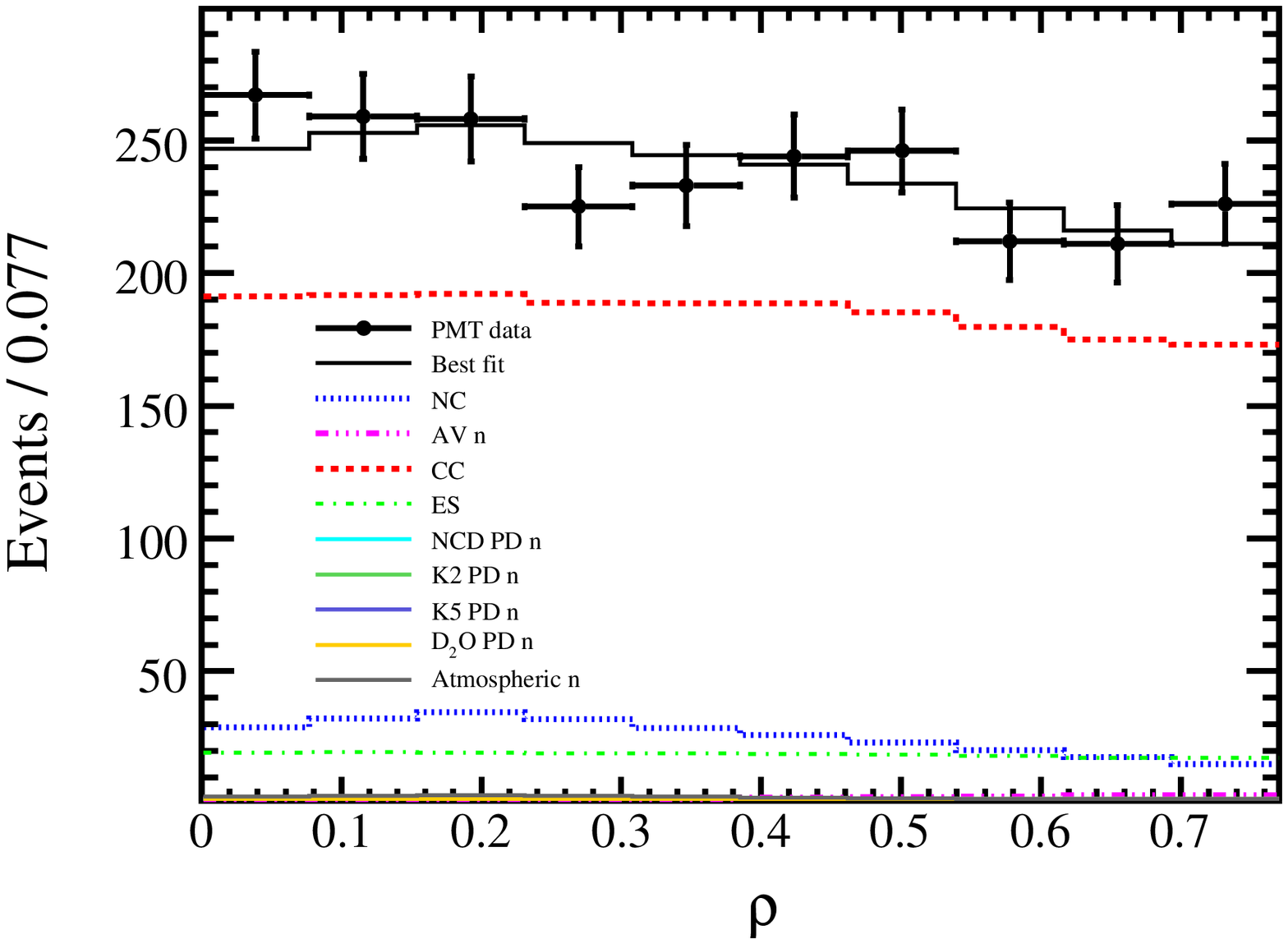}
\includegraphics[width=0.40\columnwidth]{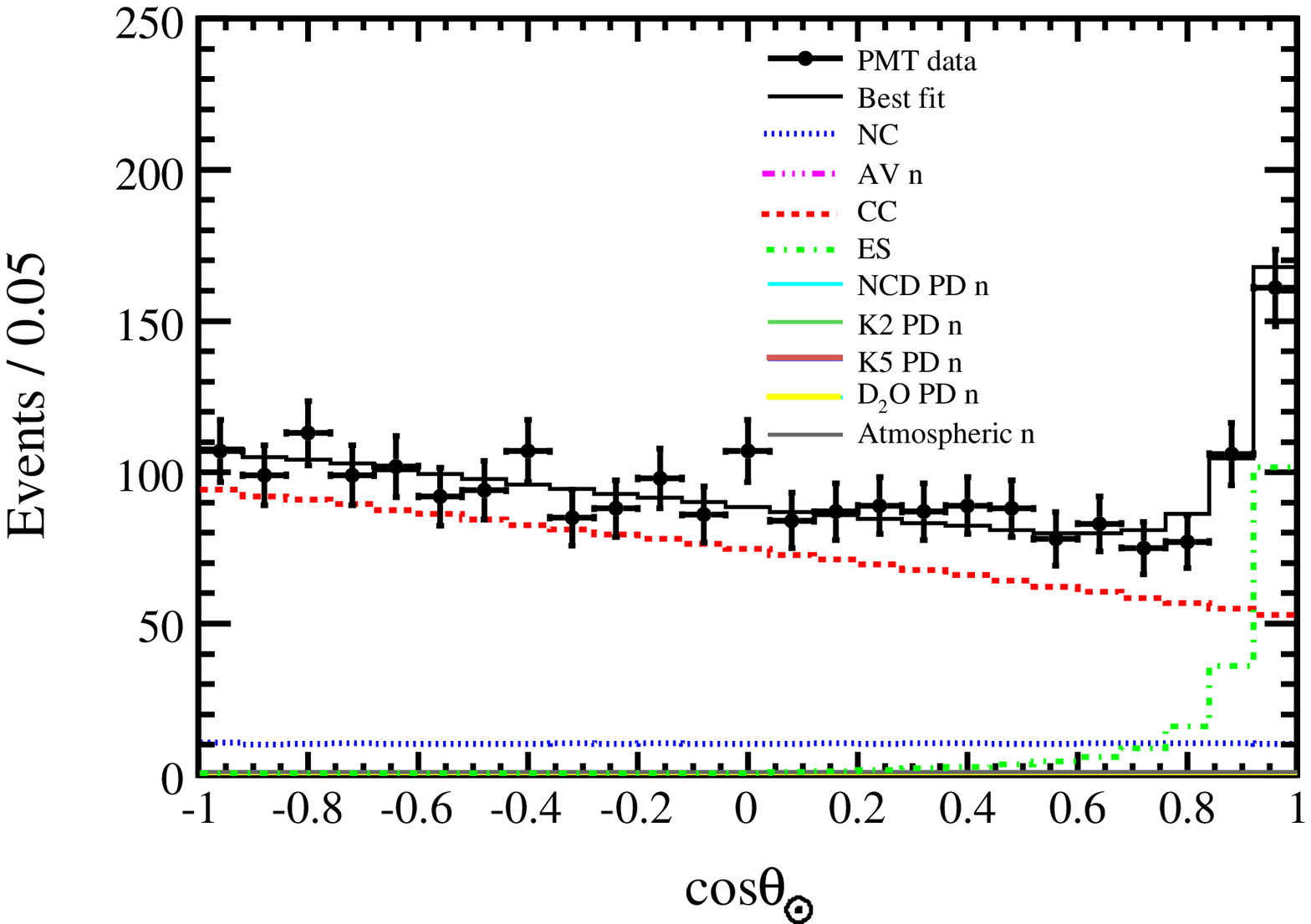}
\includegraphics[width=0.40\columnwidth]{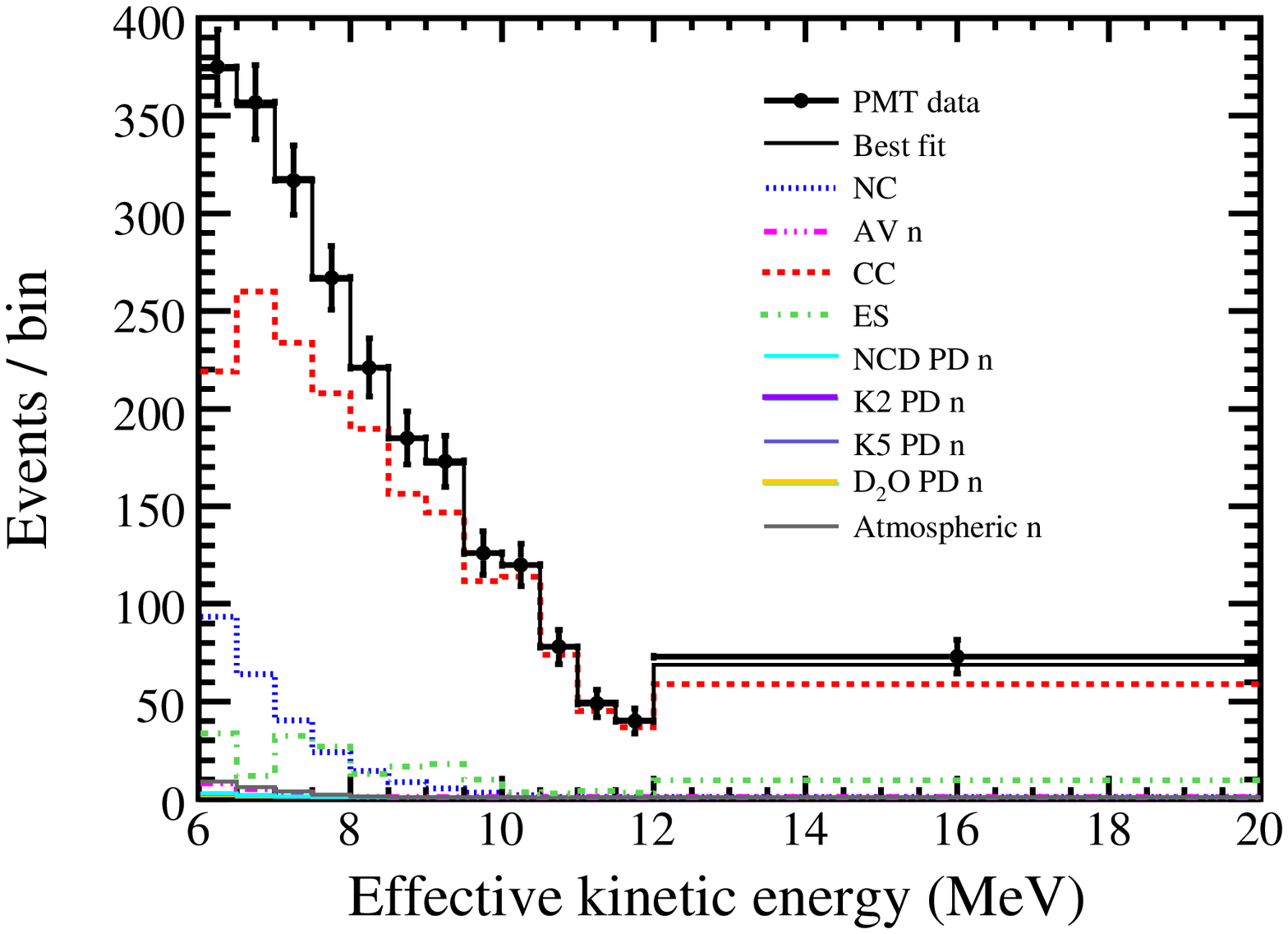}
\includegraphics[width=0.40\columnwidth]{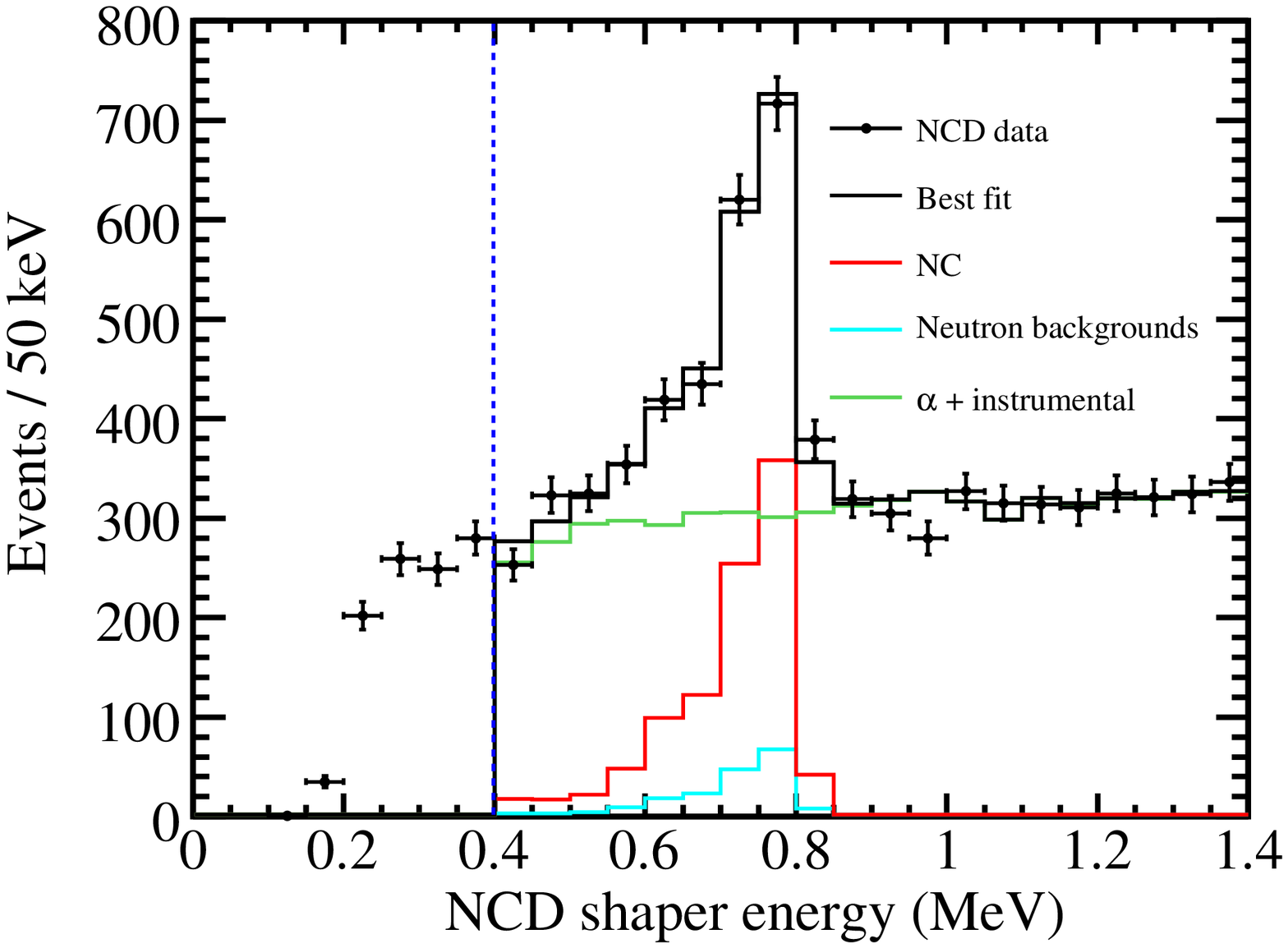}
\caption{\label{fig:mcalphafit} One-dimensional projections of PMT array and NCD array data overlaid with best-fit results to signals.  The $\chi^2$ values for the $\rho$,  $\cos{\theta_\odot}$,  $\teff$ and shaper energy distributions are 7.3, 11.9, 0.3 and 17.0, respectively.  Because these are one-dimensional projections in a multi-dimensional space, these $\chi^2$ values are quoted as a qualitative demonstration of goodness-of-fit and cannot be simply evaluated.
}
\end{figure}

Projections of the best-fit distribution with the data are shown in Fig.~\ref{fig:mcalphafit}.  The fitted systematic parameter values are provided in Table~\ref{table:mcalphasys}, while the constraints and fit results for the amplitude of different neutron and instrumental backgrounds are given in Table~\ref{table:ampfit}.  The uncertainties on the fitted values of the systematic parameters were for the most part the same as the width of the constraint that was used in the fit.  There are three systematic parameters that have fit uncertainties that are considerably narrower than the constraint. The shaper energy scale is narrower, and this appears to be a real effect of the neutron energy peak setting the energy scale.  The shaper energy resolution has fit out at $+1.2\%$, and is narrower only because of the combination of not being allowed to go negative, and having a constraint of $^{+1.0}_{-0.0}$\%.  The alpha Po depth have noticeably narrower fit uncertainties, most likely due to constraints from the spectral information of alpha backgrounds above about 0.9~MeV in the data.

\begingroup
\squeezetable
\begin{table}
\begin{center}
\caption{Systematic parameters' constraints and fit results for the full Phase-III data signal extraction.  Details of the parameterization of these nuisance parameters in the fit are described in Appendix~\ref{sec:apdxb}.  Those constraints marked with an asterisk were handled by the second term in Eqn.~\ref{eqn:loglpmtconstr}, with the covariance matrices given in the same appendix.}
\label{table:mcalphasys}
\begin{ruledtabular}
\begin{tabular}{ l l l l l }
                &                                            & \multicolumn{2}{c}{Gaussian constraint} & \\  
                       & Description                                &  Mean     & $\sigma$  &  Fit value \\             
\hline  
Systematic & \multicolumn{4}{c}{Nuisance parameters --- PMT array } \\ \hline
$f^{\textrm{PMT}}_{\textrm{NC}}$   & NC flux to PMT NC events factor            &  0.46725      &  0.00603	               &   0.46735 $\pm$ 0.00574     \\
$a_0^x$          & x coordinate shift                         &  0.0          &  4.0           	               &   1.0 $\pm$ 4.0     \\ 
$a_0^y$          & y coordinate shift                         &  0.0          &  4.0           	               &  -1.0 $\pm$ 3.9     \\
$a_0^z$          & z coordinate shift                         &  5.0          &  4.0           	               &   6.1 $\pm$ 3.6     \\
$a_1^x$          & coordinate scale                           &  0.000        &  0.006        	               &  -0.002 $\pm$ 0.008     \\
$b_0^{xy}$       & xy resolution constant term                &  0.06546      &  0.02860$^*$                   &   0.069 $\pm$ 0.029     \\
$b_1^{xy}$       & xy resolution linear term                  & -0.00005501   &  0.00006051$^*$                &  -0.000053 $\pm$ 0.000058    \\
$b_2^{xy}$       & xy resolution quadratic term               &  3.9$\times10^{-7}$ &  0.2$\times10^{-7}$ $^*$ &   0.00000038 $\pm$ 0.00000020 \\
$b_0^{z}$        & z resolution constant term                 &  0.07096      &  0.02805$^*$                   &   0.072 $\pm$ 0.027     \\
$b_1^{z}$        & z resolution linear term                   &  0.0001155    &  0.00008251$^*$                &   0.00012 $\pm$ 0.000082     \\ 
$b_0^{\theta}$\footnote{The 1$\sigma$ width of the constraint for $b_0^{\theta}$ was input incorrectly in the analysis in Ref.~\cite{ncdprl}; the correct value should be 0.12.  The fit value indicates that this error should not have any impact on the ES results.}  
	 & PMT angular resolution                     &  0.0          &  0.056 		               &   0.011 $\pm$ 0.059     \\
$a_1^\textrm{E}$          & PMT energy scale                           &  1.000        &  0.0109                	       &   1.0047 $\pm$ 0.0087     \\
$b_0^\textrm{E}$          & PMT energy resolution (neutrons)           &  0.0119       &  0.0114 	               &   0.0121 $\pm$ 0.0104     \\	
\hline  
Systematic & \multicolumn{4}{c}{Nuisance parameters - NCD array } \\ \hline											
$f^{\textrm{NCD}}_{\textrm{NC}}$   & NC flux to NCD NC events factor            &  1.7669       &  0.0590 	               &   1.7713 $\pm$ 0.0586     \\
$a_1^{\textrm{NCDE}}$     & NCD shaper energy scale                   &  1.00	      &  0.01                          &   1.0047 $\pm$ 0.0035    \\      
$b_0^{\textrm{NCDE}}$     & NCD shaper energy resolution              &  0.00         &  +0.01 -0.00                   &   0.0124 $\pm$ 0.0065     \\            
$\alpha_0$       & alpha PDF - alpha Po depth                   &  0            &  1		               &   1.21 $\pm$ 0.62     \\
$\alpha_1$       & alpha PDF - alpha bulk depth                   &  0            &  1		               &   0.25 $\pm$ 0.86     \\
$\alpha_2$       & alpha PDF - drift time                       &  0            &  1		               &  -0.11 $\pm$ 0.97     \\
$\alpha_3$       & alpha PDF - avalanche width                   &  0            &  1		               &   0.27 $\pm$ 0.94     \\
$\alpha_4$       & alpha PDF - avalanche gradient                &  0            &  1		               &  -0.06 $\pm$ 1.00     \\
$\alpha_5$       & alpha PDF - Po/bulk fraction                   &  0            &  1		               &   0.13 $\pm$ 0.97     \\
$\alpha_6$       & alpha PDF - ion mobility                     &  0            &  1                             &  -0.06 $\pm$ 0.96     \\
$\alpha_7$       & alpha PDF - data reduction cuts                    &  0            &  1                             &  -0.49 $\pm$ 0.93     \\
$p_1^{\textrm{J3}}$       & J3-type background skew Gaussian mean                 &  0.4584       &  0.0262                        &   0.4724 $\pm$ 0.0231  \\      
$p_1^{\textrm{N4}}$       & N4-type background skew Gaussian mean                 &  0.0257       &  0.0138                        &   0.0333 $\pm$ 0.0112   \\     

\end{tabular}
\end{ruledtabular}
\end{center}
\end{table}
\endgroup

\begingroup
\squeezetable
\begin{table}
\begin{center}
\caption{Constraints and fit results for the amplitude of different neutron and instrumental background classes in the full Phase-III data signal extraction.}
\label{table:ampfit}
\begin{ruledtabular}
\begin{tabular}{ l l l l l }
                &                                            & \multicolumn{2}{c}{Gaussian constraint} & \\  
                       & Description                                &  Mean     & $\sigma$  &  Fit value \\             
\hline  
Background & & & & \\ \hline
$N_{\textrm{ex}}^{\textrm{NCD}}$   & external n (AV, \hto\ backgrounds)                           & 40.9          & 20.6                           & 42.2 $\pm$ 19.3 \\ 
$N_{\textrm{ncdpd}}^{\textrm{NCD}}$& NCD bulk, cable                          & 35.6          & 12.2                           & 35.2 $\pm$ 12.1 \\
$N_{\textrm{k2pd}}^{\textrm{NCD}}$ & K2                                         & 32.8          &  5.2                           & 32.7 $\pm$ 5.1 \\ 
$N_{\textrm{k5pd}}^{\textrm{NCD}}$ & K5                                         & 31.6          &  3.7                           & 31.7 $\pm$ 3.7 \\
$N_{\textrm{d2opd}}^{\textrm{NCD}}$& \dto\ photodisintegration                          & 31.0          &  4.8                           & 30.9 $\pm$ 4.8 \\
$N_{\textrm{atmos}}^{\textrm{NCD}}$& atmospheric $\nu$ and cosmogenic muons                              &  13.6         &  2.7                           & 13.6 $\pm$ 2.7 \\
$N_{\textrm{J3}}^{\textrm{NCD}}$   & J3-type instrumental background                   &\multicolumn{2}{l}{unconstrained}             & 355.6 $\pm$ 192.3 \\
$N_{\textrm{N4}}^{\textrm{NCD}}$   & N4-type instrumental background                    &\multicolumn{2}{l}{unconstrained}             & 215.6 $\pm$ 170.5 \\
\end{tabular}
\end{ruledtabular}
\end{center}
\end{table}
\endgroup

To better understand the contributions of the systematic uncertainties to the total uncertainties, we took the quadratic difference between the fit uncertainty including systematic parameter variations and the fit uncertainty without such variations.  The equivalent neutrino fluxes with their statistical and systematic uncertainties are (in units of $10^6$~cm$^{-2}$~s$^{-1}$):
\begin{eqnarray}
\phi_{\text{CC}} & = & \snoccfluxunc \nonumber \\
\phi_{\text{ES}} & = & \snoesfluxunc \\ 
\phi_{\text{NC}} & = & \snoncfluxunc. \nonumber
\end{eqnarray} 
Table~\ref{syserror} is a summary of these system uncertainties, categorized by different sources.  

The MCMC fit results were checked against those from the other two independent methods described earlier in this section.  A comparison of the results from these three analysis methods revealed two issues.  A 10\% difference between the NC flux uncertainties was found, and subsequent investigation revealed incorrect input parameters in two methods.  After the inputs were corrected, the errors agreed, and there was no change in the fitted central values.  However, the ES flux determined from the MCMC method was 0.5$\sigma$ lower than those from the other two analyses.  This difference was found to be from the use of an inappropriate algorithm to provide a point estimation of the ES posterior distributions.  A better method of fitting the posterior distribution to a Gaussian with different widths on either side of the mode was implemented, and the ES flux results agreed with those from the other two analyses.

\begin{figure}
\begin{center}
\includegraphics[width=0.80\textwidth]{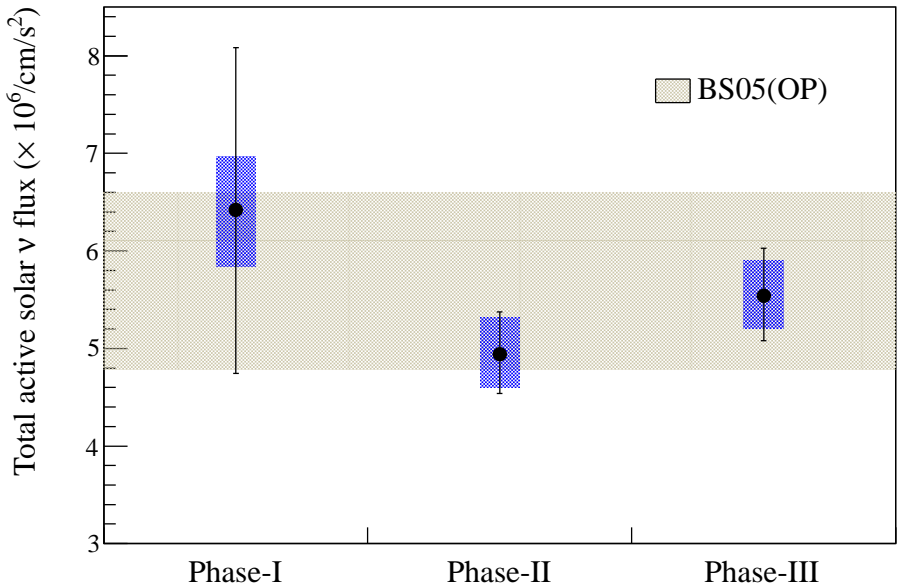}
\caption{\label{fig:nccompare}A comparison of the measured energy-unconstrained NC flux results in SNO's three phases.  The horizontal band is the 1$\sigma$ region of the expected total $^8$B solar neutrino flux in the BS05(OP) model~\cite{bs05}.}
\end{center}
\end{figure}

The ES flux presented here is 2.2$\sigma$ lower than that found by Super-Kamiokande-I~\cite{SK2}.  This is consistent with a downward statistical fluctuation in the ES signal, as evidenced in the shortfall of signals near $\cos\theta_{\odot}$ = 1 in two isolated energy bins.  The $^8$B spectral shape~\cite{wint} used here differs from that~\cite{ortiz} used in previous SNO results.  The CC, ES and NC flux results are in agreement (\textit{p} = 32.8\%~\cite{BLUE}) with the NC flux result of the first phase~\cite{snocc} and with the fluxes from the second phase~\cite{nsp}.  Table~\ref{ssmcmp} summarizes the CC, ES, and NC fluxes determined from energy-unconstrained fits in SNO's three phases, and Figure~\ref{fig:nccompare} shows a comparison of these NC measurements.

\begingroup
\begin{table}
\caption{\label{syserror}Sources of systematic uncertainties on CC, ES and NC flux measurements.  The total uncertainties differs from the individual uncertainties added in quadrature due to correlations.}
\begin{ruledtabular}
\begin{tabular}{llll}
Source       & CC uncert. & ES uncert. & NC uncert. \\ 
         & (\%) & (\%) & (\%) \\  \hline 
PMT energy scale   & $\pm2.7$ & $\pm3.6$  & $\pm0.6$ \\ 
PMT energy resolution    & $\pm0.1$ & $\pm0.3$ & $\pm0.1$ \\ 
PMT radial energy dependence   & $\pm0.9$ & $\pm0.9$ & $\pm0.0$  \\ 
PMT radial scaling    & $\pm2.7$ & $\pm2.7$ & $\pm0.1$ \\ 
PMT angular resolution    & $\pm0.2$ & $\pm2.2$ & $\pm0.0$ \\ 
Background neutrons    & $\pm0.6$ & $\pm0.7$ & $\pm2.3$ \\ 
Neutron capture    & $\pm0.4$ & $\pm0.5$ & $\pm3.3$ \\ 
Cherenkov/AV backgrounds    & $\pm0.3$ & $\pm0.3$ & $\pm0.0$ \\ 
NCD instrumentals & $\pm0.2$ & $\pm0.2$ & $\pm1.6$  \\
NCD energy scale  & $\pm0.1$ & $\pm0.1$ & $\pm0.5$ \\
NCD energy resolution & $\pm0.3$ & $\pm0.3$ & $\pm2.7$  \\
NCD alpha systematics  & $\pm0.3$ & $\pm0.4$ & $\pm2.7$ \\
PMT data reduction cuts  & $\pm0.3$ & $\pm0.3$ & $\pm0.0$ \\ \hline
Total experimental uncertainty & $\pm4.0$ & $\pm4.9$ & $\pm6.5$  \\ \hline 
Cross section~\cite{crosssection}  & $\pm 1.2$& $\pm 0.5$  & $\pm 1.1$
\end{tabular}
\end{ruledtabular}
\end{table}
\endgroup

\begingroup
\begin{table}
\caption{\label{ssmcmp}Energy-unconstrained CC, ES and NC flux results ( in units of $10^6$~cm$^{-2}$~s$^{-1}$) from three phases of SNO.  The \teff\ thresholds for the PMT array data in Phases I, II and III were 5.0, 5.5 and 6.0~MeV, respectively.  ``Energy-constrained'' flux results, in which ES and CC events were constrained to an undistorted $^{8}$B spectrum in the fit, can be found in Refs.~\cite{longd2o} and~\cite{nsp} for Phases I and II respectively. }
\begin{ruledtabular}
\begin{tabular}{lccc}
Data set & $\phi_{\text{CC}}$  & $\phi_{\text{ES}}$ & $\phi_{\text{NC}}$ \\ \hline
Phase I (306 live days)  &   ---   &  ---  &   6.42$^{+1.57}_{-1.57}$ $^{+0.55}_{-0.58}$ \\ 
Phase II (391 live days) &  1.68$^{+0.06}_{-0.06}$ $^{+0.08}_{-0.09}$  & 2.35$^{+0.22}_{-0.22}$ $^{+0.15}_{-0.15}$ & 4.94$^{+0.21}_{-0.21}$ $^{+0.38}_{-0.34}$ \\
Phase III (385 live days) &  1.67$^{+0.05}_{-0.04}$ $^{+0.07}_{-0.08}$  & 1.77$^{+0.24}_{-0.21}$ $^{+0.09}_{-0.10}$ & 5.54$^{+0.33}_{-0.31}$ $^{+0.36}_{-0.34}$ \\
\end{tabular}
\end{ruledtabular}
\end{table}
\endgroup

\section{Neutrino mixing model interpretation of results\label{sec:physint}}

The SNO measurements of the NC and CC fluxes for neutrinos originated from $^{8}$B decays inside the Sun unambiguously proved that neutrinos change their flavor while traveling to the Earth. These results can be interpreted, as in previous SNO analyses~\cite{snodn,snonc,nsp}, as neutrino flavor transitions due to the mixing of the massive neutrino states via the MSW effect~\cite{wolfenstein,ms}. Neutrino mixing parameters can be extracted by comparing the experimental data from SNO and other experiments, with the model predictions from the neutrino mixing hypothesis.  The neutrino-mixing analyses presented in the following do not include solar neutrino results from a recent low-threshold analysis that combined the Phase-I and Phase-II data sets~\cite{leta}.  

Full three-neutrino analyses are being carried out for all phases of SNO taken together and will be submitted for publication. ÊFor the purposes of the present work, a two-neutrino analysis is convenient because the mixing of the third mass eigenstate into $\nu_{e}$ is small~\cite{chooz} and $\Delta m^{2}_{\textrm{sol}}\ll\Delta m^{2}_{\textrm{atm}}$~\cite{minos}.  The two neutrino mixing parameters in this model are: the squared-mass difference of the neutrino mass eigenstates, $\Delta m^{2}\equiv \Delta m^{2}_{21}$,  and the mixing angle between the appropriate mass eigenstates, $\theta \equiv \theta_{12}$. The mixing angle is also given as $\tan^{2}\theta$ in order to compare its extracted value with our previous results, as well as with the results reported by others. The mixing model was used to propagate neutrinos inside the Sun, vacuum and the Earth, for each value of 
$\Delta m^{2}$ and $\tan^{2}\theta$. The model predictions for each experiment used in the global solar analysis were computed with the neutrino fluxes from the BS05(OP) solar model~\cite{bs05}, which is in good agreement with the helioseismological data, and the latest $^{8}$B spectrum shape with its associated uncertainties from Winter \textit{et al.}~\cite{wint}. 

The default approach in our analyses was the covariance $\chi^{2}$ method. The $\chi^{2}$ function was minimized at each point in the $\tan^{2}\theta-\Delta m^{2}$ plane with respect to the $^{8}$B neutrino flux. The least-square fit and the projection in the  $\tan^{2}\theta-\Delta m^{2}$ plane were then performed by allowing any values for the $^{8}$B neutrino flux for a given value of $\tan^{2}\theta$ and $\Delta m^{2}$. At the minimum value, $\chi^{2}_{\textrm{min}}$, the best-fit values for the mixing parameters $\tan^{2}\theta$ and $\Delta m^{2}$ were extracted, together with the corresponding value for the $^{8}$B flux. Then, the 68\%, 95\% and 99.78\% confidence level (CL) regions in the two-dimensional parameter space $\tan^{2}\theta-\Delta m^{2}$ were drawn. The uncertainties on the mixing parameters were determined by projecting the $\chi^{2}$ function passing through the best-fit point on the $\tan^{2}\theta$ and $\Delta m^{2}$ axes, separately. The one-dimensional (1D) projections were not a simple slice of the two-dimensional contour passing through $\chi^{2}_{\textrm{min}}$, but instead a projection in which $\Delta \chi^{2}=\chi^{2}-\chi^{2}_{\textrm{min}}$ was computed for each 1D axis allowing the other  parameter to take any values. From these 1D projections the uncertainties on each parameter, the 1$\sigma$ spreads were determined from the values at  $\chi^{2}_{\textrm{min}}+1$. 

The solar neutrino data used in this analyses were: SNO Phase-I (SNO-I ) summed kinetic energy spectra (CC+ES+NC+backgrounds) for day and night~\cite{snodn},   SNO Phase-II (SNO-II) CC kinetic energy spectra, ES and NC fluxes for day and night~\cite{nsp}, SNO Phase-III (SNO-III) CC, ES and NC fluxes~\cite{ncdprl}, Super-Kamiokande zenith binned energy spectra~\cite{SK2}; and the rate measurements from the Homestake~\cite{cl}, Gallex/GNO~\cite{GNO2}, SAGE~\cite{sage2} and Borexino experiments~\cite{borex}. The global solar $\chi^{2}$ function obtained by comparing these data with the corresponding model predictions were then combined with the 2881-ton-year KamLAND reactor anti-neutrino results~\cite{kam2}, assuming CPT invariance. While the results from SNO are highly sensitive to the mixing angle through the measured ratio $\phi_{\textrm{CC}} / \phi_{\textrm{NC}}$, the KamLAND measurement has a higher sensitivity to the allowed values of the $\Delta m^{2}$ parameter.

First, the constraint on the neutrino mixing parameters was placed by interpreting the measurements from the results of the three phases of SNO only.  Detailed descriptions on the use of data sets from SNO-I and SNO-II to interpret neutrino mixing can be found in Refs.~\cite{sno1url,nsp}. SNO-III data were obtained from signal extraction (Sec.~\ref{sec:sigex}) as integrated CC, ES and NC fluxes (averaged over day and night), which are tabulated in Table~\ref{table:mcalphanocor}.  The statistical correlation coefficients between the three fluxes, which were needed in building the global $\chi^2$, are tabulated in Table~\ref{table:fluxcor}.  

For the SNO-III data sample, the $\chi^{2}$ function is defined as~\cite{tesic}:
\begin{equation}
\chi^2_{} = \sum_{i,j=1}^{3} (Y_{i}^{\rm exp} - Y_{i}^{\rm th})^{T}
       [\sigma_{ij}^2(\mbox{tot})]^{-1} (Y_{j}^{\rm exp} - Y_{j}^{\rm th}), 
\label{chi2def}
\end{equation}
where $Y^{\textrm{exp}}_{i}$ is the CC, ES or NC averaged flux measurement, and  $Y^{\textrm{th}}_{i}$ is the theoretical expectation obtained from the two-neutrino mixing model. The model prediction $Y^{\textrm{th}}_{i}$ was calculated under  the assumption of the two-neutrino oscillation hypothesis, thus it depended on the number of free parameters $n$ in the fit. In the physics interpretation presented here and in Ref.~\cite{ncdprl}, the free parameters were the neutrino mixing parameters ($\Delta m^{2}$ and $\tan^{2}\theta$) and the total flux of the $^{8}$B neutrinos $\phi_{^{8}\textrm{B}}$. The shape of the $^{8}$B spectrum was fully constrained by the mixing parameters.

The covariance error matrix $\sigma^2_{ij}(\mbox{tot})$ was built as a sum of the squares of the statistical   $\sigma^2_{ij}(\mbox{exp})$ and systematic $\sigma^2_{ij}(\mbox{syst})$ uncertainties:
\begin{equation}
\sigma^2_{ij}(\mbox{tot}) = \sigma^2_{ij}(\mbox{stat}) + 
                            \sigma^2_{ij}(\mbox{syst}), 
                            \label{sigmatot}
\end{equation}
where the statistical covariance matrix is given by:
\begin{equation}
\sigma^2_{ij}(\mbox{stat}) = \rho_{ij}u_{i}u_{j}, 
\label{sigmastat}
\end{equation}
where $u_{i}$ and $\rho_{ij}$ are the statistical uncertainty for the measurement $Y^{\textrm{exp}}_{i}$  and the statistical correlation coefficient between the observables $Y^{\textrm{exp}}_{i}$ and $Y^{\textrm{exp}}_{j}$, respectively. 

The effect of each systematic uncertainty $S_k$ on the model expectation for the neutrino yield $Y^{\textrm{th}}_{i}$ was estimated by computing the change in the expectation $\Delta Y_{ik}$ with respect to the source of the uncertainty:
\begin{equation}
\gamma_{ik}=\frac{\Delta Y_{ik}}{Y_{i}}.
\label{betas}
\end{equation}
The relative uncertainties $\gamma_{ik}$ were then used to construct the systematic error matrix $
\sigma_{ij}^2\mbox{(syst)}$, which is defined as:
\begin{equation}
\sigma_{ij}^2\mbox{(syst)} 
      = Y^{\textrm{th}}_{i}Y^{\textrm{th}}_{j}\sum_{k=1}^{K}
       r^{k}_{ij} \gamma_{ik}\gamma_{jk},
               \label{systmatrix}
\end{equation}
where $K$ is the number of systematic uncertainties affecting the observables $i$ and $j$.   A coefficient $r^{k}_{ij}$ describes a correlation between the observables $i$ and $j$ induced by the systematic uncertainty $k$, within a single phase of a given experiment.  Values of the correlation coefficients $r^{k}_{ij}$ are summarized in Table~\ref{systcorr2} for the SNO-III data sample. In our analyses, the correlations among the systematic uncertainties between the three phases of SNO was also accounted for. However, these correlations had little impact on the allowed regions in the $\tan^{2}\theta-\Delta m^{2}$ plane.  The relative errors for the most important energy related systematic uncertainties, such as PMT energy scale and resolution, and also for the $^{8}$B spectrum shape uncertainty, were computed for each value of the mixing parameters as $\gamma_{ik}=\Delta Y^{\textrm{th}}_{ik}/Y^{\textrm{th}}_{i}$. 
   
\begin{table}[htdp]
\caption{Correlation coefficients for the same sources of systematic uncertainties affecting different types of signals for the SNO-III data set.}
\begin{center}
\begin{minipage}[t]{5in}
\begin{ruledtabular}
\begin{tabular}{lccc}
Source of systematic & NC-CC &NC-ES & CC-ES \\
\hline
 PMT energy scale & +1&+1 & +1\\
 PMT energy resolution & +1&+1&+1\\
 PMT radial energy dependence  & +1 &+1&+1\\
 PMT vertex resolution  &+1&+1&+1\\
 PMT vertex accuracy &+1 &+1&+1\\
 PMT angular resolution  &+1 &-1&-1\\
 Background neutrons &+1&+1&+1 \\
 Neutron capture  &+1&+1&+1\\
 Cherenkov/AV backgrounds  &+1 &+1 &+1\\
 NCD instrumentals &+1 &+1&+1 \\
 NCD energy scale & +1&+1&+1\\
 NCD energy resolution & +1&+1&+1\\
 NCD alpha systematics & +1 &+1&+1\\
 PMT data reduction cuts  & 0 & 0&+1\\
 \end{tabular}
 \end{ruledtabular}
 \end{minipage}
\end{center}
\label{systcorr2}
\end{table}%

\begin{figure}[htbp]
\begin{center}
\unitlength1cm
 \includegraphics[width=0.8\textwidth]{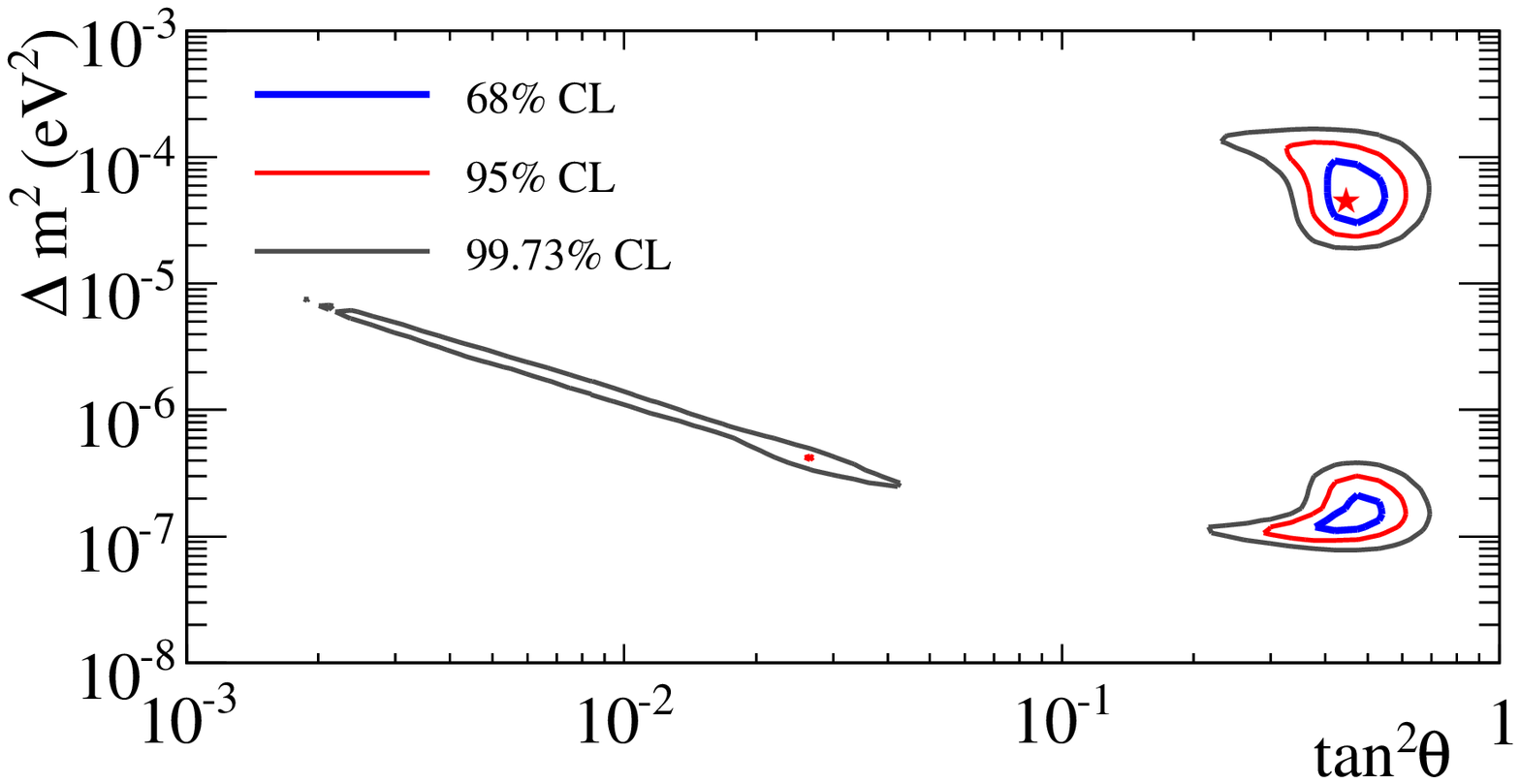}
 \caption[After SNO-III: SNO-only neutrino oscillation confidence level contours.]{After SNO-III: SNO-only neutrino oscillation confidence level contours published in Reference~\cite{ncdprl}. This analysis includes the summed kinetic energy spectra from Phase I (day and night); NC and ES fluxes, and CC kinetic energy spectra from Phase II (day and night); and CC, NC and ES fluxes from Phase III. The best-fit point is at: $ \Delta m^{2}=4.57 \times 10^{-5}$ eV$^{2}$, $\tan ^{2}\theta = 0.447$, $\phi_{^{8}\textrm{B}}=5.12 \times 10^{6}$ cm$^{-2}$s$^{-1}$. The $hep$ neutrino flux was fixed at $7.93 \times 10^{3}$ cm$^{-2}$s$^{-1}$.}
        \label{fig:snomsw2}
\end{center}
\end{figure}

With the inclusion of SNO Phase-III results, the following best-fit neutrino mixing parameters were found for the SNO-only analyses: $\Delta m^{2}=4.57^{+2.30}_{-1.22} \times 10^{-5}$ eV$^{2}$ and  $\tan ^{2}\theta = 0.447^{+0.045}_{-0.048}$.  The flux of the $^{8}$B neutrinos was floated with respect to the BS05(OP) model prediction of $5.69 \times 10^{6}$ cm$^{-2}$s$^{-1}$, and the best-fit value was $\phi_{^{8}\textrm{B}}=5.12 \times 10^{6}$ cm$^{-2}$s$^{-1}$. The flux of $hep$ neutrinos was fixed at the BP05(OP) model value of $7.93 \times 10^{3}$ cm$^{-2}$s$^{-1}$. The minimum $\chi ^{2}$ at the best-fit point was 73.77 for 72 degrees of freedom. The allowed regions at 68\%, 95\% and 99.73\% confidence level (CL) in the $\Delta m^{2}-\tan ^{2}\theta$ plane from this fit, shown in Figure \ref{fig:snomsw2}, were significant improvements compared to the Phase-II result~\cite{nsp}. The vacuum (``VAC'') oscillation  region was ruled out at the 99.73\% CL for the first time using SNO data only. The remaining regions in the oscillation plane are significantly smaller than those presented in Ref.~\cite{nsp}, with reduced marginalized 1$\sigma$ uncertainties. The two best-fit results are given in Table \ref{snomswtab}, with a comparison of the effects of including the SNO-III data sample in the SNO-only oscillation analysis.

\begin{table}[htdp]
\caption{SNO-only neutrino oscillation best-fit parameters.}
\begin{center}
\renewcommand{\arraystretch}{1.3}
\begin{minipage}[t]{5in}
\begin{ruledtabular}
\begin{tabular}{lccc}
Analysis & $\Delta m^{2}$ ($10^{-5}$ eV$^{2}$) & $\tan ^{2} \theta$  & $^{8}$B flux ($10^{6}$ cm$^{-2}$s$^{-1}$)\\
\hline
Before SNO-III & $5.0^{+6.2}_{-1.8}$ & $0.45^{+0.11}_{-0.10}$ & 5.11\\
After SNO-III & $4.57^{+2.30}_{-1.22}$ & $0.45^{+0.05}_{-0.05}$ & 5.12\\
\end{tabular}
\end{ruledtabular}
\end{minipage}
\end{center}
\label{snomswtab}
\end{table}%

\begin{figure}[htbp]
\begin{center}
\unitlength1cm
 \includegraphics[width=0.8\textwidth]{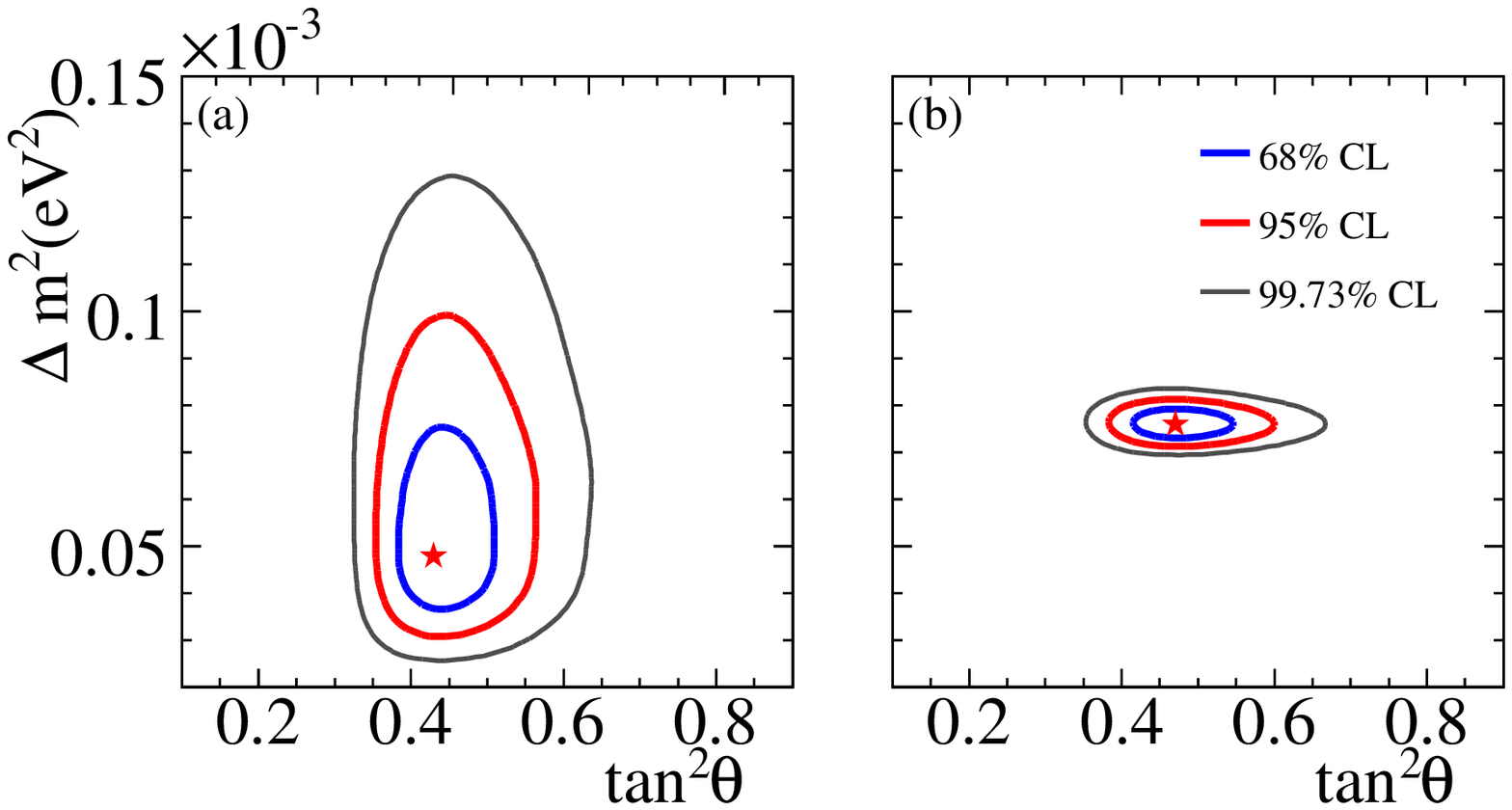}
 \caption[After SNO-III: Global solar and global solar+KamLAND neutrino oscillation confidence level contours.]{After SNO-III: Neutrino oscillation parameters confidence level contours. (a) Global solar analysis including the rate measurements from Homestake, Gallex/GNO, SAGE and Borexino; SK-I zenith-energy spectra from~\cite{SK2}, summed kinetic energy spectra from SNO-I (day and night); NC and ES fluxes, and CC kinetic energy spectra from SNO-II (day and night); and CC, ES and NC fluxes from SNO-III. The best-fit point was at: $ \Delta m^{2}=4.90 \times 10^{-5}$ eV$^{2}$, $\tan ^{2}\theta = 0.437$, $\phi_{^{8}\textrm{B}}=5.21 \times 10^{6}$ cm$^{-2}$s$^{-1}$. The $hep$ neutrino flux was fixed at $7.93 \times 10^{3}$ cm$^{-2}$s$^{-1}$.  (b) Including KamLAND data from~\cite{kam2}, the best-fit point was at: $\Delta m^{2}=7.59 \times 10^{-5}$ eV$^{2}$, $\tan ^{2}\theta = 0.468$, $\phi_{^{8}\textrm{B}}=4.92 \times 10^{6}$ cm$^{-2}$s$^{-1}$.}
        \label{fig:globmsw2}
\end{center}
\end{figure}

In our first report of Phase-III results~\cite{ncdprl}, the following changes in the global solar neutrino analysis were made with respect to our Phase-II analysis in Ref.~\cite{nsp}: the model predictions for all solar neutrino experiments were computed using the BS05(OP) model and the $^{8}$B neutrino spectrum shape from Ref.~\cite{wint}, the inclusion of the 192 live-day results from Borexino~\cite{borex}, update of data from SK-1 using results from Ref.~\cite{SK2}, and, most importantly, the new measurements from the third phase of SNO were incorporated. The global fit to these solar neutrino data led to the following neutrino mixing parameters: $\Delta m^{2}=4.90^{+1.64}_{-0.93} \times 10^{-5}$ eV$^{2}$ and  $\theta = 33.5^{+1.3}_{-1.3}$, with $^{8}$B flux of $\phi_{^{8}\textrm{B}}=5.21 \times 10^{6}$ cm$^{-2}$s$^{-1}$. The minimum $\chi ^{2}$ at the best-fit point was 130.29 for 120 degrees of freedom. The allowed regions from this analysis are shown on the left panel in Fig.~\ref{fig:globmsw2}.  The constraint on both neutrino mixing parameters was much better than our results in Phase II~\cite{nsp}.  When the 2881-ton-year KamLAND results were included in this analysis~\cite{kam2}, the best-fit parameters became: $\Delta m^{2}=7.59^{+0.21}_{-0.19} \times 10^{-5}$ eV$^{2}$ and  $\theta = 34.4^{+1.3}_{-1.2}$ degrees, and the $^{8}$B flux of $\phi_{^{8}\textrm{B}}=4.92 \times 10^{6}$ cm$^{-2}$s$^{-1}$.  The improvement in comparison with the former analysis was observed in the allowed regions from the combined fit shown in the right panel in Fig.~\ref{fig:globmsw2}.  A summary of these results is given in Table \ref{globmswtab}. 
\begin{table}[htdp]
\caption[Global solar-only and global solar+KamLAND best-fit parameters.]{Global solar only and global solar+KamLAND best-fit parameters. The global solar results before the SNO-III phase do not include data from Borexino. The global solar+KamLAND results after SNO-III include the latest data from Borexino~\cite{borex} and KamLAND~\cite{kam2}. }
\begin{center}
\renewcommand{\arraystretch}{1.3}
\begin{minipage}[t]{6in}
\begin{ruledtabular}
\begin{tabular}{lccc}
Analysis & $\Delta m^{2}$ ($10^{-5}$ eV$^{2}$) & $\tan^{2}\theta$  & $^{8}$B flux ($10^{6}$ cm$^{-2}$s$^{-1}$)\\
\hline
\multicolumn{4}{c} {Before SNO-III phase} \\
\hline
Global solar& $6.5^{+4.4}_{-2.3}$ & $0.45^{+0.09}_{-0.08}$ & 5.06\\
with KamLAND& $8.0^{+0.6}_{-0.4}$ & $0.45^{+0.09}_{-0.07}$ & 4.93\\
\hline
\multicolumn{4}{c} {After SNO-III phase} \\
\hline
Global solar& $4.90^{+1.64}_{-0.93}$ & $0.44^{+0.05}_{-0.04}$ & 5.21\\
with KamLAND& $7.59^{+0.21}_{-0.19}$ & $0.47^{+0.05}_{-0.04}$ & 4.92\\
\end{tabular}
\end{ruledtabular}
\end{minipage}
\end{center}
\label{globmswtab}
\end{table}%

In comparison to Phase-II results, this combined fit of the solar neutrino data and the 2881-ton-year results from KamLAND improved the constraints on the neutrino mixing parameters: mixing angle $\theta$ and $\Delta m^{2}$ by 45\% and 60\%, respectively, at the time of the publication of Ref.~\cite{ncdprl}. This improvement on the mixing angle was dominated by the SNO experiment and the Phase-III results. The fitted values for the $^{8}$B neutrino flux were in agreement with recent predictions from solar models.

\section{Summary\label{sec:conclusion}}
We have presented a detailed description of the SNO Phase-III results that were published in Ref.~\cite{ncdprl}.  Neutrons from the NC reaction were detected predominantly by the NCD array.  The use of this technique, which was independent of the neutron detection methods in previous phases, resulted in reduced correlations between the CC, ES and NC fluxes and an improvement in the mixing angle uncertainty.  

Several techniques to reliably calibrate the PMT and NCD arrays were developed and are detailed in this paper.  The presence of the NCD array changed the optical properties, and hence the energy response, of the PMT array.  Extensive studies and evaluation of the techniques used in calibrating the PMT array in previous phases were performed and reported.   Radioactive backgrounds associated with the NCD array, the \dto\ target and other detector components were precisely quantified by \textit{in situ} and \textit{ex situ} measurements.  These measurements provided the constraints for the respective nuisance parameters in the determination of the $\nu_e$ and the total active neutrino fluxes.

The total flux of active neutrinos was measured to be $\snoncfluxunc \times 10^{6}$~cm$^{-2}$~s$^{-1}$, and was consistent with previous measurements and standard solar models.  A global analysis of neutrino mixing parameters using solar and reactor neutrino results yielded the best-fit values of the neutrino mixing parameters of $\Delta m^2 = \snodmsquared$ and $\theta= \snothetaonetwo$.  

A detailed paper that describes an analysis of data combined from all three phases of SNO is in preparation.


\begin{acknowledgments}
This research was supported by: Canada: Natural Sciences and Engineering Research Council, Industry Canada, National Research Council, Northern Ontario Heritage Fund, Atomic Energy of Canada, Ltd., Ontario Power Generation, High Performance Computing Virtual Laboratory, Canada Foundation for Innovation, Canada Research Chairs program; US: Department of Energy, National Energy Research Scientific Computing Center, Alfred P. Sloan Foundation; UK: Science and Technology Facilities Council (formerly Particle Physics and Astronomy Research Council); Portugal: Funda\c{c}\~{a}o para a Ci\^{e}ncia e a Tecnologia.  We thank the SNO technical staff for their strong contributions and David Sinclair for careful review of the neutron efficiency analysis.  We thank INCO (now Vale, Ltd.) for hosting this project in their Creighton mine.
\end{acknowledgments}

\appendix
\section{Instrumental background cuts for NCD array data\label{sec:apdxa}}

Two independent sets of cuts were developed to remove instrumental backgrounds in the NCD array data.  These cuts exploited the differences in the characteristics between ionization and non-ionization events.   One of these two sets inspected the characteristics of the digitized waveforms in the time domain, while the other in the frequency domain.   Cuts in both sets were used in the selection of the candidate event data set in the solar neutrino analysis.  A summary of these cuts is provided in the following.

\begin{figure}[htbp]
\includegraphics[width=0.60\columnwidth]{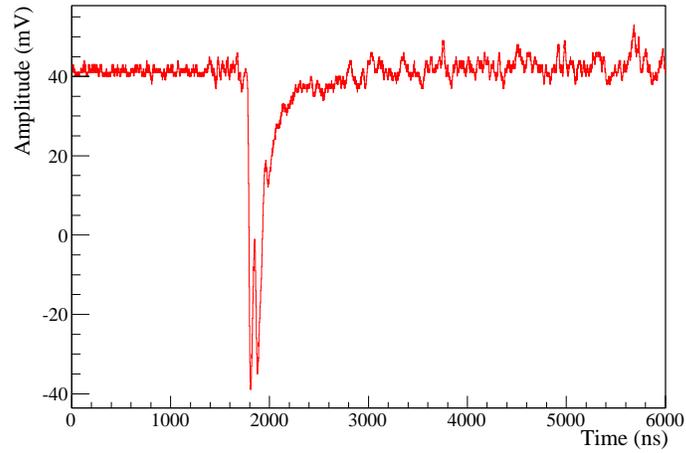}
\includegraphics[width=0.60\columnwidth]{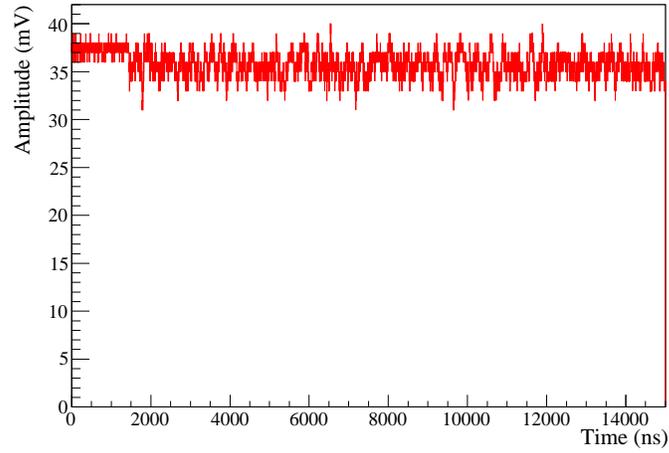}
\includegraphics[width=0.60\columnwidth]{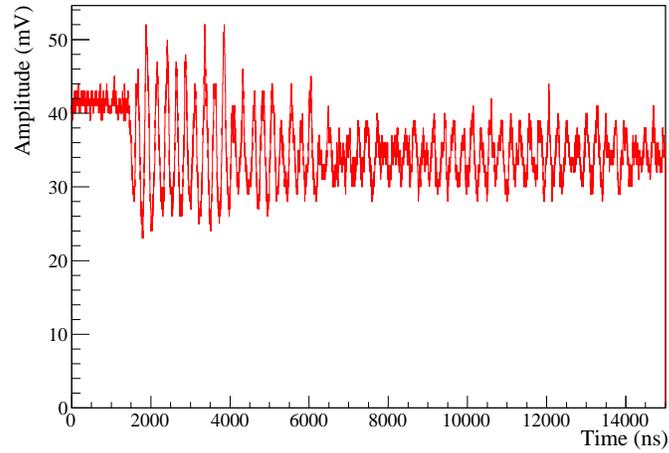}
\caption{\label{fig:badpulses} Examples of instrumental background pulses in the NCD array data.  Top:  a fork event, in which a small third reflection is seen in the tail of the pulse.  Middle: a flat trace.  Bottom: an oscillatory pulse. }
\end{figure}

\subsection{Bursts and overflow cuts}

Two burst cuts were developed to remove events that occurred within a very short time window.  If there were four or more shaper events within 100~ms, all events within this sequence were removed in the shaper burst cut.  Similarly, if there were four or more MUX-scope events within 100~ms, such events were removed in the MUX burst cut.  In the shaper-overflow cut, shaper events that arrived within a short time (15~$\mu$s to 5~ms) after a previous event had saturated the shaper were removed.

\subsection{Time correlation of shaper-ADC and MUX-scope events}

For each real ionization in the proportional counter, there was a time-correlated pair of shaper and MUX-scope events.  Instrumental background events exhibited a shorter time difference between the two events.  A time-correlation cut was developed to remove shaper-MUX-scope event pairs from the same string that showed such an anomalous timing characteristic.

\subsection{Fork event cuts}

The end of each counter string was attached to an open delay line.  Pulses were reflected at this open termination; thus, some physics pulses exhibited a double-peak structure.  There were also instrumental background pulses that exhibited similar double-peak characteristics, but with much different pulse width, time separation and amplitude ratio between the peaks.  The time-domain fork cut removed these ``fork'' events by exploiting these differences.  Some fork events also featured a third reflection at the tail of the waveform.  These  events were removed by a cut that was specifically designed to search for such reflection in the time domain.  The frequency-domain fork cut removed events with a peak around 12~MHz, which was the characteristic frequency of the fork events, in the power spectrum.  A fork event example is shown in the top panel of Fig.~\ref{fig:badpulses}.

\subsection{Flat, oscillatory and narrow pulse event cuts}

The flat-trace cut removed events that did not have a well-defined pulse profile.  These pulses were mostly noise that crossed the trigger threshold.   The dominant type of noise events during normal data taking was oscillatory pulses.  These events were identified and removed by a cut on the number of times the waveform crossed the baseline.  In the narrow pulse cut, pulses with widths that were too narrow to be ionization events were removed.  These pulses were mostly discharges and some of them carried a large amount of charge.  Spike events were also removed by identifying those with an abnormal ratio of their area and maximum amplitude.   In the frequency domain, waveforms with unusual symmetry had a non-characteristic zero-frequency intercept of the phase in their Fourier transform.  Waveforms with little power at low frequency were mostly flat or oscillation events. Waveforms with a large peak in the power spectrum at a frequency above 8 MHz (3.7 MHz) were mostly fork (oscillation) events.  Three cuts were implemented to reject events with such anomalous characteristics in the power spectrum.  Examples of flat and oscillatory trace events are shown in Fig.~\ref{fig:badpulses}.

\section{Parameterization of systematic uncertainties for the PMT and NCD array data\label{sec:apdxb}}

The handling of systematic uncertainties in the negative log-likelihood (NLL) for the PMT and NCD array data are described in Secs.~\ref{sec:pmtnllsys} and~\ref{sec:ncdnllsys} respectively.   In this Appendix, their parameterizations in the NLL are discussed.

\subsection{Parameterization of systematic for the PMT array data\label{sec:apdxbpmt}}

As discussed in Sec.~\ref{sec:pmtnllsys}, systematic effects were assessed by remapping observables in the simulated PMT array data.  These remapped variables were the effective kinetic energy, reconstructed vertex position and reconstructed direction of the event.

For the simulated values of the event kinetic energy ($T_g$), volume-weighted radius $\rho_g$, and angle relative to the vector from the Sun $\theta_g$, the remapping functions are:
\begin{equation}
T_{\textrm{remap}} = a_1^{\textrm{E}} T_0 + b_0^{E}(T_0-T_g),
\label{eqn:tremap}
\end{equation}
\begin{equation}
x_{\textrm{remap}} = a_0^{x} + (1+a_1^{x})x + (b_0^{xy}+b_1^{xy}z+b_2^{xy}z^2)(x-x_g),
\label{eqn:xremap}
\end{equation}
\begin{equation}
y_{\textrm{remap}} = a_0^{y} + (1+a_1^{x})y + (b_0^{xy}+b_1^{xy}z+b_2^{xy}z^2)(y-y_g),
\label{eqn:yremap}
\end{equation}
\begin{equation}
z_{\textrm{remap}} = a_0^{z} + (1+a_1^{x})z + (b_0^{z}+b_1^{z}z)(z-z_g),
\label{eqn:zremap}
\end{equation}
\begin{equation}
\rho_{\textrm{remap}} = (\sqrt{ x^2_{\textrm{remap}} + y^2_{\textrm{remap}}+z^2_{\textrm{remap}} }/600\textrm{cm})^3 ,\ \textrm{ and}
\label{eqn:rremap}
\end{equation}
\begin{equation}
\cos{\theta_{\textrm{remap}}} = 1 + (1+ b_0^{\theta})(\cos{\theta}-1).
\label{eqn:thremap}
\end{equation}
where $T_0$, ($x$, $y$, $z$), and $\theta$ are the nominal fitted effective kinetic energy, position and angle of the simulated event.  The remapping of $\cos\theta$ was applied to the ES channel only.  In these expressions, the nuisance parameters were modeled as Gaussian distributions with means and widths, which are given in Table~\ref{table:mcalphasys}.  The nuisance parameters $b_i^{xy}$ ($i={0,1,2}$) and $b_i^{z}$ ($i={0,1}$) are correlated, and their covariance matrices are:
\begin{equation}
V_{b^{xy}} = 
\left( \begin{array}{ccc}
 0.000818124  & -2.24984\times10^{-7} & -4.19131\times10^{-9} \\
 -2.24984\times10^{-7} & 3.66098\times10^{-9}  & 3.71423\times10^{-12} \\
 -4.19131\times10^{-9} & 3.71423\times10^{-12}  & 3.92118\times10^{-14} \end{array} \right),\ \textrm{ and}
\end{equation}
\begin{equation}
V_{b^{z}} = 
\left( \begin{array}{cc}
 0.00078696  & 3.47188\times10^{-7} \\ 
 3.47188\times10^{-7} & 6.80761\times10^{-9} \end{array} \right).
\end{equation}

\subsection{Parameterization of systematic for the NCD array data}

There were ten systematic uncertainties associated with the shaper energy PDFs.   For the uncertainties associated with the shaper-ADC energy $E_{\textrm{NCD}}$ and resolution, the energy PDFs were rebuilt using Eqn.~\ref{eqn:eadcremap}.  

For the alpha-background-related PDFs,  the simulated alpha background PDF was rebuilt by multiplying the unmodified PDF  ($Q^{\alpha}$)  by a re-weighting factor ($\alpha_i$) and a multiplicative function in shaper energy $s_i(E_{\textrm{NCD}})$:
\begin{equation}
Q_{\textrm{MC}}^{\alpha}(E_{\textrm{NCD}}) = Q^{\alpha}(E_{\textrm{NCD}}) (1+\sum_{i=0}^{7}\alpha_i\,s_i(E_{\textrm{NCD}})).
\end{equation}

The reweighting functions, $s_i$, are:
\begin{eqnarray}
{\rm Po~alpha~depth~variation} &s_0&= -2.06+6.58\,E_{\textrm{NCD}}-6.56\,E_{\textrm{NCD}}^2 \nonumber \\
& & \hspace{4cm}+2.11\,E_{\textrm{NCD}}^3 \\
{\rm Bulk~alpha~depth~variation} &s_1& = -0.0684+0.0892\,E_{\textrm{NCD}} \\
{\rm Drift~time~variation} &s_2&=-0.131+0.252\,E_{\textrm{NCD}}-0.117\,E_{\textrm{NCD}}^2 \\
{\rm Avalanche~width~offset~variation} &s_3&=-0.0541+0.0536\,E_{\textrm{NCD}} \\
{\rm Avalanche~gradient~offset~variation} &s_4&=-0.0138 \\
{\rm Ion~mobility~variation} &s_5&=-0.00930 \\
{\rm Po/bulk~fraction~variation} &s_6&= -0.00405+0.0386\,E_{\textrm{NCD}} \\
{\rm Data~reduction~cut~systematic} &s_7&=0.861-2.77\,E_{\textrm{NCD}}+2.72\,E_{\textrm{NCD}}^2 \nonumber \\
& & \hspace{4cm}-0.870\,E_{\textrm{NCD}}^3 .
\end{eqnarray}

In addition two instrumental background PDFs, based on events on strings J3 and N4, were parameterized as:
\begin{equation}
Q_{\textrm{MC}}^{\textrm{J3}} = \exp \left[  -\frac{1}{2} \left(  \frac{E_{\textrm{NCD}}-p_1^{\textrm{J3}}}{0.34\,p_1^{\textrm{J3}}} \right)^2 \right]
           \left[    1+{\rm erf} \left(-2.0\,(E_{\textrm{NCD}}-p_1^{\textrm{J3}} \right)  \right], \text{ and}
\end{equation}
\begin{equation}
Q_{\textrm{MC}}^{\textrm{N4}} = \exp \left[  -\frac{1}{2} \left(  \frac{E_{\textrm{NCD}}-p_1^{\textrm{N4}}}{19.6\,p_1^{\textrm{N4}}} \right)^2 \right]
           \left[    1+{\rm erf} \left(-1.59\,(E_{\textrm{NCD}}-p_1^{\textrm{N4}} \right)  \right].
\end{equation}
In both cases the number of instrumental background events was allowed to float freely in the fit.  

The numerical values of the constraints $\alpha_i$ and the instrumental background parameters ($p_1^{\textrm{J3}}$ and $p_1^{\textrm{N4}}$) are tabulated in Table~\ref{table:mcalphasys}.

\bibliographystyle{apsrev}

\begin{thebibliography}{999}
\bibitem{ft_footnote}
\bibinfo{note}{\fecal\ is derived for events in which only a single electron, of energy \te, produces Cherenkov light.  The energy of an event that follows from \fecal\ is therefore referred to as \teff\ since it is the effective energy of the event assuming that it resulted from a single fast electron.}
\end{thebibliography}

\end{document}